\documentclass[11pt]{article}
\usepackage{amssymb, amsmath,amsfonts,amsthm,euscript,mathrsfs,  MnSymbol,verbatim, enumerate,multirow,bbding, slashed, multicol,
color,array,tikz, tikz-cd,tikz-3dplot,ytableau,multicol}
  \usepackage[pdftex]{hyperref}

\csname @addtoreset\endcsname{equation}{section}

\usepackage{ifpdf}
\ifpdf
   \pdfoutput=1
   \pdftrue
  \usepackage[pdftex]{hyperref}
\else
   \pdffalse
   \usepackage{cite}
\fi

\begin{document}

\unitlength = .8mm

\begin{titlepage}
~\\
\begin{center}
\baselineskip=14pt{\LARGE
 Singularities and Gauge Theory Phases  \\
}
\vspace{2 cm}
{\large  Mboyo Esole$^{\spadesuit,\heartsuit}$,  Shu-Heng Shao$^{\heartsuit}$,  and Shing-Tung Yau$^{\spadesuit,\clubsuit}$ } \\
\vspace{1 cm}
${}^\spadesuit$Department of Mathematics, \ Harvard University, Cambridge, MA 02138, U.S.A.\\
${}^\heartsuit$Jefferson Physical Laboratory, Harvard University, Cambridge, MA 02138, U.S.A.\\
${}^\clubsuit$Taida Institute for Mathematical Science, National Taiwan University, 
 Taipei, 
Taiwan.
\end{center}

\begin{center}
\end{center}

\abstract{Motivated by M-theory compactification on elliptic Calabi-Yau threefolds, we present a correspondence between networks of small resolutions for singular elliptic fibrations and Coulomb branches of five-dimensional $\mathcal{N}=1$ gauge theories. While resolutions correspond to subchambers of the Coulomb branch, partial resolutions correspond to higher codimension loci at which the Coulomb branch intersects the Coulomb-Higgs   branches. 
Flops between different resolutions are identified with reflections on the Coulomb branch. Physics aside, this correspondence provides an interesting link between elliptic fibrations and representation theory.}

\vfill
Email:{\tt    \{esole , yau\} at  math.harvard.edu, shshao  at  physics.harvard.edu
}
\end{titlepage}

\eject

\tableofcontents

\newpage

\section{Introduction}
\ytableausetup
{mathmode, boxsize=0.6em}

  M-theory compactifications have always been a rich setup for  exploring the  interplay between  gauge theory and geometry.   
Compactification of M-theory on Calabi-Yau threefolds gives rise to  five-dimensional $\mathcal{N}=1$ theories with  vector multiplets and hypermultiplets \cite{Cadavid:1995bk}. 
The vacuum expectation values (vevs) of scalars in the vector multiplets parametrize the Coulomb branch of the theory while those of the scalars in the hypermultiplets parametrize the Higgs branch. There are also mixed branches, which we will call the Coulomb-Higgs branches, where parts of both the vector multiplet scalars and hypermultiplet scalars have nonzero vevs.
Different crepant resolutions of the same singular Calabi-Yau threefold  correspond to different subchambers of the Coulomb branch \cite{Witten:1996qb, Morrison:1996xf, Intriligator:1997pq}.
Here we present a pedagogical and detailed  demonstration of  this correspondence between the network of resolutions with the subchambers of the Coulomb branch of the quantum field theory. 
The analogous story for M-theory compactifications on Calabi-Yau fourfolds has been considered in \cite{Becker:1996gj, Mayr:1996sh,Diaconescu:1998ua,Gukov:1999ya, Haack:2001jz, Grimm:2011fx,Intriligator:2012ue,Hayashi:2013lra,Hayashi:2014kca}.

We would like to emphasize that our correspondence  goes \textit{beyond} the context of M-theory.  
 In particular, the total space does not have to be Calabi-Yau and it can be either a threefold or a fourfold. For this reason, we will also study the codimension three fibers for our resolutions.
On the gauge theory side, the Coulomb branch can be solely described by the representation theory. 
Thus our correspondence, from a pure mathematical point of view, provides an interesting link between small resolutions for singular Weierstrass models and representation theory.

On the geometry side, we focus on  elliptically fibered threefolds {or fourfolds} with a section over the base $B$. Such elliptic fibrations always admit a  (singular) Weierstrass model \cite{Formulaire,MumfordSuominen}. 
We will use the Weierstrass model as our starting point and consider those given by  the ``Tate forms" {with \textit{general} coefficients $a_{i,j}$}. {Specifically, we will consider the Tate form of type I$_N^s$, which has the explicit gauge groups  $SU(N)$ after resolving the singularities \cite{Bershadsky:1996nh,Katz:2011qp}}.       
 The base $B$ is assumed to be nonsingular and of complex dimension two or three.    
 We present a simple derivation for small resolutions of the $SU(N)$ Weierstrass model with $N=2,3,4$ by giving a unified description that can be summarized by  a network of successive blow ups. 
 Flop transitions between different resolutions can be visualized from the ramification of branches in the {network of resolutions}. Some of the flops are induced by the $\mathbb{Z}_2$ automorphism in the Mordell-Weil group of the original Weierstrass model. The same feature was also observed in the case of the $SU(5)$ model \cite{Esole:2011sm}. 
  We also study the fiber enhancements in codimension two and three for each resolution. Over the codimension two loci, we recover the standard enhancements $SU(N)\rightarrow SU(N+1)$ and $SU(N)\rightarrow SO(2N)$ \cite{Katz:1996xe,Morrison:2011mb}.  In the $SU(4)$ model, we find a non-Kodaira type fiber of type I$^{\star+}_0$ in codimension three.

On the gauge theory side, we consider the low energy quantum field theory by compactifying M-theory on an elliptic Calabi-Yau threefold of the $SU(N)$ Weierstrass model type. This theory is the five-dimensional $\mathcal{N}=1$ gauge theory with gauge group $SU(N)$ and hypermultiplets in the fundamental representation (\,\ydiagram{1}\,) and antisymmetric representations (\raisebox{2pt}{\,\ydiagram{1,1}\,}). These are the representations arising from the rank one enhancements $SU(N)\rightarrow SU(N+1)$ and $SU(N)\rightarrow SO(2N)$ of the Weierstrass model in codimension two. This theory has a Coulomb branch in its vacuum moduli space parametrized by the vev of the real scalar field $\phi$ in the vector multiplet. It also has a number of Coulomb-Higgs branches in its vacuum moduli space parametrized by both the vevs of some components of the real scalar $\phi$ and the vevs of some massless matter scalars $Q,\tilde Q$. From a representation-theoretic perspective, we consider the partitioning of the Coulomb branch into several \textit{subchambers} separated by certain codimension one \textit{walls} $W_w$. Each wall $W_w$ is labeled by a weight in the fundamental or antisymmetric representation. 
The hypermultiplet scalars $Q_w,\tilde Q_{\bar w}$ with weight $w$ become massless at the wall $W_w$ and we can activate their vevs to go to the Coulomb-Higgs branch. {These walls are sometimes called the Higgs branch roots in the physics literature where the Coulomb-Higgs branches and the Coulomb branch intersect.}

After collecting the necessary data on the geometry and the gauge theory side, we present a one-to-one correspondence between the {network of resolutions} for the Weierstrass model with the Coulomb branch of the corresponding gauge theory. Starting from the bulk of the Coulomb branch, each subchamber of the Coulomb branch corresponds to a  resolution in the {network}. Next going to codimension one, each wall $W_w$ corresponds to a partial resolution in the {network}.  Intersections of walls are also matched with partial resolutions that appear  in earlier branches of the  {network of resolutions} for the Weierstrass model. In addition, flops between different resolutions are realized as reflections\footnote{These are reflections with respect to certain walls on the Coulomb branch, not to be confused with the Weyl reflections. We will restrict ourselves to the fundamental chamber, so we will not talk about the Weyl reflections in this paper.} with respect to certain walls on the Coulomb branch. The $SU(3)$ and $SU(4)$ cases are demonstrated in Figure \ref{intro3} and \ref{intro}. We end up with the following dictionary between the Coulomb branch (left) and the network of resolutions (right):
\begin{center}
\begin{tabular}{ccl}
Coulomb branch & $\iff$ & {Network of Resolutions} \\
Subchamber & $\iff$  & Resolution \\
Walls and their intersections & $\iff$ & Partial resolutions \\
Moving on to the walls or their intersections 
& $\iff$ &Blowing down \\
Reflection &$\iff$ & Flop
\end{tabular}\\
\end{center}

The vanishing nodes (cycles) in the fiber of the Weierstrass model can also be read off from this correspondence. For example, one of the four nodes in the fiber of the $SU(4)$ model shrinks in the partial resolution $\mathscr{E}_1$ (see Figure \ref{SU4tree}\footnote{In Figure \ref{SU4tree}, the affine node $C_0$ is omitted so only three out of the four affine Dynkin nodes are shown there.} or (\ref{SU4E1})). On the gauge theory side, the corresponding  line $L$ indeed lies on the boundary of the Coulomb branch (see Figure \ref{SU4Coulomb}), where part of the non-abelian gauge symmetry is restored, signaling vanishing nodes on the geometry side.

\textbf{Note added}~
While this work was finalized, a closely related paper \cite{Hayashi:2014kca} appeared on arXiv. The authors introduced a powerful graphical tool, called the box graph, to classify all the subchambers on the Coulomb branch from the representation theory input. We give the box graph descriptions for the resolutions studied in this paper in Sec \ref{sec:networkbox}. We also generalize the box graphs to \textit{partial} resolutions.

\begin{figure}[h!t]

\begin{tabular}{lcr}
\raisebox{-2.cm}{
\begin{tikzpicture}[scale=1.5]
\coordinate (O) at (0,0);
\coordinate (A1)  at (30:2);
\coordinate (A2)  at (-30:2);
\draw[ultra thick, -stealth,black] (O)--(A1);
\draw[ultra thick, -stealth,black] (O)--(A2);
\draw[line width=1mm,blue] (O)--(0:1.75);
\draw[fill=red, opacity=.1] (O)--(A1)--(A2);
\draw[fill=green, opacity=1] (0,0) circle (.05);
\node at (-.2,0) {\color{green!} \scalebox{1.3}{$\mathcal{O}$~}};
\node at (15:1.5) {\color{red} \scalebox{1.3}{$\mathcal{C}^+$}};
\node at (-15:1.5) {\color{red} \scalebox{1.3}{$\mathcal{C}^-$}};
\node at (30:2.2) {\color{black} \scalebox{1.1}{$\mu^1$}  };
\node at (-30:2.2) {\color{black} \scalebox{1.1}{$\mu^2$}};
\node at (0:1.9) {\color{blue} \scalebox{1.3}{~~~~~$W_{w_2}$}};
\end{tikzpicture}}

\phantom{~~~~~~~~~~}

\begin{tikzcd}[column sep=huge, row sep=tiny,scale=1.5]
& & \scalebox{1.4}{\color{red}$ \mathscr{T}^+$}  \arrow[bend left=75, leftrightarrow, dashed, color=red]{dd}[right] {\quad \text{\large flop}}\\{\color{green}
 \scalebox{1.4}{ $\mathscr{E}_0$ }}\arrow[leftarrow]{r} & \scalebox{1.4}{{\color{blue}$ \mathscr{E}_1$} }\arrow[leftarrow]{ur} \arrow[leftarrow]{dr}& \\\
&    &  \scalebox{1.4}{{\color{red}$\mathscr{T}^-$}}
\end{tikzcd} 

\end{tabular}

\caption{Left: The $SU(3)$ Coulomb branch. It is spanned non-negatively by the two vectors $\mu^1$ and $\mu^2$. The Coulomb branch is divided by the line $W_{w_2}$ into two subchambers $\mathcal{C}^\pm$. The line $W_{w_2}$ is the codimension one \textit{wall} where the Coulomb-Higgs branch intersects the Coulomb branch. Right: The network of small resolutions for the $SU(3)$ model. Each letter stands for a (partial) resolution of the original singular Weierstrass model $\mathscr{E}_0$ and each arrow represents a blow up. By going along (against) an arrow, we blow down (up) a variety.
 The identifications between the Coulomb branch with the (partially) resolved varieties are given by $\color{red}\mathscr{T}^\pm = \mathcal{C}^\pm$, $\color{blue}\mathscr{E}_1= W_{w_2}$, and $\color{green}\mathscr{E}_0=\mathcal{O}$. The flop is realized as the reflection with respect to the line (wall) $W_{w_2}$.}\label{intro3}

\begin{tabular}{ccc}

\raisebox{-3cm}{
\scalebox{.85}{
\begin{tikzpicture}[scale=4]
\coordinate (O) at (0,0,0);
\coordinate   (P1) at (0.707107,0,0.707107);
\coordinate (P2) at (-0.707107,0,-0.707107);
\coordinate (P3) at (0 , 1.414,0);
\coordinate (w-) at (0.493561 ,0.427,0.493561);
\coordinate (w+) at (-0.493561 ,0.427,-0.493561 );

\coordinate (C) at (17:1.6);

\draw [black, ultra thick] (P1)--(P2)--(P3)--(P1);
\draw [orange,ultra thick](O)--(w+);
\draw [purple, ultra thick](O)--(w-);
\draw [red,ultra thick](O)--(P3);
\draw[blue,ultra thick,dashed] (C)--(O);
\draw [ultra thick,black] (P1)--(C);
\draw [purple, ultra thick](w-)--(C);
\draw [orange,  thick,dashed] (w+)--(C);
\draw [black,ultra thick] (P3)--(C);
\draw[dashed, black, thick] (P2)--(C);
\draw[fill=orange, opacity=.2] (w+)--(O)--(C);
\draw[fill=red, opacity=.2] (O)--(P3)--(C);
\draw[fill=purple, opacity=.4] (w-)--(O)--(C);

 \draw[ultra thick,-stealth] (.1 , 1.351)--(.05, 1.384);
   \coordinate[label={\color{black} \scalebox{.8}{$\mu^2$}}] (MU2) at ( 0.15 ,1.35);

    \draw[ultra thick,-stealth] (.738, 0,0.53)--(.730, 0,0.59);
   \coordinate[label={\color{black} \scalebox{.8}{$\mu^3$}}] (MU3) at ( 0.937 ,0,0.787);

   \draw[ultra thick,-stealth] (-.66 , 0,   -0.724)--(-.67, 0,   -0.72);
   \coordinate[label={\color{black} \scalebox{.8}{$\mu^3$}}] (MU3) at ( -0.587 ,0,-0.707);

\coordinate[label={\color{blue} \scalebox{.8}{$\ell$}}] (ell) at ( 270 : .15);
\draw[fill=blue, opacity=1] (0 ,0) circle (.02);

\coordinate[label={\color{purple} \scalebox{.8}{$p_-$}}] (p-) at ( 0.43 ,0.38,0.493561);
\draw[fill=purple, opacity=1] (0.493561 ,0.427,0.493561)
 circle (.02);
 
\coordinate[label={\color{orange} \scalebox{.8}{$p_+$}}] (p+) at (-0.553561 ,0.427,-0.493561 );
\draw[fill=orange, opacity=1] (-0.493561 ,0.427,-0.493561 )
 circle (.02);

\coordinate[label={\color{red} \scalebox{.8}{$p_0$}}] (p0) at (0 , 1.44,0);
\draw[fill=red, opacity=1] (0 ,1.414,0)
 circle (.02);

 \coordinate[label={\color{green} \scalebox{.8}{$O$}}]  (c) at (17:1.65);
\draw[fill=green, opacity=1] (17:1.6)
 circle (.02);

 \coordinate[label={\color{purple} \scalebox{.8}{$W^-$}}] (W-) at ( 0.65 ,0.32,0.493561);
 
 \coordinate[label={\color{orange} \scalebox{.8}{$W^+$}}] (W+) at (-0.303561 ,0.265,-0.493561 );
 
 \coordinate[label={\color{red} \scalebox{.8}{$W^0$}}] (W0) at (0.4 , .81,0);

  \coordinate[label={\color{blue} \scalebox{.8}{$L$}}] (L) at ( 0.9 ,0.26,0.493561);

 \draw[fill=gray, opacity=1] (0.707107,0,0.707107)
 circle (.02);
  \coordinate[label={\color{gray} \scalebox{.8}{$\ell_-$}}] (ell-) at ( 0.757 ,0,   .99);

 \draw[fill=gray, opacity=1] (-0.707107,0, -0.707107)
 circle (.02);
  \coordinate[label={\color{gray} \scalebox{.8}{$\ell_+$}}] (ell+) at ( -0.705 ,0,   -.5);

     \coordinate[label={\color{black} \scalebox{1}{$\mathcal{C}^+_-$}}] (C+-) at ( -0.337 ,0,-0.377);
     
      \coordinate[label={\color{black} \scalebox{1}{$\mathcal{C}^+_+$}}] (C++) at (-0.303561 ,0.535,-0.493561 );
      
        \coordinate[label={\color{black} \scalebox{1}{$\mathcal{C}^-_+$}}] (C-+) at (-0.1 ,0.455,-0.493561 );

        \coordinate[label={\color{black} \scalebox{1}{$\mathcal{C}^-_-$}}] (C--) at ( 0.45 ,0.07,0.493561);
 
\end{tikzpicture}
}}

\phantom{}

\begin{tikzcd}[column sep=.7cm, row sep= .07 cm]
& & &\scalebox{1.3}{$ \mathscr{T}^+_- $}   \arrow[bend left=65, leftrightarrow, dashed, color=red]{dddddd}{\text{flop}} \\
& & \scalebox{1.3}{ \color{orange}{ $\mathscr{T}^+$} }\arrow[leftarrow, thick]{ru}\arrow[leftarrow, thick]{rd}& \\
& & & \scalebox{1.3}{$\mathscr{T}^+_+ $}    \arrow[bend left=65, leftrightarrow, dashed, color=red]{dd}{\color{red} \text{flop}}\\
\scalebox{1.3}{\color{green}{$\mathscr{E}_0$}}\arrow[leftarrow,thick]{r}& \scalebox{1.3}{ \color{blue}{$\mathscr{E}_1$ }}\arrow[leftarrow,thick]{r} \arrow[leftarrow, thick]{uur}\arrow[leftarrow,thick]{ddr} &\scalebox{1.3}{ \color{red}{$\mathscr{B}$}} \arrow[leftarrow,thick]{ru}\arrow[leftarrow,thick]{rd}
& \\
& & &\scalebox{1.2}{ $\mathscr{T}^-_+$} \\
& &\scalebox{1.2}{ \color{purple}{$\mathscr{T}^-$}}  \arrow[leftarrow,thick]{ru}\arrow[leftarrow,thick]{rd}  &   \\
& & &\scalebox{1.2}{$ \mathscr{T}^-_-$}
\end{tikzcd}

\end{tabular}
\caption{Left: The $SU(4)$ Coulomb branch. It is the three-dimensional cone spanned non-negatively by the three vectors $\mu^1,\mu^2,\mu^3$. There are three triangles (walls) $W^+$, $W^0$, $W^-$ with vertices $(p_+,\ell, O)$, $(p_0, \ell,O)$, and $(p_-, \ell, O)$, respectively, extending infinitely from the apex $O$. The three walls divide the Coulomb branch into four subchambers $\mathcal{C}^\pm_\pm$. The four subchambers are tetrahedrons in the above figure with vertices $\mathcal{C}^+_-:(\ell_+, \ell, p_+,O)$, $\mathcal{C}^+_+:(p_+, \ell, p_0,O)$, $\mathcal{C}^-_+:(p_0, \ell, p_-,O)$, $\mathcal{C}^-_-:(\ell_-, \ell, p_-,O)$ extending infinitely from the apex $O$. The three triangles intersect at a single line $L:(\ell,O)$. The point $O$ is the origin of the Coulomb branch.   Right: The network of resolutions for the $SU(4)$ Weierstrass model. One needs to blow up three times to completely resolve the singularity, leading to four resolved varieties $\mathscr{T}^\pm_\pm$. The identifications with the Coulomb branch are given by $\mathscr{T}^\pm_\pm = \mathcal{C}^\pm_\pm$, $\color{orange}\mathscr{T}^+= W^+$, $\color{red}\mathscr{B} = W^0$, $\color{purple}\mathscr{T}^-= W^-$, $\color{blue}\mathscr{E}_1=L$, and $\color{green}\mathscr{E}_0=O$. The flops are realized as reflections with respect to the wall $W^0$.}\label{intro}
\end{figure}

\clearpage

\section{Geometry: Small Resolutions of Weierstrass Models}

We first  fix our convention and spell out some basic definitions.

\subsubsection*{ Resolution of singularities}
A {\em resolution of singularities} is a map $f: X'\rightarrow X$ between a nonsingular variety $X'$ and a singular variety $X$ such that the following conditions are satisfied:
\begin{enumerate}
\item $X'$ is a nonsingular variety.
\item $f$ is a surjective birational map. 
\item $f$ is a proper map.
\item $f$  is  an isomorphism away from the singular locus of $X$.
\end{enumerate}

\subsubsection*{ Small birational map, crepant birational  map}

A birational map is  said to be {\em small} when  the exceptional locus has codimension two or higher. 
 A birational map is  said to be {\em crepant} when  $X$ is normal and $f$ preserves the canonical class, that is $f^\star K_X=K_{X'}$.
 A small resolution is always crepant, but a crepant resolution is not necessary small. 
One way to construct a small resolution is to give a sequence of blowups with centers that are non-Cartier  Weil divisors.

When working over $\mathbb{C}$, a morphism $\pi: Y\rightarrow B$ is  {\em  flat } if and only if  the fibers are all equidimensional. We will require our resolutions to be small, crepant, and flat.  
 
\subsubsection*{Notations for  blow ups}
After a blow up, the center of the blowup becomes a Cartier divisor called the exceptional divisor. 
We denote the exceptional divisor by $E$.
Since $E$ is a Cartier divisor, it admits a local equation $e=0$ that is a rational section of $\mathscr{O}(E)$. 
If we blow up $X$ along  an ideal $(g_1,\cdots,  g_n)$ to arrive at a new space $X'$ we use the notation 
$$\displaystyle{
\begin{tikzcd}[column sep=4.5 cm]
X  \arrow[leftarrow]{r}{\displaystyle  (g_1,\cdots, g_n\ | \  \bar g_1:\cdots :\bar g_n)} &  X',
\end{tikzcd}}$$
where $[\bar g_1:\cdots:\bar g_n]$ are projective coordinates of the exceptional locus and 
are related to the generators $(g_1,\cdots, g_n)$ by the condition 
$$
\mathbf{rank}
\begin{pmatrix}
g_1 & \cdots &  g_n\\
\bar g_1& \cdots & \bar g_n
\end{pmatrix}= 1,
$$
which is equivalent to asking all the minors to vanish: 
$$
\bar g_i g_j-\bar g_j g_i=0, \quad i,j=1,\cdots, n.
$$
If we blowup an ideal generated by $g_i$, we express the blowup with the following notation \cite{Lawrie:2012gg}: 
$$\displaystyle{
\begin{tikzcd}[column sep=4cm]
X  \arrow[leftarrow]{r}{\displaystyle  (g_1,\cdots, g_n|\  e)} &  X',
\end{tikzcd}}$$
 where $e$ defines a generator of the  principal ideal corresponding to the exceptional locus of the blowup. 
Such a blowup is induced by the rescaling 
 $$g_k=e \bar g_k, \quad k=1,\cdots, n.$$ 
 We can think of $e$ as a section of $\mathscr{O}(E)$, where $E$ is the exceptional divisor of the blowup of  $( g_1, \cdots, g_n)$. Then $\bar g_k$ are projective coordinates of the projective bundle generated by the blowup.   If $g_i$ is a section of $\mathscr{O}(D_i)$, then $\bar g_i$ is a section of $\mathscr{O}(D_i-E)$.

\par Since we will often need successive blowups, we will denote by  $E_k$ the exceptional divisor of the $k$-th blowup and by $e_k$ a rational section of $\mathscr{O}(E_k)$.

\subsection{Weierstrass models}
A Weierstrass model \cite{ Formulaire,MumfordSuominen,Nakayama.Global, Nakayama.Local} is an elliptic fibration over a base variety $B$, where over  each point on the base, the fiber is an elliptic curve described by a plane cubic algebraic curve with  equation 
\begin{align}
\mathscr{E}_0:~ y^2z + a_1 xyz + a_3 yz^2 -( x^3+a_2 x^2z+a_4 xz^2 +a_6z^3) = 0,\label{tate}
\end{align}
where $[x:y:z]$ are the  homogeneous coordinates  of $\mathbb{P}^2$ and the coefficients $a_i$ are sections of certain line bundles over the base $B$ described below.  The cubic curve is a projective curve of genus one. It has a clear choice of a rational point given by $x=z=0$. 
The tangent to the curve at that point is $z=0$ and it has a triple intersection with the curve.

Globally, a Weierstrass model over a base $B$ requires a choice of a line bundle  $\mathscr{L}\rightarrow B$ so that the equation \eqref{tate} is the zero locus of a section of 
the line bundle 
 \begin{align}
\mathscr{O}(3) \otimes \pi^* \mathscr{L}^6
\end{align}
inside the projective bundle 
\begin{align}
\pi:~\mathbb{P}[ \mathscr{O}_B\oplus \mathscr{L}^2 \oplus \mathscr{L}^3]\rightarrow B.
\end{align}
The Weierstrass  model is Calabi-Yau only when $c_1(B)=c_1(\mathscr{L})$ as can be seen by applying the adjunction formula.   
 The homogeneous coordinates $x,y,z$ of the $\mathbb{P}^2$-bundle and the coefficients $a_i$ are sections of the following line bundles:
 $$\left\{
\begin{tabular}{l}
 $z$ is a section of $\mathscr{O}(1)$,\\
$x$ is  a section of $\mathscr{O}(1)\otimes\pi^\star\mathscr{L}^2$,\\
 $y$ is a section of $\mathscr{O}(1)\otimes\pi^\star \mathscr{L}^3$,\\
 $a_i$ is a section of $\pi^\star \mathscr{L}^i$.
\end{tabular}
\right.
$$ 
  In the following we will take the base variety $B$ to be a nonsingular algebraic variety of complex dimension two or three. 
Some comments on our notation:
\begin{itemize}
\item
We use the classical convention for  the projectivization $\pi: \mathbb{P}(\mathscr{E})\rightarrow B$ of a locally free  sheaf $\mathscr{E}$ over $B$: the fibers of $\mathbb{P}(\mathscr{E})$ are the lines of  $\mathscr{E}$ passing through the origin and not the  hyperplanes\footnote{ The  convention we  use for projective bundles  is the opposite of the convention used in Hartshorne but matches the convention used in most papers in F-theory, the conventions of  Fulton's  book on  intersection theory, and in the (coming) book of Eisenbud and  Harris on intersection theory.}. 
\item
We denote the tautological line bundle of the projective bundle $\mathbb{P}(\mathscr{E})$ by $\mathscr{O}_{\mathbb{P}(\mathscr{E})}(-1)$. 
Its dual is the canonical line bundle  $\mathscr{O}_{\mathbb{P}(\mathscr{E})}(1)$. 
 When the context is clear, we will  abuse the notation and write    $\mathscr{O}(-1)$ and $\mathscr{O}(1)$ respectively for  $\mathscr{O}_{\mathbb{P}(\mathscr{E})}(-1)$ and  $\mathscr{O}_{\mathbb{P}(\mathscr{E})}(1)$.  
We also write  $\mathscr{O}(-n)$ (for $n>0$) for the $n$th tensor product of $\mathscr{O}(-1)$. Its dual is $\mathscr{O}(n)$, the $n$th tensor product of   $\mathscr{O}(1)$. 
\end{itemize}

\paragraph{Mordell-Weil group}

A Weierstrass model is a true elliptic fibration in the sense that the generic fiber is a genus one curve endowed with a choice of a rational point. As we move over the base, that rational point becomes a  section of the fibration. Here the section is given by the point $x=z=0$  on every fiber. The Mordell-Weil group of the elliptic fibration is the group of sections of the elliptic fibration. For a Weierstrass model, we take its origin to be the section $x=z=0$. 
 Given a point on the base $B$ in the Weierstrass model, the opposite of a point $[x:y:z]$ under Mordell-Weil group is $[x:-y-a_1 x-a_3 z:z]$. 
 This defines a fiberwise $\mathbb{Z}_2$ automorphism of $\mathscr{E}_0$,
 \begin{align}
 \iota:~ &\mathscr{E}_0\rightarrow \mathscr{E}_0 :\quad [x:y:z]\mapsto [x:-y-a_1 x-a_3 z:z]. \label{MW}
 \end{align}
 If the Weierstrass model is singular, after a resolution $\iota$ is not necessarily an  automorphism of the resolved space. However, the mapping it induces, which will be called the \textit{inverse action},
  can map a resolution of $\mathscr{E}_0$  to another one and can even be a flop transition.

\paragraph{Singular fibers and Tate forms}

An elliptic curve given by a Weierstrass equation is singular if and only if its discriminant  $\Delta$ is zero.  If a  Weierstrass equation is defined over $k$ and  let  $\bar{k}$ be the algebraic closure of $k$, then two nonsingular elliptic curves are isomorphic over $\bar{k}$ if and only if they have the same $j$-invariant.  We can write the discriminant and the $j$-invariant in terms of variables $(b_2,b_4,b_6)$ or $(c_4,c_6)$ which are defined as follows \cite{Formulaire, Miranda.Lecture}:
\begin{align}\label{eq.formulaire}
b_2 &= a_1^2+ 4 a_2,\\
b_4 &= a_1 a_3 + 2 a_4 ,\\
b_6  &= a_3^2 + 4 a_6 , \\
b_8  &=b_2 a_6 -a_1 a_3 a_4 + a_2 a_3^2-a_4^2,\\
c_4 &= b_2^2 -24 b_4, \\
c_6 &= -b_2^3+ 36 b_2 b_4 -216 b_6,\\
\Delta &= -b_2^2 b_8 -8 b_4^3 -27 b_6^2 + 9 b_2 b_4 b_6=\frac{1}{1728} (c_4^3-c_6^2), \\
  j &=\frac{c_4^3}{\Delta}.
\end{align}
These quantities are related by the following relations:
\begin{equation}
4 b_8 =b_2 b_6 -b_4^2 \quad \text{and}\quad 1728 \Delta=c_4^3 -c_6^2. 
\end{equation}

 A nonsingular  Weierstrass model only has nodal and cuspidial curves as  singular fibers. In order to have more interesting singular fibers, we have to consider singular Weierstrass models. 
The singularity of an elliptic fibration over divisors of the base are classified by Kodaira and N\'eron \cite{Kodaira,Neron}  and can be predicted by manipulating the coefficients of the Weierstrass equation following Tate's algorithm \cite{Tate}. 
 We can force a given singularity over a hypersurface (a Cartier divisor) cut by an equation: 
\begin{align}
e_0=0
\end{align}
by allowing the coefficients $a_i$ to vanish on $e_0$ with certain multiplicities.  Given the order of $e_0$ for each of the sections $a_i$, the types of singularity are given by Tate's algorithm. If $a_i$ has vanishing order $k$, we will write
\begin{align}
a_i = a_{i, k } e_0^k.
\end{align}
(If $k=0$ we will simply write $a_{i,k}$ as $a_i$.) 

In this paper we will consider the type I$_N^s$ Weierstrass model corresponding to gauge group $SU(N)$. 
For $N$ being even $N=2n$ or odd $N=2n+1$, the vanishing orders for I$_{2n}^s:SU(2n)$ and I$_{2n+1}^s:SU(2n+1)$ are \cite{Bershadsky:1996nh,Katz:2011qp}:
\begin{align}
SU(2n):~&a_1 = a_1 ,~a_2 = a_{2,1} e_0,~a_{3} = a_{3,n }e_0^n,~a_4 = a_{4,n}e_0^n,~a_6 = a_{6,2n} e_0^{2n},\\
SU(2n+1):~&a_1 = a_1 ,~a_2 = a_{2,1} e_0,~a_{3} = a_{3,n }e_0^n,~a_4 = a_{4,n+1}e_0^{n+1},~a_6 = a_{6,2n+1} e_0^{2n+1}.
\end{align}
In the case of $SU(2n)$, the  discriminant factorizes as follows
\begin{equation}
\Delta=e_0^{2n} \Big[-a_1^4 P_{2n}+\mathcal{O}(e_0)\Big], \quad P_{2n}:=-a_1a_{3,n} a_{4,n} -a_{4,n}^2 +a_1^2a_{6,2n}\label{P2n}.
\end{equation}
The first component $e_0^{2n}$ is the locus over which we have the fiber of type I$_{2n}^s$ after resolution of singularities. The second component corresponding to the bracket is the locus over which we have the nodal curves I$_1$. 
These two divisors intersect in codimension two in the base along $e_0=a_1=0$ and $e_0=P_{2n}=0$. They intersect further in codimension three along $e_0=a_1=a_{4,n}=0$.  
We see that $e_0=a_1=0$ is on the cuspidal locus $c_4=c_6=0$ while $e_0=P_{2n}=0$ is not. We will see in later sections that there are rank one enhancements in the codimension two loci $e_0=a_1=0$ and $e_0 = P_{2n}=0$.

In the case of $SU(2n+1)$ we have 
\begin{equation}
\Delta=e_0^{2n+1} \Big[-a_1^4 P_{2n+1}+\mathcal{O}(e_0)\Big], \quad  P_{2n+1}:=a_{2,1} a_{3,n}^2 - a_1 a_{3,n} a_{4,n+1} + a_1^2 a_{6,2n+1}.\label{P2n+1}
 \end{equation}
The discriminant again  contains two components. They intersect in codimension two along $e_0=a_1=0$ and $e_0=P_{2n+1}=0$. These two codimension two loci intersect further in codimension three along $e_0=a_1=a_{2,1}=0$ and $e_0=a_1=a_{3,n}=0$. We will see in later sections that there are rank one enhancements in codimension two loci $e_0=a_1=0$ and $e_0 = P_{2n+1}=0$.

\subsection{I$_2^s$: The $SU(2)$ model}

The Tate form for the $SU(2)$ model is \cite{Bershadsky:1996nh,Katz:2011qp}
\begin{align}
\mathscr{E}_0:~
Y:= y^2  + a_1 xy + a_{3,1} e_0 y - ( x^3 + a_{2,1} e_0 x^2  + a_{4,1} e_0x  + a_{6,2} e_0^2 ) = 0,\label{tatesu2}
\end{align}
where we are in the patch $z\neq 0$.
It is easy to see that there is no singularity at $z=0$, so we will henceforth stay in this patch and set $z=1$. In fact, the total space is singular at
\begin{align}
x=y=e_0=0,
\end{align}
where all the partial derivatives of $Y$ vanish.
The singularity is sitting at a point $x=y=0$ over the divisor $e_0=0$ in the base. 
Above the divisor $e_0=0$, the elliptic curve becomes 
\begin{equation}
y^2+ a_1 x y - x^3=0
\end{equation}
which can be written explicitly as a nodal curve: 
\begin{equation}
(y+ \frac{1}{2}a_1 x  )^2-x^2(x+ \frac{1}{4}a_1^2)=0.
\end{equation}
In particular, we see that over $e_0=a_1=0$, the nodal curve becomes a cuspidal curve 
\begin{equation}
(y+\frac{1}{2} x)^2-x^3=0.
\end{equation}
This suggests a possibility of fiber enhancement for the resolved variety over $e_0=a_1=0$.

To resolve the singularity, we will blow up the singular locus $x=y=e_0=0$.

\subsubsection*{Resolution $\mathscr{E}_1:(x,y,e_0|e_1)$}

To blow up the center $(x,y,e_0)$, we introduce a $\mathbb{P}^2$ with homogeneous coordinates $[\bar x: \bar y: \bar e_0]$ such that they are collinear with $x,y,e_0$. That is,
\begin{align}
x= e_1 \bar x,\quad 
y= e_1 \bar y,\quad 
e_0 = e_1 \bar e_0,\label{collinear}
\end{align} 
where $e_1=0$ is the exceptional divisor. Note that $e_1$ is always defined since at least one of $\bar x, \bar y, \bar e_0$ is nonzero. To simplify our notations, we will henceforth drop the bar for the new projective coordinates and forget about the original unbarred coordinates $x,y,e_0$. The collinear condition (\ref{collinear}) is then rewritten as the replacement,
\begin{align}
&(x,y,e_0)\rightarrow (e_1 x,e_1 y,e_1 e_0)\label{replace}.
\end{align}
The blow up will be denoted as
\begin{equation}
\displaystyle{
\begin{tikzcd}[column sep=3cm]
\displaystyle{ \mathscr{E}_0}\arrow[leftarrow]{r}{\displaystyle (x,y,e_0|e_1)} &  \mathscr{E}_1,
\end{tikzcd}}
\end{equation}
with the last entry $e_1$ in parentheses being the ideal of the exceptional divisor.
By doing the replacement (\ref{replace}) in $\mathscr{E}_0$ and factoring out $e_1^2$  (which shows that the singularity has multiplicity two), we arrive at the resolved variety $\mathscr{E}_1$:
\begin{align}
\mathscr{E}_1:~y^2 + a_1 xy + a_{3,1} e_0 y = e_1 x^3 + a_{2,1} e_1 e_0x^2 + a_{4,1} e_0x +a_{6,2} e_0^2.\label{propersu2}
\end{align}
Since the blow up introduces an extra $\mathbb{P}^2$, now the ambient space is parametrized by
\begin{align}
\, [e_1 x:e_1 y:z=1][x:y:e_0].
\end{align}

As one can easily check, $\mathscr{E}_1$ is a nonsingular variety if $\dim_{\mathbb{C}} B \le 3$. Therefore for the $SU(2)$ model, we need only one blow up to fully resolve the singularity.

\paragraph{Fiber enhancements}

In $\mathscr{E}_0$, the fibers are singular over $e_0=0$. Now after the blow up, the divisor $e_0=0$ is replaced by $e_1e_0=0$, over which the fiber is still singular even though the total space is nonsingular. The fiber over the codimension one hypersurface $e_0e_1=0$ consists of the following two nodes, which are both isomorphic to $\mathbb{P}^1$,
\begin{align}
\begin{split}
&C_0:~e_0=y^2+a_1 xy-e_1 x^3=0,\\
&C_1:~e_1=y^2 +a_1xy +a_{3,1}e_0y-a_{4,1} e_0x-a_{6,2} e_0^2=0.
\end{split}
\end{align}
   Over $e_0=0$, the ambient space is described by a  fibration of Hirzebruch surfaces $\mathbb{F}_1$. This can be seen  by introducing the variables $X=e_1 x$ and $Y=e_1 y$. {Indeed we then have the following ambient space parametrized by the projective coordinates 
\begin{equation}
[X:Y:z][x:y:0],
\end{equation}
together with the relation 
\begin{equation}
x Y-y X=0,
\end{equation}
which is the definition for the Hirzebruch surface $\mathbb{F}_1$.}

The equation of $C_0$ is better understood  by putting back the projective variable $z$ into the defining equations for the nodes,
\begin{align}
\begin{split}
&C_0:~e_0=zy^2+a_1 xyz-e_1 x^3=0,\\
&C_1:~e_1=zy^2 +a_1xy z+a_{3,1}e_0yz^2-a_{4,1} e_0xz^2-a_{6,2} e_0^2z^3=0.
\end{split}
\end{align}
We see that the equation for $C_0$ fixes the value of $e_1$ and hence fixes to a point in the first $\mathbb{P}^2$ in the ambient space. It follows that $C_0$ is parametrized by $[x:y]$. At $x=0$, even though $e_1$ is not fixed by the equation, the equation implies $z=0$ since $y\neq0$ if $x=0$. Hence the equation also fixes to a point, i.e. $[e_1x=0, e_1y=0,z=1]$, in the first $\mathbb{P}^2$.
In other words, $C_0$ describes a $\mathbb{P}^1$-bundle over the divisor $e_0 e_1=0$ in the base.

Over $e_1=0$, the ambient space is just a $\mathbb{P}^2$-bundle with projective coordinates $[x:y:e_0]$. It follows that 
 $C_1$ is a quadric in $\mathbb{P}^2:[x:y:e_0]$. In particular it means that $C_1$ defines a quadric bundles over the divisor $e_0 e_1=0$. A quadric bundle, in contrast to a $\mathbb{P}^1$-bundle, can have singular fibers. 
 These singular fibers are located at the zero locus of the  discriminant of the quadric as we will see later.  All together, the nodes $C_0$ and $C_1$  intersect at two points
\begin{align}
C_0\cap C_1:~[0:0:1][1:0:0]+[0:0:1][1:-a_1:0].
\end{align}
This the I$_2$ fiber in Kodaira's classification. In the gauge theory language, this is interpreted as the affine Dynkin diagram for $SU(2)$.

Now let us move on to some special codimension two loci on $B$ where more interesting fibers appear.
The fiber formed by $C_0$ and $C_1$ can degenerate in two different ways: the two intersection points can coincide so that the fiber becomes a fiber of Kodaira type III, or $C_1$ can degenerate into two lines, giving in this way a fiber of Kodaira type I$_3$. The latter would happen when the discriminant of the quadric describing $C_1$ vanishes. This discriminant is precisely the $P_2$ introduced in \eqref{P2n}.

Over ${e_0e_1=0}$ and
\begin{align}
P_2=-a_1a_{3,1} a_{4,1} -a_{4,1}^2 +a_1^2a_{6,2}=0\label{P2}
\end{align}
 but $a_1,a_{4,1}\neq0$, the quadric $C_1$ splits into two lines,
\begin{align}
y^2 +a_1xy +a_{3,1}e_0y-a_{4,1} e_0x-a_{6,2} e_0^2 = {1\over a_1a_{4,1}} \left[  a_{4,1}y+a_1a_{4,1} x+a_1a_{6,2}e_0\right]\left( a_1 y-a_{4,1} e_0\right)=0.
\end{align}
That is, the node $C_1$ splits as
\begin{align}
C_1\rightarrow &C_1^{(1)}: ~e_1=a_{4,1}y+a_1a_{4,1} x+a_1a_{6,2}e_0=0,\   
&&~~[0:0:1][a_{4,1}x: -a_1 x -{a_1a_{6,2}}e_0:a_{4,1}e_0],\\
&C_1^{(2)}:~e_1=a_1 y-a_{4,1} e_0=0, \
 &&~~[0:0:1][a_1x:{a_{4,1}}e_0:a_1e_0].
\end{align}
Right next to  each node we write their explicit parametrizations. Note $C_1^{(1)}$ and $C_1^{(2)}$ intersect at $a_1^2a_{4.1} x + (a_1^2a_{6,2} +a_{4,1}^2)e_0=0$. Hence the fiber enhances from I$_2$ to the I$_3$ fiber over $P_2=0$ on the divisor $e_0e_1=0$. In the gauge theory language, this is the rank one enhancement from $SU(2)$ to $SU(3)$.

Over $e_0e_1=a_1=0$, the two fibers $C_0$ and $C_1$ meet at a double point, so the fiber enhances from I$_2$ to the type III fiber there.

\begin{figure}[htb]
\begin{center}
\includegraphics[scale=.7]{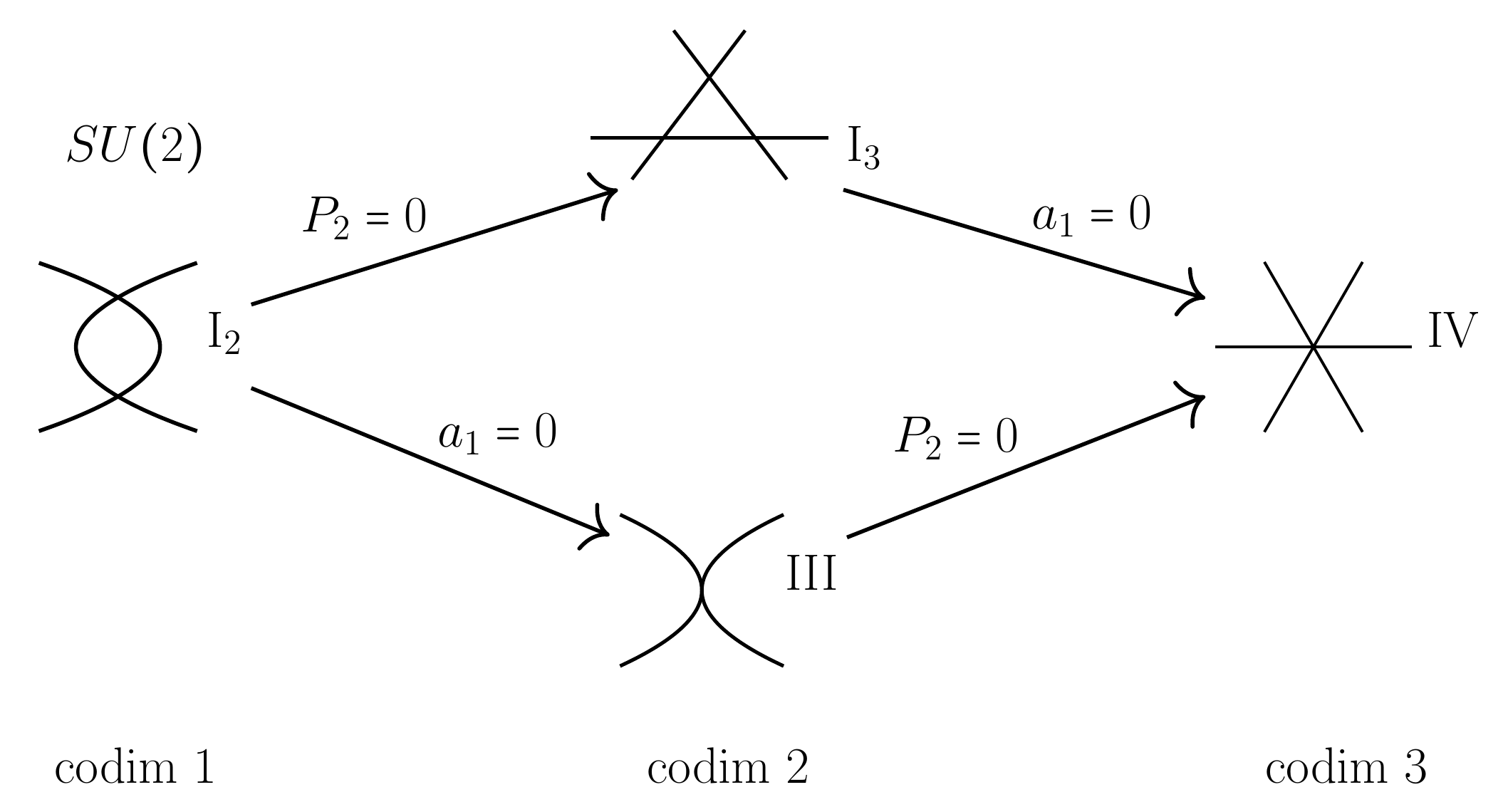}
\end{center}
\caption{The fiber enhancements over the divisor $e_0e_1=0$ for the $SU(2)$ model. Note that the codimension three locus $e_0e_1=a_1=P_2=0$ is the same as $e_0e_1=a_1= a_{4,1}=0$ (see \eqref{P2}). }\label{SU2enhancement}
\end{figure}

Over $e_0e_1 = a_1 =a_{4,1}=0$, the node $C_1$ becomes
\begin{align}
C_1:~ e_1 = y^2 + a_{3,1} e_0 y -a_{6,2} e_0^2=0.
\end{align}
Hence $C_1$ splits into two nodes $C_1^{(1)'}$ and $C_1^{(2)'}$ parametrized by
\begin{align}
C_1 \rightarrow &C_1^{(i)'}:~[0:0:1] [ x: y^{(i)} :e_0^{(i)}] ,~~i=1,2,
\end{align}
where $y^{(i)},e^{(i)}$, $i=1,2$, are the two roots of $y^2 + a_{3,1} e_0 y -a_{6,2} e_0^2=0$. The three nodes $C_0$, $C_1^{(1)'}$, and $C_1^{(2)'}$ meet at a point $[0:0:1][1:0:0]$, so the fiber is of type IV.
The fiber enhancements for $\mathscr{E}_1$ in the $SU(2)$ model are summarized in Table \ref{SU2} and Figure \ref{SU2enhancement}.

\paragraph{{Network of resolutions}}
The  blow up for the $SU(2)$ model is summarized in the following {(degenerate) network},
$$\displaystyle{
\begin{tikzcd}[column sep=3cm]
\displaystyle{ \mathscr{E}_0}\arrow[leftarrow]{r}{\displaystyle (x,y,e_0|e_1)} &  \mathscr{E}_1,
\end{tikzcd}}$$
where the arrow represents the blow up. This network will be the key data we extract from the geometry side. We will see more nontrivial {networks} in the following.

\begin{table}[htb]
\begin{center}
\scalebox{1}{\begin{tabular}{|c|c|c|c|}
\hline
$e_0e_1=0$& $e_0e_1=P_2=0$ &  $e_0e_1=a_1=0$ & $e_0e_1 =a_1=a_{4,1}=0$ \\
\hline
& {\footnotesize $C_1\to C_{1}^{(1)}+C_{1}^{(2)}$ }&&{\footnotesize $C_1\to C_{1}^{(1)'}+C_{1}^{(2)'}$ }
\scalebox{.8}{ \begin{tabular}{l}
\end{tabular} }

 \\
\hline
 I$_2$& I$_3$& III &IV\\
\hline
&&&\\
\scalebox{.95}{
\begin{tikzpicture}
\draw[scale=1,domain=-1.2:1.2,smooth,thick,variable=\t]
plot ({-.6+1.2*(\t)^2},{(\t) });
\draw[scale=1,domain=-1.2:1.2,smooth,thick,variable=\t]
plot ({.6-1.2*(\t)^2},{(\t) });
\end{tikzpicture}
}& 
\scalebox{1.7}{\begin{tikzpicture} [scale=.9] \draw (-1,0.2)--(1,0.2);
\draw  (-.65,-.15)--(.3,1.1);
\draw  (.65,-.15)--(-.3,1.1); \end{tikzpicture}
}
& 
 \scalebox{1}{\begin{tikzpicture}
\draw[scale=1,domain=-1.2:1.2,smooth,thick,variable=\t]
plot ({.9*(\t)^2},{(\t) });
\draw[scale=1,domain=-1.2:1.2,smooth,thick,variable=\t]
plot ({-.9*(\t)^2},{(\t) });
\end{tikzpicture}
}&
\scalebox{.7}{\begin{tikzpicture}
\draw [ thick] (60:-2cm) -- (60:2cm);
\draw [ thick] (60*2:-2cm) -- (60*2:2cm);
\draw [ thick] (60*3:-2cm) -- (60*3:2cm);
\end{tikzpicture}}
\\ 
\hline
\end{tabular}}
\end{center}
\caption{The fiber enhancements for $\mathscr{E}_1$ in the $SU(2)$ model. Here $P_2=-a_1a_{3,1} a_{4,1} -a_{4,1}^2 +a_1^2a_{6,2}=0$.}\label{SU2}
\end{table}

\subsection{I$_3^s$: The $SU(3)$ model}

Let us move on to the $SU(3)$ Weierstrass model \cite{Bershadsky:1996nh,Katz:2011qp},
\begin{align}
y^2  + a_1 xy + a_{3,1}e_0 y = x^3 + a_{2,1} e_0 x^2  + a_{4,2} e_0^2 x + a_{6,3} e_0^3 .
\end{align}
Again the total space is singular at
\begin{align}
x=y=e_0=0.
\end{align}

To resolve the singularity, we proceed as before by blowing up along the ideal $(x,y,e_0)$.

\subsubsection{First blow up and conifold singularity:~ $\mathscr{E}_1:(x,y,e_0|e_1)$}

By blowing up along the ideal $(x,y,e_0)$
\begin{align}
(x,y,e_0) \rightarrow (e_1 x,e_1 y, e_1 e_0),
\end{align}
we obtain the resolved variety $\mathscr{E}_1$,
\begin{align}
\mathscr{E}_1:y(y  + a_1 x + a_{3,1} e_0 ) 
- e_1 \left(
x^3 + a_{2,1} e_0x^2  + a_{4,2} e_0^2 x  + a_{6,3} e_0^3 
\right)=0.\label{3E1}
\end{align}
The ambient space is parametrized by the following projective coordinates
\begin{align}
\,
[e_1 x: e_1 y: z=1][x:y:e_0].
\end{align}

\paragraph{Description of the fiber}

Generally over the divisor ${e_0e_1=0}$, we have three nodes in the fiber,
\begin{align}
\begin{split}
&C_0:~ e_0 =  y^2 + a_1 xy -e_1 x^3 =0,\\
&C_{1}:~e_1 = y+a_1x+a_{3,1}e_0=0,\\
&~~[0:0:1][x: -a_1 x-a_{3,1}e_0 :e_0],\\
&C_{1 }' :~ e_1= y= 0,\\
&~~[0:0:1][x:0:e_0].\label{affineSU3}
\end{split}
\end{align}
Below each node is its explicit parametrization. For example, $C_1'$ is parametrized by $x,e_0$ in the last $\mathbb{P}^2:[x:0:e_0]$. The three nodes intersect pairwise at three different points and we identify them as the I$_3$ fiber.
It should be emphasized at this point that even though $\mathscr{E}_1$ is still singular as we will see shortly, we already obtain the full affine Dynkin diagram for $SU(3)$. If we consider the K\"ahler moduli here, while the size of the affine node $C_0$ is set by the size of the original $\mathbb{P}^2$ for the projective bundle, the sizes of $C_1$ and $C_1'$ are not independent and are controlled by the size of the $\mathbb{P}^2$ we introduced to perform the first blow up. 
This can be seen by noting that $C_1$ and $C_1'$ are both complex lines inside the new $\mathbb{P}^2:[x:y:e_0]$ and scale uniformly with the new $\mathbb{P}^2$. It is only after the second blow up that the two nodes $C_1$ and $C_1'$ acquire independent K\"ahler parameters controlling their sizes. This is quite in contrast with the usual blow up of a complex surface with $A-D-E$ singularity, where we obtain new nodes at each step of the blow up.

\paragraph{Conifold singularity}

In contrast to the $SU(2)$ model, the variety $\mathscr{E}_1$ after the first blow up is still singular. To see this, we define
\begin{align}
&s(x,y,e_0)= y + a_1x+a_{3,1}e_0,\\
&Q(x,e_0)=x^3+a_{2,1}e_0x^2 + a_{4,2} e_0^2x+a_{6,3}e_0^3,
\end{align}
and rewrite $\mathscr{E}_1$ (\ref{3E1}) as
 \begin{align}
\mathscr{E}_1:y s = e_1 Q.\label{proper1}
\end{align}
In this expression, it is clear there is a conifold singularity at 
\begin{align}
y=e_1 = s = Q=0.\label{conifold3}
\end{align}
Note that $y=e_1=s=0$ is precisely the intersection $C_{1}'\cap C_{1}$. Over a general point on the base $B$, $Q$ can not be zero at the same time as $y=e_1=s=0$. Only at
\begin{align}
P_3 = a_{3,1}^3 - a_1 a_{2,1} a_{3,1}^2 + a_1^2a_{3,1} a_{4,2} - a_1^3 a_{6,3}=0\label{P3}
\end{align} 
is there a solution to (\ref{conifold3}). Hence the conifold singularity only occurs at a codimension two locus on the base $B$ defined by $P_3= e_1e_0=0$. Note that $P_3$ was first introduced as the leading term in the second component of the discriminant \eqref{P2n+1}. As we will see in Appendix \ref{app1}, after the second blow up, there will be a fiber enhancement at this codimension two locus.

\subsubsection{Second blow ups and flop: $\mathscr{T}^+:(y,e_1|e_2)$ and $\mathscr{T}^-:(s,e_1|e_2)$}

Next we wish to blow up the conifold singularity of $y s = e_1 Q$. As usual for the conifold singularity, there are two possible blow ups one can do: we can either blow up along the ideal $(y,e_1)$ or the ideal $(s,e_1)$. The two resolutions $\mathscr{T}^+$ and $\mathscr{T}^-$ are related by the flop exchanging $y$ with $-s = -y -a_1 x -a_{3,1}e_0$, which is the inverse action (\ref{MW}) induced by the $\mathbb{Z}_2$ automorphism in the Mordell-Weil group. 
Geometrically, $\mathscr{T}^+$ and $\mathscr{T}^-$ are obtained by blowing up along $C_1':~e_1=y=0$ and $C_1:~e_1 = s =0$, respectively. Hence the flop exchanges the two nodes $C_1'$ and $C_1$ in the $SU(3)$ Dynkin diagram.  Also, since we blow up along divisors $C_1$ or $C_1'$, the resolutions are guaranteed to be crepant.
One can check that after the second blow up, $\mathscr{T}^\pm$ are both nonsingular varieties for $\dim_{\mathbb{C}}B\le 3$. We therefore arrive at the network of resolutions for the $SU(3)$ model in Figure \ref{tree3}.

\begin{figure}[htb]
\begin{center}
\includegraphics[scale=.75]{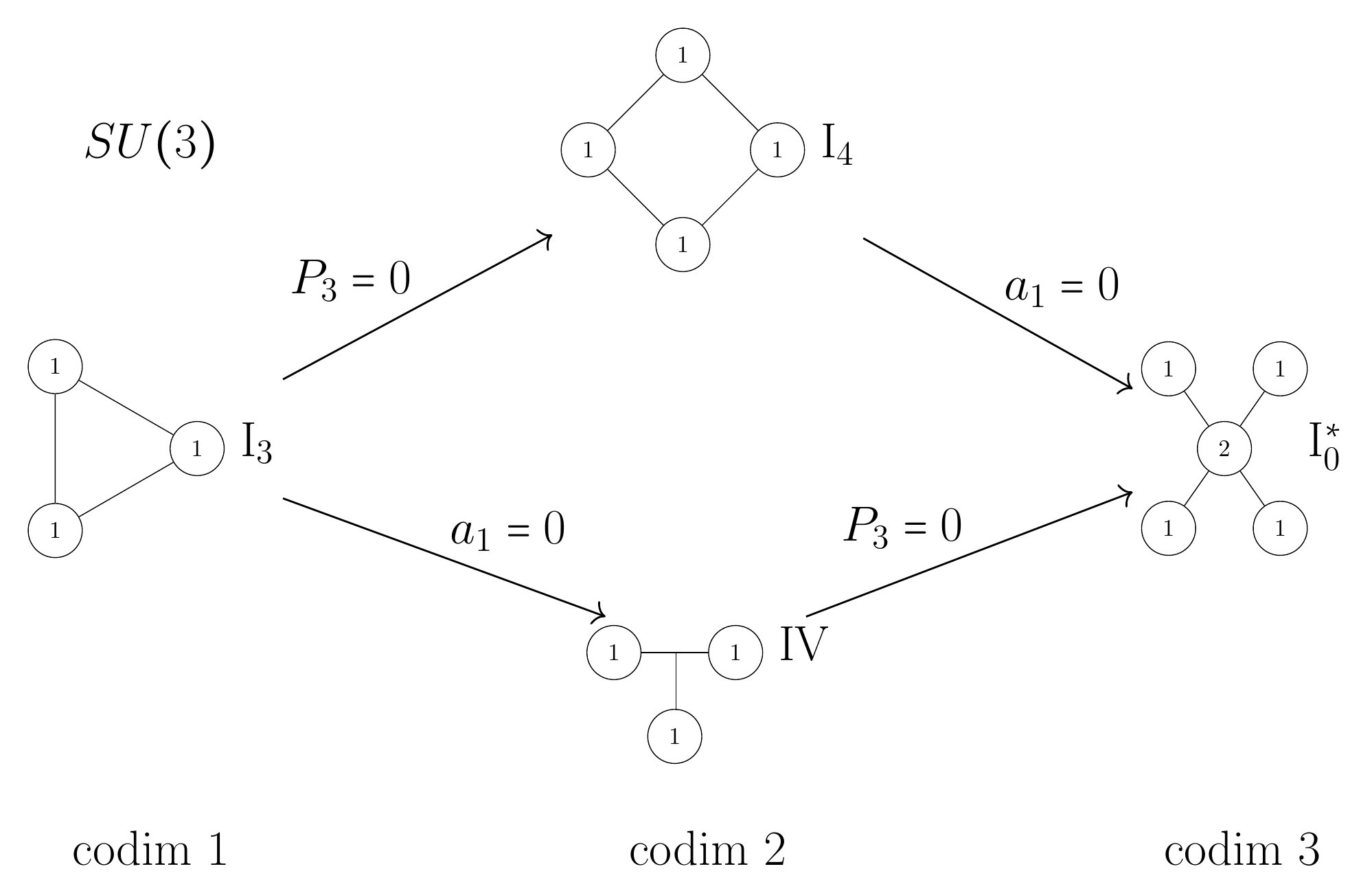}
\end{center}
\caption{The fiber enhancements over the divisor $e_0e_1e_2=0$ for the resolved $SU(3)$ model $\mathscr{T}^\pm$. The fiber enhancements are the same for both resolutions up to relabeling. The trivalent point for IV means that the three nodes meet at the same point. Note that the codimension three locus $e_0e_1e_2=a_1=P_3=0$ is the same as $e_0e_1e_2=a_1= a_{3,1}=0$ (see \eqref{P3}). Here $P_3 = a_{3,1}^3 - a_1 a_{2,1} a_{3,1}^2 + a_1^2a_{3,1} a_{4,2} - a_1^3 a_{6,3}=0$. }\label{SU3enhancement}
\end{figure}

A detailed analysis of $\mathscr{T}^+$ and $\mathscr{T}^-$ can be found in Appendix \ref{app1}. The fiber enhancements over codimension two and three loci are summarized in Figure \ref{SU3enhancement}.

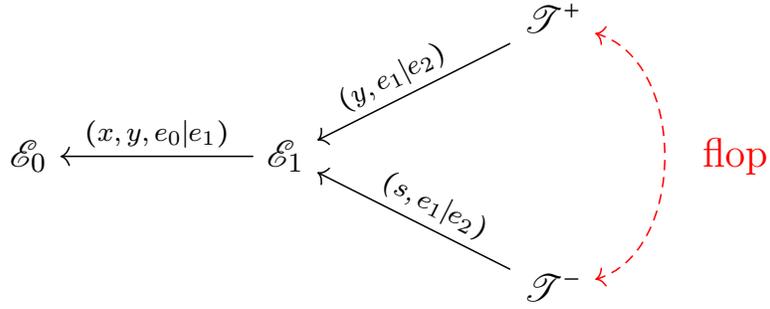
\begin{figure}[thb]
\begin{center}
\scalebox{1.3}{
\begin{tikzcd}[column sep=huge,  ampersand replacement=\&]
\& \& \mathscr{T}^+  \arrow[bend left=75, leftrightarrow, dashed, color=red]{dd}[right] {\quad \text{\large flop}} \\
\mathscr{E}_0\arrow[leftarrow]{r}{\displaystyle (x,y,e_0|e_1)} \& \mathscr{E}_1 \arrow[leftarrow,sloped, near end ]{ur}{\displaystyle(y,e_1|e_2)} \arrow[leftarrow, near start, sloped]{dr}{\displaystyle (s,e_1|e_2)} \& \\\
\&    \& \mathscr{T}^-
\end{tikzcd}}
\end{center}
\caption{The network of resolutions for the $SU(3)$ model. Each letter stands for a (partial) resolution and each arrow represents a blow up. Starting from $\mathscr{E}_0$, there is a unique (crepant) blow up $(x,y,e_0|e_1)$ to go to the partial resolution $\mathscr{E}_1$. For the second blow ups, there are two inequivalent blow ups leading to $\mathscr{T}^\pm$. The two resolutions $\mathscr{T}^\pm$ are related by a flop induced by the $\mathbb{Z}_2$ automorphism \eqref{MW} in the Mordel-Weil group. Here $s=y+a_1x+a_{3,1}e_0$.}\label{tree3}
\end{figure}

\clearpage

\subsection{I$_4^s$: The $SU(4)$ model}

\begin{figure}[htb]
\begin{center}
\includegraphics[scale=.6]{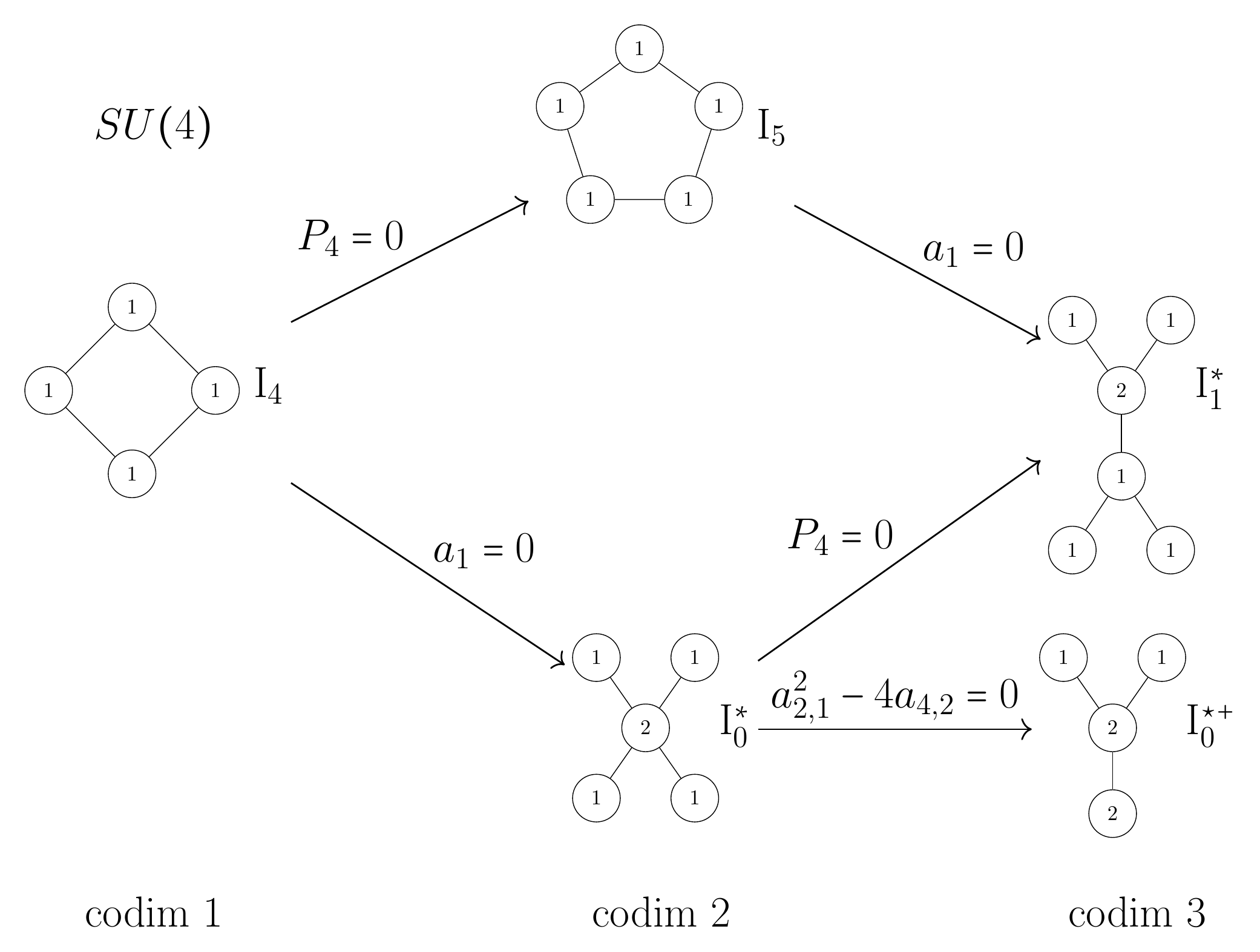}
\end{center}\caption{  {The fiber enhancements over the divisor $e_0e_1e_2e_3=0$ for the resolved $SU(4)$ model $\mathscr{T}^\pm_\pm$. Even though the splittings of the nodes are different for the four resolutions, the fiber enhancements are the same. See Table \ref{T++} and \ref{T+-} for the splittings of the nodes. Over  
$e_0e_1e_2e_3=a_1=a_{2,1}^2 -4a_{4,2}=0$, we found a non-Kodaira type fiber I$_0^{*+}$, which is a degeneration of I$_0^*$. Note that the codimension three locus $e_0e_1e_2e_3=a_1=P_4=0$ is the same as $e_0e_1e_2e_3=a_1= a_{4,2}=0$ (see \eqref{P4}). Here $P_4=
-a_{4,2}^2 -a_1 a_{3,2}a_{4,2}+a_1^2a_{6,4}=0$.}}\label{SU4enhancement}
\end{figure}

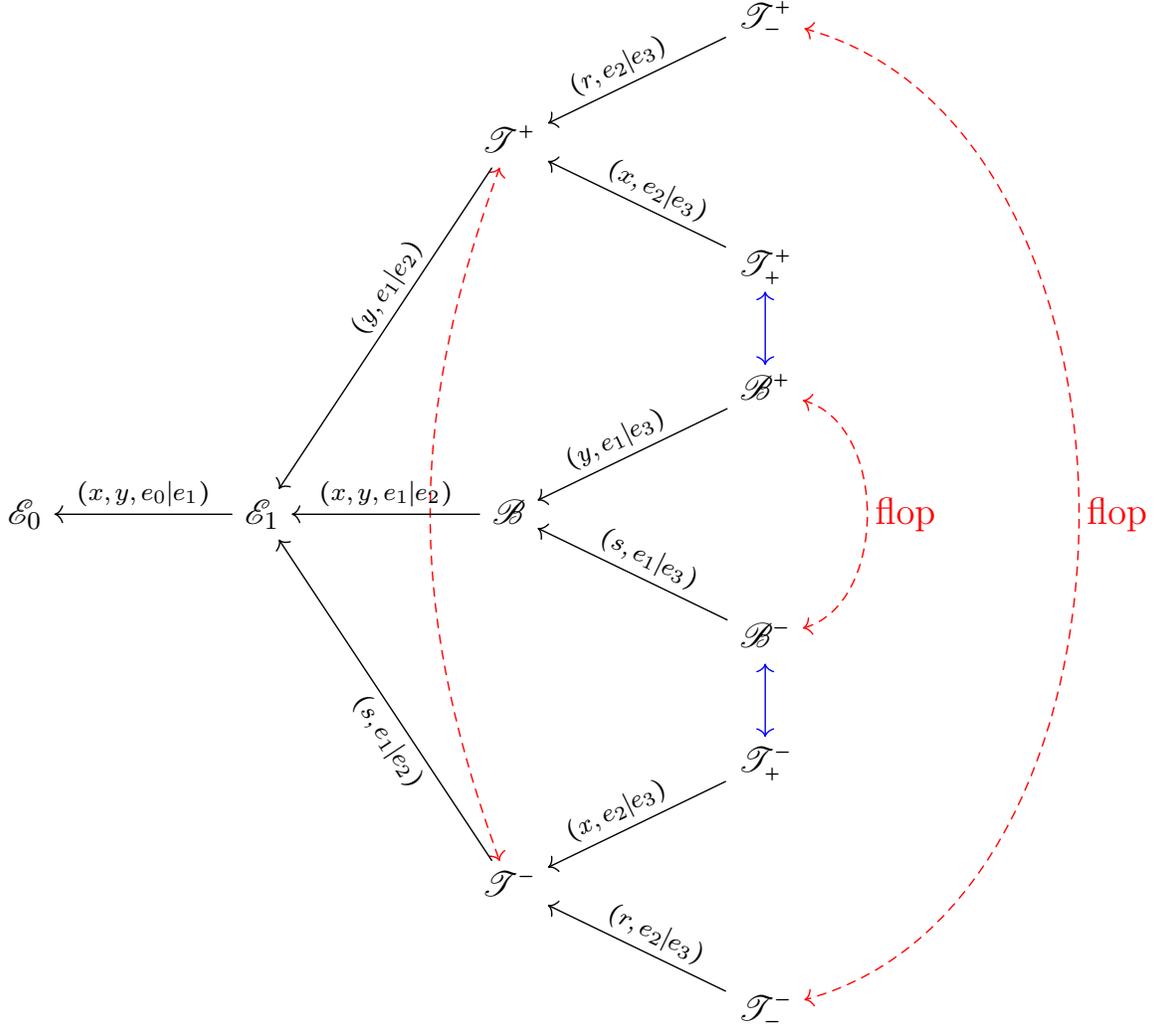
\begin{figure}
\begin{center}
\scalebox{1.2}{
\begin{tikzcd}[column sep=huge, ampersand replacement=\&]
\& \& \& \mathscr{T}^+_-   \arrow[bend left=75, leftrightarrow, dashed, color=red]{dddddddd}{\color{red} \text{\large flop}}  \\
\& \& \mathscr{T}^+  \arrow[bend right=20, leftrightarrow, dashed, color=red]{dddddd} \arrow[leftarrow, ,sloped, near end]{ru}{\displaystyle (r,e_2|e_3)} \arrow[leftarrow, sloped , near start ]{rd}{\displaystyle (x,e_2|e_3)} \& \\
\& \& \& \mathscr{T}^+_+  \arrow[leftrightarrow, color=blue]{d}\\
\& \& \& \mathscr{B}^+  \arrow[bend left=75, leftrightarrow, dashed, color=red]{dd}{\color{red}\text{\large flop}}
\\
\mathscr{E}_0\arrow[leftarrow]{r}{\displaystyle (x,y,e_0|e_1)} \& \mathscr{E}_1 \arrow[leftarrow]{r}{\displaystyle(x,y,e_1|e_2)}  \arrow[leftarrow, ,sloped, near end]{uuur}{\displaystyle (y,e_1|e_2)} \arrow[leftarrow, ,sloped, near end]{dddr}[swap]{\displaystyle (s,e_1|e_2)} \&\mathscr{B} \arrow[leftarrow, ,sloped, near end]{ru}{\displaystyle (y,e_1|e_3)} \arrow[leftarrow, ,sloped, near start]{rd}{\displaystyle (s,e_1|e_3)} \& \\
\& \& \& \mathscr{B}^- \arrow[leftrightarrow, color=blue]{d} \\
\& \& \& \mathscr{T}^-_+ \\
\& \& \mathscr{T}^-  \arrow[leftarrow,sloped, near end]{ru}{\displaystyle (x,e_2|e_3)} \arrow[leftarrow,sloped, near start]{rd}{\displaystyle (r,e_2|e_3)}  \&   \\
\& \& \& \mathscr{T}^-_-
\end{tikzcd}}
\end{center}
\caption{The network of resolutions for the $SU(4)$ model. Each letter stands for a (partial) resolution and each arrow stands for a blow up. For the $SU(4)$ model, one needs to blow up three time to completely resolve the singularity. The red lines are the flops induced by the $\mathbb{Z}_2$ automorphism (\ref{MW}) in the Mordell-Weil group. The blue line indicates that the two resolutions are identified as a single one (see Section \ref{isomorphism}). After the identifications, we end up with four resolutions $\mathscr{T}^\pm_\pm$ for the $SU(4)$ model. Later on we will identify this network of resolutions with the $SU(4)$ Coulomb branch (see Figure \ref{SU4Coulomb}).}\label{tree4} \end{figure}

For the $SU(4)$ model we need three blow ups to completely resolve the singularity for  a base of dimension two or three. 
The details  can be found in Appendix \ref{app2}.
The $SU(4)$ model is \cite{Bershadsky:1996nh,Katz:2011qp}
\begin{align}
\mathscr{E}_0:~y^2  + a_1 xy + a_{3,2} e_0^2 y = x^3 + a_{2,1} e_0x^2  + a_{4,2} e_0^2 x  + a_{6,4} e_0^4.
\end{align}
After three blow ups, we end up with four  resolutions $\mathscr{T}^\pm_\pm$. The network of resolutions
 is given in Figure \ref{tree4}. The fiber enhancements in codimension two and three loci are summarized in Figure \ref{SU4enhancement}.

In the $SU(4)$ network of resolutions, the red lines are the flops induced by the $\mathbb{Z}_2$ automorphism (\ref{MW}) of $\mathscr{E}_0$. The blue line indicates that the two varieties are isomorphic to each other and will therefore be identified as one resolution. See section \ref{isomorphism} for a detailed discussion. The fiber enhancements are summarized in Appendix \ref{app2}, Tables \ref{T++} and \ref{T+-}.

\clearpage

\section{Gauge Theory: Coulomb Branches of 5d $\mathcal{N}=1$ Gauge Theories}

In this section we discuss the 5d $\mathcal{N}=1$ theories arising from M-theory compactification on elliptic Calabi-Yau threefolds of the type I$_N^s$, corresponding to gauge group $SU(N)$. We will not discuss the explicit prepotential or the 5d Chern-Simons term, but focus on the  Coulomb branch from a purely representation theory aspect. We will mainly follow \cite{Intriligator:1997pq}.

Consider 5d $\mathcal{N}=1$ gauge theory with a vector multiplet in gauge group $G=SU(N)$ 
and massless hypermultiplets in representations $R_a$, where $a$ labels different representations. We will restrict ourselves to the case where the $R_a$ are the fundamental representation (\,\ydiagram{1}\,), the two-index antisymmetric representation (\raisebox{2pt}{\,\ydiagram{1,1}\,}), and their conjugates. These are the representation arising from the rank one enhancements $SU(N)\rightarrow SU(N+1)$ and $SU(N) \rightarrow SO(2N)$, respectively. The numbers for each such representations will be assumed to be nonzero but otherwise unconstrained.

In the 5d $SU(N)$ $\mathcal{N}=1$ vector multiplet, 
there is a real adjoint scalar $\phi$ parametrizing the Coulomb branch. 
Modding out the residual gauge symmetry, the Coulomb branch is described by the fundamental chamber, i.e. the dual of Cartan subalgebra modulo Weyl reflections, which we will denote by $\mathcal{C}$,
\begin{align*}
\mathcal{C}:& =\text{fundmanetal chamber} =\big\{ \phi ~\big|~ \phi \cdot \alpha^i \ge 0,~ \alpha^i=\text{simple root}\big\}\\
&= \text{Coulomb branch}.
\end{align*}
 In $\mathcal{C}$ we will associate to each weight $w$ in the representation $R_a$ a codimension one \textit{wall} $W_{w}\subset \mathcal{C}$ defined by\footnote{The wall defined here is a codimension one hypersurface on the Coulomb branch where some matter scalars become massless. This is not to be confused with the boundary of the Coulomb branch where some of the $W$-bosons become massless.}
\begin{align}
\text{wall}:~~W_{w } : = \Big\{ \phi \in \mathcal{C} ~\Big| ~\phi\cdot w =0\Big\}\subset \mathcal{C}.\label{wall}
\end{align}

On the Coulomb branch where $\phi$ acquires a vev, the 5d supersymmetry induces the following mass terms to the hypermultiplet 
\begin{align}
(\phi\cdot  w)^2 |Q_w|^2 + (\phi \cdot \bar w)^2 |\tilde Q_{\bar w}|^2 
\end{align}
where $Q_w$ and $\tilde Q_{\bar w}$ are the two complex scalars in the hypermultiplet with weights $w$ and $\bar w$ (the conjugate of $w$). Therefore at the wall $W_w$, the matter scalars $Q_w$ and $\tilde Q_{\bar w}$ become massless and we can activate their vevs to go to the Coulomb-Higgs branch. That is, the codimension one walls are the intersections of the Coulomb and Coulomb-Higgs branches. 

The main object we will study on the gauge theory side is the partitioning of the Coulomb branch $\mathcal{C}$ into several subchambers separated by the walls $W_w$ (\ref{wall}). In the following we will consider three explicit examples.

\subsection{$SU(2)$ with $\bf 2$}

The fundamental chamber in this case is a half line,
\begin{align}
\mathcal{C} = \mathbb{R}_{\ge0}.
\end{align}
The relevant representation is the fundamental representation $\bf 2$ from the rank one enhancement $SU(2)\rightarrow SU(3)$. There are two weights in $\bf 2$. For rank one, $\phi \cdot w$ is a scalar product so it is zero if and only if $\phi$ or $w$ is zero. Since the weights for $\bf 2$ are both nonzero, the walls are just the origin $O$ of the fundamental chamber, $\phi=0$. The $SU(2)$ Coulomb branch is shown in Figure \ref{SU2Coulomb}.

\subsection{$SU(3)$ with $\bf 3$}

Let us denote the simple roots by $\alpha^i$, $i=1,2$, normalized such that $|\alpha^i|^2=2$. Let $\mu^i$ be the fundamental weights such that $\mu^i \cdot \alpha^j = \delta^{ij}$. The fundamental chamber $\mathcal{C}$ is spanned by the two fundamental weights $\mu^i$ with non-negative coefficients,
\begin{align}
\mathcal{C} :~\mathbb{R}_{\ge 0} \,\mu^1 + \mathbb{R}_{\ge 0} \,\mu^2.
\end{align}

The relevant representation here is the fundamental representation $\bf 3$ from the rank one enhancement $SU(3)\rightarrow SU(4)$, with weights
\begin{align}
w_1 = [ 1~0],~w_2 = [-1~1],~w_3=[0-1].
\end{align}
Let 
\begin{align}
\phi = \phi_1 \mu^1 + \phi_2 \mu^2 \in \mathcal{C},~~~\phi_{1,2}\ge 0,
\end{align}
be a general point in the fundamental chamber $\mathcal{C}$. The inner products $\phi \cdot w$ can then be computed as\footnote{Recall that the inner product between fundamental weights is given by the inverse of the Cartan matrix, $\mu^i \cdot \mu^j = (A^{-1})_{ij}$. Here our normalization is $|\alpha^i|^2=2$ and the Cartan matrix is defined by $A_{ij} = 2{\alpha^i \cdot \alpha^j/ |\alpha^i|^2}$. }
\begin{align}
\begin{split}
&\phi \cdot w_1 = {1\over 3} (2\phi_1 +\phi_2),\\
&\phi \cdot w_2 = {1\over 3} (-\phi_1 + \phi_2),\\
&\phi \cdot w_3 = {1\over 3} (-\phi_1 - 2\phi_2).
\end{split}
\end{align}
Since $\phi_{1,2}\ge 0$, the inner products $\phi\cdot w_1$ and $\phi\cdot w_3$ are never zero except at the origin $O$. It follows that these walls $W_{w_1}$, $W_{w_3}$ do not divide the fundamental chamber $\mathcal{C}$. The only nontrivial wall is
\begin{align}
 W_{w_2}: ~\phi_1 = \phi_2,
\end{align}
dividing the $SU(3)$ Coulomb branch $\mathcal{C}$ into two subchambers, which we will call $\mathcal{C}^+$ and $\mathcal{C}^-$. At the wall $W_{w_2}$, some hypermultiplet scalars becomes massless so $W_{w_2}$ is the intersection between the Coulomb and the Coulomb-Higgs branch. The $SU(3)$ Coulomb branch is shown in Figure \ref{SU3Coulomb}.

\subsection{$SU(4)$ with $\bf 4$ and $\bf 6$}

Let us denote the simple roots by $\alpha^i$, $i=1,2,3$ and the fundamental weights by $\mu^i$. The fundamental chamber $\mathcal{C}$ is spanned by $\mu^i$ with non-negative coefficients,
\begin{align}
\mathcal{C} = \mathbb{R}_{\ge 0} \mu^1 + \mathbb{R}_{\ge 0} \mu^2 + \mathbb{R}_{\ge 0}\mu^3.
\end{align}

The relevant representations are $\bf 4$ and $\bf 6$ from the rank one enhancements $SU(4)\rightarrow SU(5)$ and $SU(4)\rightarrow SO(8)$, respectively. Out of the ten weights $w_p^{\bf 4}, ~w_q^{\bf 6},~p=1,\cdots ,4,~q=1,\cdots ,6$, in $\bf 4$ and $\bf 6$, there are two weights $w_2^{\bf 4},w_3^{\bf 4}$ from $\bf 4$ and two weights $w_3^{\bf 6}, w_4^{\bf 6}$ from $\bf 6$ giving vanishing $\phi \cdot w$ in the bulk of the fundamental chamber $\mathcal{C}$. Their Dynkin labels are
\begin{align}
\begin{split}
&w_2^{\bf 4}= [-1~ 1~0],~~w_3^{\bf 4}= [0~-1~~1],\\
&w_3^{\bf 6}  =[-1~ 0~1] ,~~w_4 ^{\bf 6} = [1~~0 ~-1].\label{su4wt}
\end{split}
\end{align}
Note that $w_4^{\bf 6} = -w_3^{\bf 6}$ so they define the same wall.

If we parametrize $\phi$ by
\begin{align}
\phi = \phi_1  \mu^1 + \phi_2 \mu^2 +\phi_3 \mu^3 \in \mathcal{C}
\end{align}
with $\phi_{1,2,3}\ge0$, the four weights (\ref{su4wt}) define the following three nontrivial walls in the fundamental chamber $\mathcal{C}$,
\begin{align}
\begin{split}
& W^+:= W_{w_2^{\bf 4}}:~\phi \cdot w_2^{\bf 4} = {1\over 4} ( -\phi_1 +2\phi_2 + \phi_3)=0,\\
& W^0:=  W_{w_3^{\bf 6}}:~\phi \cdot w_3^{\bf 6} = {1\over 4} ( -\phi_1  + \phi_3) =0 ,\\
& W^-:=  W_{w_3^{\bf 4}}:~\phi \cdot w_3^{\bf 4} = {1\over 4} ( -\phi_1 -2\phi_2 + \phi_3) = 0.
\end{split}
\end{align}
For notational simplicity, we have renamed $W_{w_2^{\bf 4}},~W_{w_3^{\bf 6}},~W_{w_3^{\bf 4}}$ as $W^+,~W^0,~W^-$, respectively. As shown in Figure \ref{SU4Coulomb}, these three walls divide the Coulomb branch $\mathcal{C}$ into four subchambers, $\mathcal{C}^+_-,~\mathcal{C}^+_+,~\mathcal{C}^-_+,~\mathcal{C}^-_-$. 

The three walls intersect at a single line, which we will denote by $L$, 
\begin{align}
L= W^+ \cap W^0\cap W^-:~ \phi_2 = -\phi_1+\phi_3 =0.
\end{align}
Since $\phi_2=0$, $L$ lies on the boundary of the fundamental chamber $\mathcal{C}$. This will be a crucial fact as we study the vanishing nodes of the fiber.

To summarize, the $SU(4)$ Coulomb branch $\mathcal{C}$ is divided by three walls $W^+,~W^0,~W^-$ into four subchambers $\mathcal{C}^+_-,~\mathcal{C}^+_+,~\mathcal{C}^-_+,~\mathcal{C}^-_-$, and the three walls intersect at a line $L$. The $SU(4)$ Coulomb branch is shown in Figure \ref{SU4Coulomb}.

In the next section, we will see the partitioning of the Coulomb branch exactly matches with the topology  of the network of resolutions.

\section{The Correspondence: Networks of Resolutions and Coulomb Branches}

We will now demonstrate the one-to-one correspondence between  resolutions in the network and the subchambers in the Coulomb branch. This correspondence also holds between partial resolutions with walls and their intersections. Furthermore, flops transitions between different resolutions are realized as transitions between different subchambers by reflections with respect to certain walls on the Coulomb branch (not to be confused with the Weyl reflections).  We have the following dictionary between the Coulomb branch (left) and the network of resolutions (right):
\begin{center}
\begin{tabular}{ccl}
Coulomb branch & $\iff$ & Network of Resolutions \\
Subchamber & $\iff$  & Resolution \\
Walls and their intersections & $\iff$ & Partial resolutions \\
 Moving on to the walls or their intersections 
& $\iff$ &Blowing down\\
Reflection &$\iff$ & Flop
\end{tabular}
\end{center}

In the following we will study three explicit examples to demonstrate this correspondence.

\subsection{$SU(2)$}

\begin{figure}[h]
\begin{center}
\begin{tikzpicture}[scale=1.7]
\draw[line width=1mm, color=blue,-stealth] (0,0)--(0:1.75);
\draw[fill=green, opacity=1] (0,0) circle (.05);
\node at (-.2,0) {\color{green!} \scalebox{1.5}{$O$}};
\node at (0:1.9) {\color{blue}  \scalebox{1.5}{$\mathcal{C}$}};
\end{tikzpicture}
\end{center}
\caption{The $SU(2)$ Coulomb branch.}\label{SU2Coulomb}
\begin{center}
\begin{tikzcd}[column sep=huge,row sep=.4cm]
{\color{green} \scalebox{1.6} {  $\mathscr{E}_0 $ } } \scalebox{1.2} {\arrow[leftarrow, thick]{r}} & {\color{blue} \scalebox{1.6} {  $\mathscr{E}_1  $ } }\\
\raisebox{0.1cm}{\scalebox{.3}{\begin{tikzpicture}

 \draw[scale=2,domain=-1.2:1.2,smooth,  thick,variable=\t]
plot ({(\t)^2-1},{(\t)^3-\t});
\end{tikzpicture}}}
 
&\scalebox{1.4}{$\bigcirc$}\\
\scalebox{1.1}{0d}&\scalebox{1.1}{1d}
\end{tikzcd}
\end{center}
\caption{The $SU(2)$ network of resolutions. The singular fiber of the resolution is drawn in the second row where the affine node $C_0$ is always ignored. In this case we have the nodal curve for $\mathscr{E}_0$ and the (affine) $SU(2)$ Dynkin diagram as the fiber for $\mathscr{E}_1$. The identifications with the Coulomb branch are given by $\color{blue}\mathscr{E}_1 = \mathcal{C}$ and $\color{green}\mathscr{E}_0=O$.}
\end{figure}

In the $SU(2)$ model, we only need to do one blow up and this is consistent with the fact that the $SU(2)$ Coulomb branch has real dimension one. The Coulomb branch $\mathcal{C}$ is a half line and there is no nontrivial wall dividing it. This corresponds to the fact that the crepant resolution $\mathscr{E}_1\rightarrow \mathscr{E}_0$ is unique. We hence have the following identification,
\begin{align}
\mathcal{C} =\mathscr{E}_1.
\end{align}
By blowing down $\mathscr{E}_1$ to the singular Weierstrass model $\mathscr{E}_0$,
correspondingly on the gauge theory side we move from the bulk of the Coulomb branch $\mathcal{C}$ to the origin $O$. Hence the origin $O$ is identified with $\mathscr{E}_0$,
\begin{align}
O = \mathscr{E}_0.
\end{align} 
We summarize the identifications for the $SU(2)$ model in Table \ref{id2}.
\begin{table}[h!]
\begin{center}
$SU(2)$:~~~~\begin{tabular}{|c|c|c|}
\hline
&Network of Resolutions & Coulomb Branch\\
\hline
1d &$\mathscr{E}_1$&$\mathcal{C}$\\
\hline
0d & $\mathscr{E}_0$&$O$\\
\hline
\end{tabular}
\end{center}
\caption{identifications between the resolution and the (in this case only one) subchamber on the Coulomb branch of the $SU(2)$ model.}\label{id2}
\end{table}

\subsection{$SU(3)$}

In the $SU(3)$ model, we need to do two blow ups and the Coulomb branch is indeed of real dimension two. While the first blow up $\mathscr{E}_1\rightarrow \mathscr{E}_0$ is unique, there are two options for the second blow up leading to $\mathscr{T}^+$ and $\mathscr{T}^-$. On the gauge theory side, there are two subchambers $\mathcal{C}^+$ and $\mathcal{C}^-$ on the Coulomb branch and these are thus identified with the two resolutions,
\begin{align}
\mathcal{C}^\pm = \mathscr{T}^\pm.
\end{align}
 This identification is consistent with the intersection of the two subchambers $\mathcal{C}^\pm$ in the following sense. On the gauge theory side, the two subchambers $\mathcal{C}^\pm$ intersect at a line (wall) $W_{w_2}$ (see Figure \ref{SU3Coulomb}),
 \begin{align}
 W_{w_2}  =\mathcal{C}^+\cap \mathcal{C}^-.
 \end{align}
Correspondingly on the geometry side, the two resolutions $\mathscr{T}^\pm$ can meet with each other at $\mathscr{E}_1$ by blowing down (see Figure \ref{SU3tree}). We thus have the identification in codimension one,
\begin{align}
W_{w_2} = \mathscr{E}_1.
\end{align}

Finally, blowing down $\mathscr{E}_1$ to $\mathscr{E}_0$ corresponds to going along the line (wall) $W_{w_2}$ to the origin $O$ of the Coulomb branch. Hence
\begin{align}
O=  \mathscr{E}_0.
\end{align} 
We summarize the identifications for the $SU(3)$ model in Table \ref{id3}.

\begin{table}[thb]
\begin{center}
$SU(3)$:~~~~\begin{tabular}{|c|c|c|}
\hline
&Network of Resolutions & Coulomb Branch\\
\hline
2d& $\mathscr{T}^+$ &$ \mathcal{C}^+$\\
\hline
2d& $\mathscr{T}^-$ &$ \mathcal{C}^-$\\
\hline
1d &$\mathscr{E}_1$&$W_{w_2}$\\
\hline
0d & $\mathscr{E}_0$&$O$\\
\hline
\end{tabular}
\end{center}
\caption{identifications between (partial) resolutions and  subchambers $\mathcal{C}^\pm$ or the wall $W_{w_2}$ on the Coulomb branch of the $SU(3)$ model.}\label{id3}
\end{table}

As a further consistency check, we first note that the singular fiber for the partial resolution $\mathscr{E}_1$ is already the full affine $SU(3)$ Dynkin diagram (see (\ref{affineSU3})). This implies the corresponding line (wall) $W_{w_2}$ should \textit{not} lie on the boundary of the Coulomb branch (the two black lines in Figure \ref{SU3Coulomb}) where part of the non-abelian gauge symmetries is restored. This is indeed the case as $W_{w_2}$ lies in the bulk of the Coulomb branch $\mathcal{C}$ (see Figure \ref{SU3Coulomb}).

The flop transition is also beautifully identified as the reflection on the Coulomb branch. The flop induced by the $\mathbb{Z}_2$ automorphism (\ref{MW}) in the Mordell-Weil group exchanges $\mathscr{T}^+_\pm$ with $\mathscr{T}^-_\pm$,
\begin{center}
\begin{tikzcd}[column sep=huge]
\mathscr{T}^+ \arrow[leftrightarrow,red,dashed, thick]{r}{\text{flop}} &\mathscr{T}^-,
\end{tikzcd}
\end{center}
which corresponds to the reflection with respect to the wall $W^0$ on the Coulomb branch,
\begin{center}
\begin{tikzcd}[column sep=huge]
\mathcal{C}^+ \arrow[leftrightarrow,red,dashed, thick]{r}{\text{reflection}}[swap]{\color{blue}W_{w_2}} &\mathcal{C}^-.
\end{tikzcd}
\end{center}

\subsection{$SU(4)$}

\begin{figure}

\begin{center}
\begin{tabular}{lcr}
\raisebox{.7cm}{
\begin{tikzpicture}[every node/.style={circle,draw, minimum size= 5 mm}, scale=.8]
\node [thick] (T+-) at (0,0) { $\mathcal{C}^+_-$};
\node [thick] (T++) at (2.2,0) {$\mathcal{C}^+_+$} ;
\node [thick] (T-+) at (4.4,0) {$\mathcal{C}^-_+$} ;
\node [thick] (T--) at (6.6,0) {$\mathcal{C}^-_-$} ;
\draw [orange,thick] (T+-)--(T++);
\draw [red,thick] (T++)--(T-+);
\draw [purple,thick] (T-+)--(T--);
\coordinate [label= {\color{orange} $W^+$ } ] (w+) at (1.2,  - .2);
\coordinate [label= {\color{red} $W^0$ } ] (w0) at (3.4,  - .2);
\coordinate [label= {\color{purple} $W^-$ } ] (w-) at (5.6,  - .2);

\end{tikzpicture}}

\phantom{~~~~~~~~~~~~~~~}

 \scalebox{1}{\begin{tikzpicture}[every node/.style={circle,draw, minimum size= 4 mm}, scale=.8]
\node (w-) [purple,thick] at (0,0) {$\color{purple}W^-$};
\node (w+) [xshift=-1.8cm,thick,orange]  at (90:-1.8*0 cm) {$\color{orange}W^+$};
\draw [blue ,thick](- 1.1cm,0)--+(0,-.7cm);
\node  (w0) [xshift=-.9 cm,thick,red] at (90:-1.35cm) { $\color{red}W^0$};
\draw[blue,thick] (w-)--(w+);
\coordinate [label= {\color{blue} $L$ } ] (L) at (-1.1,   -.1);
\end{tikzpicture}}
\end{tabular}
\end{center}

\caption{Intersections in the $SU(4)$ Coulomb branch. Left: Intersections in codimension zero for the subchambers $\mathcal{C}^\pm_\pm$. Right: Intersections in codimension one for the walls $W^+$, $W^0$, $W^-$. The trivalent point means that the three walls intersect at a single line $L$. Intersections in higher codimensions are trivial.}\label{int4}
\end{figure}
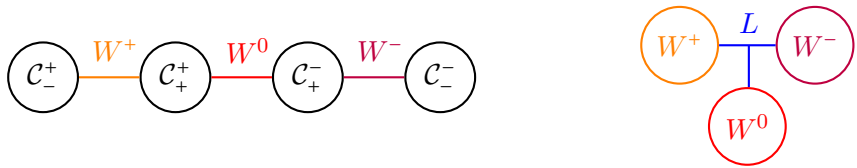

In the $SU(4)$ model, we need three blow ups and the Coulomb branch is indeed of real dimension three. There are four subchambers $\mathcal{C}^\pm_\pm$ on the Coulomb branch shown as tetrahedrons in Figure \ref{SU4Coulomb} with vertices 
\begin{align}
\mathcal{C}^+_-:(\ell_+, \ell, p_+,O),~~ \mathcal{C}^+_+:(p_+, \ell, p_0,O),~~ \mathcal{C}^-_+:(p_0, \ell, p_-,O),~ ~\mathcal{C}^-_-:(\ell_-, \ell, p_-,O),
\end{align}
extending infinitely from the apex $O$. They are identified with the four resolutions $\mathscr{T}^\pm_\pm$ in Figure \ref{SU4tree},
\begin{align}
\mathcal{C}^\pm_\pm = \mathscr{T}^\pm_\pm.
\end{align}
This identifications are consistent with the intersections between $\mathcal{C}^\pm_\pm$. For example on the geometry side, we can blow down $\mathscr{T}^+_+$ and $\mathscr{T}^+_-$ to the partial resolution $\mathscr{T}^+$ or blow down $\mathscr{T}^+_+$ and $\mathscr{T}^-_+$ to the partial resolution $\mathscr{B}$. However, there is no way to blow down once so that $\mathscr{T}^+_+$ can meet with $\mathscr{T}^-_-$. Correspondingly on the gauge theory side, while the subchamber $\mathcal{C}^+_+$ share walls with $\mathcal{C}^+_-$ and $\mathcal{C}^-_+$, it is not adjacent to $\mathcal{C}^-_-$ by a codimension one wall. The intersections for the subchambers $\mathcal{C}^\pm_\pm$ are summarized in the left figure in Figure \ref{int4}.

\begin{figure}[h!]
\begin{tabular}{cc}

\begin{tikzpicture}[scale=2.2]
\coordinate (O) at (0,0);
\coordinate (A1)  at (30:2);
\coordinate (A2)  at (-30:2);
\draw[ultra thick, -stealth,black] (O)--(A1);
\draw[ultra thick, -stealth,black] (O)--(A2);
\draw[line width=1mm,blue] (O)--(0:1.75);
\draw[fill=red, opacity=.1] (O)--(A1)--(A2);
\draw[fill=green, opacity=1] (0,0) circle (.05);
\node at (-.2,0) {\color{green!} \scalebox{1.5}{$O$~}};
\node at (15:1.5) {\color{red} \scalebox{1.5}{$\mathcal{C}^+$}};
\node at (-15:1.5) {\color{red} \scalebox{1.5}{$\mathcal{C}^-$}};
\node at (30:2.2) {\color{black} \scalebox{1.3}{$\mu^1$}  };
\node at (-30:2.2) {\color{black} \scalebox{1.3}{$\mu^2$}};
\node at (0:1.9) {\color{blue} \scalebox{1.5}{~~~~$W_{w_2}$}};
\end{tikzpicture}

\phantom{~~~~~~~~~~~~~~~~~}

\raisebox{2.3cm}{
\begin{tikzpicture}[every node/.style={circle,draw, minimum size= 5 mm}, scale=2]
\node [ultra thick,red] (C-) at (0,0) { \scalebox{1.2}{\color{red}$\mathcal{C}^-$}};
\node [ultra thick,red] (C+) at (3,0) { \scalebox{1.2}{\color{red}$\mathcal{C}^+$}};

\draw [blue,ultra thick] (C-)--(C+);

\coordinate [label= \scalebox{1.2}{{\color{blue} $W_{w_2}$} } ] (w+) at (1.5,  - .
2);

\end{tikzpicture}}

\end{tabular}
\caption{Left: The $SU(3)$ Coulomb branch. It is spanned non-negatively by the two vectors $\mu^1$ and $\mu^2$. The wall $W_{w_2}$ divides the Coulomb branch into two subchambers $\mathcal{C}^\pm$. Right: The two subchambers $\mathcal{C}^\pm$ intersect at a line $W_{w_2}$.}\label{SU3Coulomb}

\begin{center}
\begin{tikzcd}[column sep=huge,row sep=tiny]
& & {\color{red} \scalebox{1.6}{$\mathscr{T}^+$}} \arrow[bend left=75, leftrightarrow, dashed, color=red]{dd}[right] {\quad \text{\large flop}}  \\
{\color{green} \scalebox{1.6} {$\mathscr{E}_0$ }}\arrow[leftarrow, thick]{r} & {\color{blue} \scalebox{1.6}{$\mathscr{E}_1$}} \arrow[leftarrow,  thick]{ur} \arrow[leftarrow,  thick]{dr}& \\\
&    &{\color{red}\scalebox{1.6}{$ \mathscr{T}^-$}}\\
\raisebox{0.1cm}{\scalebox{.4}{\begin{tikzpicture}

 \draw[scale=2,domain=-1.2:1.2,smooth,ultra thick,variable=\t]
plot ({(\t)^2-1},{(\t)^3-\t});
\end{tikzpicture}}}&\scalebox{1.4}{$\bigcirc \text{---} \bigcirc$}  &\scalebox{1.4}{$\bigcirc \text{---} \bigcirc$}\\
\scalebox{1.1}{0d} &\scalebox{1.1}{ 1d} &\scalebox{1.1}{2d}
\end{tikzcd}
\end{center}
\caption{The $SU(3)$ network of resolutions. The singular fiber for each (partial) resolution is shown in the second row, where the affine node $C_0$ is always ignored. The identifications with the Coulomb branch are given by $\color{red}\mathscr{T}^\pm = \mathcal{C}^\pm$, $\color{blue}\mathscr{E}_1= W_{w_2}$, and $\color{green}\mathscr{E}_0=O$. The flop is realized as the reflection with respect to the line $W_{w_2}$ on the Coulomb branch.}\label{SU3tree}
\end{figure}

\clearpage

The intersections of the four subchambers give three walls $W^+$, $W^0$, $W^-$ (see the left figure of Figure \ref{int4}). They are shown as triangles in Figure \ref{SU4Coulomb} with vertices,
\begin{align}
\begin{split}
&W^+ = \mathcal{C}^+_- \cap \mathcal{C}^+_+ :~(p_+ , \ell ,O),\\
&W^0 = \mathcal{C}^+_+ \cap \mathcal{C}^-_+:~(p_0 , \ell , O),\\
&W^- = \mathcal{C}^-_+ \cap \mathcal{C}^-_-:~ (p_- , \ell ,O),
\end{split}
\end{align} 
extending infinitely from the apex $O$. They are identified as the three partial resolutions in the network in Figure \ref{SU4tree},
\begin{align}
W^+ = \mathscr{T}^+,~~
W^0 = \mathscr{B},~~
W^- = \mathscr{T}^-.
\end{align}
Again the identifications are consistent with the intersections of the walls in the following sense. The three walls intersect at a \textit{single line} $L$ rather than pairwise at three lines (see Figure \ref{SU4Coulomb} or the right figure of Figure \ref{int4}),
\begin{align}
L= W^+ \cap
W^0 \cap
W^-.
\end{align}
 On the other hand, by  blowing down the three partial resolutions $\mathscr{T}^+$, $\mathscr{B}$, and $\mathscr{T}^-$, they indeed meet at a single partial resolution $\mathscr{E}_1$ (see Figure \ref{SU4tree}). Hence we reach the following identification,
 \begin{align}
 L = \mathscr{E}_1.
 \end{align}

\begin{figure}

\begin{tabular}{ccc}

\scalebox{.8}{
\begin{tikzpicture}[scale=4]
\coordinate (O) at (0,0,0);
\coordinate   (P1) at (0.707107,0,0.707107);
\coordinate (P2) at (-0.707107,0,-0.707107);
\coordinate (P3) at (0 , 1.414,0);
\coordinate (w-) at (0.493561 ,0.427,0.493561);
\coordinate (w+) at (-0.493561 ,0.427,-0.493561 );

\coordinate (C) at (17:1.6);

\draw [black, ultra thick] (P1)--(P2)--(P3)--(P1);
\draw [orange,ultra thick](O)--(w+);
\draw [purple, ultra thick](O)--(w-);
\draw [red,ultra thick](O)--(P3);
\draw[blue,ultra thick,dashed] (C)--(O);
\draw [ultra thick,black] (P1)--(C);
\draw [purple, ultra thick](w-)--(C);
\draw [orange,  thick,dashed] (w+)--(C);
\draw [black,ultra thick] (P3)--(C);
\draw[dashed, black, thick] (P2)--(C);
\draw[fill=orange, opacity=.2] (w+)--(O)--(C);
\draw[fill=red, opacity=.2] (O)--(P3)--(C);
\draw[fill=purple, opacity=.4] (w-)--(O)--(C);

 \draw[ultra thick,-stealth] (.1 , 1.351)--(.05, 1.384);
   \coordinate[label={\color{black} \scalebox{.8}{$\mu^2$}}] (MU2) at ( 0.15 ,1.35);

    \draw[ultra thick,-stealth] (.738, 0,0.53)--(.730, 0,0.59);
   \coordinate[label={\color{black} \scalebox{.8}{$\mu^3$}}] (MU3) at ( 0.937 ,0,0.787);

   \draw[ultra thick,-stealth] (-.66 , 0,   -0.724)--(-.67, 0,   -0.72);
   \coordinate[label={\color{black} \scalebox{.8}{$\mu^3$}}] (MU3) at ( -0.587 ,0,-0.707);

\coordinate[label={\color{blue} \scalebox{.8}{$\ell$}}] (ell) at ( 270 : .15);
\draw[fill=blue, opacity=1] (0 ,0) circle (.02);

\coordinate[label={\color{purple} \scalebox{.8}{$p_-$}}] (p-) at ( 0.43 ,0.38,0.493561);
\draw[fill=purple, opacity=1] (0.493561 ,0.427,0.493561)
 circle (.02);
 
\coordinate[label={\color{orange} \scalebox{.8}{$p_+$}}] (p+) at (-0.553561 ,0.427,-0.493561 );
\draw[fill=orange, opacity=1] (-0.493561 ,0.427,-0.493561 )
 circle (.02);

\coordinate[label={\color{red} \scalebox{.8}{$p_0$}}] (p0) at (0 , 1.44,0);
\draw[fill=red, opacity=1] (0 ,1.414,0)
 circle (.02);

 \coordinate[label={\color{green} \scalebox{.8}{$O$}}]  (c) at (17:1.65);
\draw[fill=green, opacity=1] (17:1.6)
 circle (.02);

 \coordinate[label={\color{purple} \scalebox{.8}{$W^-$}}] (W-) at ( 0.65 ,0.32,0.493561);
 
 \coordinate[label={\color{orange} \scalebox{.8}{$W^+$}}] (W+) at (-0.303561 ,0.265,-0.493561 );
 
 \coordinate[label={\color{red} \scalebox{.8}{$W^0$}}] (W0) at (0.4 , .81,0);

  \coordinate[label={\color{blue} \scalebox{.8}{$L$}}] (L) at ( 0.9 ,0.26,0.493561);

 \draw[fill=gray, opacity=1] (0.707107,0,0.707107)
 circle (.02);
  \coordinate[label={\color{gray} \scalebox{.8}{$\ell_-$}}] (ell-) at ( 0.757 ,0,   .99);

 \draw[fill=gray, opacity=1] (-0.707107,0, -0.707107)
 circle (.02);
  \coordinate[label={\color{gray} \scalebox{.8}{$\ell_+$}}] (ell+) at ( -0.705 ,0,   -.5);

     \coordinate[label={\color{black} \scalebox{1}{$\mathcal{C}^+_-$}}] (C+-) at ( -0.337 ,0,-0.377);
     
      \coordinate[label={\color{black} \scalebox{1}{$\mathcal{C}^+_+$}}] (C++) at (-0.303561 ,0.535,-0.493561 );
      
        \coordinate[label={\color{black} \scalebox{1}{$\mathcal{C}^-_+$}}] (C-+) at (-0.1 ,0.455,-0.493561 );

        \coordinate[label={\color{black} \scalebox{1}{$\mathcal{C}^-_-$}}] (C--) at ( 0.45 ,0.07,0.493561);
 
\end{tikzpicture}
}

\phantom{~~~~~}

\scalebox{.8}{

\begin{tikzpicture}[scale=4]

\coordinate (O) at (0,0,0);

\coordinate  (mu1) at (-1, 0,0);
\coordinate (mu3) at ( 1, 0,0);
\coordinate (mu2) at (0 , 1.414,0);
\draw[ultra thick,black] (mu2)--(mu1);
\draw[ultra thick,black] (mu3)--(mu2);
\draw[ultra thick,black] (mu1)--(mu3);

\coordinate (w+) at (-0.698 , 0.427,0);
\coordinate (w-) at (+0.698 , 0.427,0);

\draw[ultra thick,orange] (O)--(w+);
\draw[ultra thick,purple] (O)--(w-);

\draw[ultra thick,red] (O)--(mu2);

\coordinate[label={\color{orange} \scalebox{1}{$W^+$}}] (W+) at (143 : .35);
\coordinate[label={\color{red} \scalebox{1}{$W^0$}}] (W0) at (97 : .95);
\coordinate[label={\color{purple} \scalebox{1}{$W^-$}}] (W-) at (35 : .5);

\coordinate[label={\color{black} \scalebox{1.2}{$\mathcal{C}^-_-$}}] (Cmm) at (6 : .7);
\coordinate[label={\color{black} \scalebox{1.2}{$\mathcal{C}^-_+$}}] (Cmp) at (65 : .65);
\coordinate[label={\color{black} \scalebox{1.2}{$\mathcal{C}^+_+$}}] (Cpp) at (115 : .65);
\coordinate[label={\color{black} \scalebox{1.2}{$\mathcal{C}^+_-$}}] (Cpm) at (174 : .5);
\draw[fill=red, opacity=1] (0,1.414) circle (.02);
\coordinate[label ={\color{red} \scalebox{.8} {$p_0$} }] (p0) at (92:1.41);

\draw[fill=purple, opacity=1] (+0.698 , 0.427)
 circle (.02);
\coordinate[label={\color{purple} \scalebox{.8}{$p_-$}}] (p-) at (30 : .87);

\draw[fill=orange, opacity=1] (-0.698 , 0.427)
 circle (.02);
\coordinate[label={\color{orange} \scalebox{.8}{$p_+$}}] (p+) at (150 : .85);

\draw[fill=blue, opacity=1] (0 ,0)
 circle (.02);
\coordinate[label={\color{blue} \scalebox{.8}{$\ell$}}] (ell) at ( 270 : .15);

\coordinate[label={\color{gray} \scalebox{.8}{$\ell_+$}} ] (ell+) at (-1, - .15);
\draw[fill=gray, opacity=1] (-1 , 0)
 circle (.02);
 
 \coordinate[label={\color{gray} \scalebox{.8}{$\ell_-$}} ] (ell-) at (1, - .15);
\draw[fill=gray, opacity=1] (1 , 0)
 circle (.02);
\end{tikzpicture}

}

\end{tabular}

\caption{Left: The $SU(4)$ Coulomb branch. It is the three-dimensional cone spanned non-negatively by the vectors $\mu^1$, $\mu^2$, $\mu^3$. The three walls $W^+$, $W^0$, $W^-$ are triangles in the above figure with vertices $(p_+,\ell, O)$, $(p_0, \ell,O)$, and $(p_-, \ell, O)$, respectively, extending infinitely from the apex $O$. The three walls divide the Coulomb branch into four subchambers $\mathcal{C}^\pm_\pm$. The four subchambers are tetrahedrons in the above figure with vertices $\mathcal{C}^+_-:(\ell_+, \ell, p_+,O)$, $\mathcal{C}^+_+:(p_+, \ell, p_0,O)$, $\mathcal{C}^-_+:(p_0, \ell, p_-,O)$, $\mathcal{C}^-_-:(\ell_-, \ell, p_-,O)$ extending infinitely from the apex $O$. The three walls intersect at a semi-infinite line $L:(\ell,O)$ lying on the bottom of the Coulomb branch, which is spanned by $\mu^1$ and $\mu^3$. Right: The two-dimensional projection along $L$. }\label{SU4Coulomb}

\begin{center}
\begin{tikzcd}[column sep=.7cm, row sep= .07 cm]
& & &\scalebox{1.3}{$ \mathscr{T}^+_- $}   \arrow[bend left=65, leftrightarrow, dashed, color=red]{dddddd}{\text{flop}} \\
& & \scalebox{1.3}{ \color{orange}{ $\mathscr{T}^+$} }\arrow[leftarrow, thick]{ru}\arrow[leftarrow, thick]{rd}& \\
& & & \scalebox{1.3}{$\mathscr{T}^+_+ $}    \arrow[bend left=65, leftrightarrow, dashed, color=red]{dd}{\color{red} \text{flop}}\\
\scalebox{1.3}{\color{green}{$\mathscr{E}_0$}}\arrow[leftarrow,thick]{r}& \scalebox{1.3}{ \color{blue}{$\mathscr{E}_1$ }}\arrow[leftarrow,thick]{r} \arrow[leftarrow, thick]{uur}\arrow[leftarrow,thick]{ddr} &\scalebox{1.3}{ \color{red}{$\mathscr{B}$}} \arrow[leftarrow,thick]{ru}\arrow[leftarrow,thick]{rd}
& \\
& & &\scalebox{1.2}{ $\mathscr{T}^-_+$} \\
& &\scalebox{1.2}{ \color{purple}{$\mathscr{T}^-$}}  \arrow[leftarrow,thick]{ru}\arrow[leftarrow,thick]{rd}  &   \\
& & &\scalebox{1.2}{$ \mathscr{T}^-_-$}\\
\raisebox{0.2cm}{\scalebox{.3}{\begin{tikzpicture}

 \draw[scale=2,domain=-1.2:1.2,smooth, thick,variable=\t]
plot ({(\t)^2-1},{(\t)^3-\t});
\end{tikzpicture}}}
&\scalebox{1.3}{$\bigcirc \text{--} \bigcirc$}&\scalebox{1.3}{$\bigcirc \text{--} \bigcirc\text{--}\bigcirc$}&\scalebox{1.3}{$\bigcirc \text{--} \bigcirc\text{--}\bigcirc$}\\
\scalebox{1}{0d} & \scalebox{1}{1d}& \scalebox{1}{2d}&\scalebox{1}{3d}   
\end{tikzcd}
\end{center}
\caption{The $SU(4)$ network of resolutions. The singular fiber for each (partial) resolution is shown in the second row, where the affine node $C_0$ is always ignored. The resolutions are identified with the Coulomb branch as $\mathscr{T}^\pm_\pm = \mathcal{C}^\pm_\pm$, $\color{orange}\mathscr{T}^+= W^+$, $\color{red}\mathscr{B} = W^0$, $\color{purple}\mathscr{T}^-= W^-$, $\color{blue}\mathscr{E}_1=L$, and $\color{green}\mathscr{E}_0=O$. The flops are realized as reflections with respect to the wall $W^0$. }\label{SU4tree}
\end{figure}

Note that the fiber for each of the three partial resolutions is a full affine $SU(4)$ Dynkin diagram (see (\ref{affineSU4+}), (\ref{affineSU4-}), and (\ref{affineSU4B})). This is consistent with the fact that the three walls $W^+,~W^0$, $W^-$ lie in the bulk of the Coulomb branch rather than on the boundary.

On the other hand, the fiber for $\mathscr{E}_1$ is only an affine $SU(3)$ Dynkin diagram (see (\ref{SU4E1})). That is, one of the four nodes in the affine $SU(4)$ Dynkin diagram shrinks when we blow down to the partial resolution $\mathscr{E}_1$. Correspondingly on the gauge theory side, the line $L$ indeed lies on the boundary of the Coulomb branch. This provides a nontrivial check for the correspondence.

 Finally as before, the origin $O$ is identified as the original singular Weierstrass model $\mathscr{E}_0$,
 \begin{align}
 O = \mathscr{E}_0.
  \end{align}
  We summarized the identifications for the $SU(4)$ model in Table \ref{id4}.
  
The flop is realized as reflection as follows. The flop induced by the $\mathbb{Z}_2$ automorphism (\ref{MW}) in the Mordell-Weil group exchanges $\mathscr{T}^+_\pm$ with $\mathscr{T}^-_\pm$,
\begin{center}
\begin{tikzcd}[column sep=huge]
\mathscr{T}^+_\pm \arrow[leftrightarrow,red,dashed, thick]{r}{\text{flop}} &\mathscr{T}^-_\pm.
\end{tikzcd}
\end{center}
It corresponds to the reflection with respect to the wall $W^0$,
\begin{center}
\begin{tikzcd}[column sep=huge]
\mathcal{C}^+_\pm \arrow[leftrightarrow,red,dashed, thick]{r}{\text{reflection}}[swap]{W^0} &\mathcal{C}^-_\pm.
\end{tikzcd}
\end{center}

  \begin{table}[h!tb]
\begin{center}
$SU(4)$:~~~~\begin{tabular}{|c|c|c|}
\hline
&Network of Resolutions & Coulomb Branch\\
\hline
3d& $\mathscr{T}^+_+$ &$ \mathcal{C}^+_+$\\
\hline
3d& $\mathscr{T}^+_-$ &$ \mathcal{C}^+_-$\\
\hline
3d& $\mathscr{T}^-_+$ &$ \mathcal{C}^-_+$\\
\hline
3d& $\mathscr{T}^-_-$ &$ \mathcal{C}^-_-$\\
\hline
2d& $\mathscr{T}^+$ &$ W^+$\\
\hline
2d& $\mathscr{B}$ &$W^0$\\
\hline
2d& $\mathscr{T}^-$ &$W^-$\\
\hline
1d &$\mathscr{E}_1$&$L$\\
\hline
0d & $\mathscr{E}_0$&$O$\\
\hline
\end{tabular}
\end{center}
\caption{identifications between (partial) resolutions and  subchambers $\mathcal{C}^\pm_\pm$, walls $W^\pm,W^0$, or the intersection of walls $L$ on the Coulomb branch of the $SU(4)$ model.}\label{id4}
\end{table}

\section{Network of Boxes}\label{sec:networkbox}

In \cite{Hayashi:2014kca} the authors introduce a powerful graphical tool called the box graph to classify all the subchambers on the Coulomb branch from the representation theory side. Some of the fibers for the corresponding geometries can also be predicted from the box graphs.\footnote{Not all the fibers can be predicted from the box graphs. For example, the codimension two fiber of type III in Figure \ref{SU2enhancement} for the $SU(2)$ model cannot be distinguished from the type I$_2^s$ fiber in the box graph.}  In this section we describe our (partial) resolutions using the box graph technology and confirm the fibers predicted from the box graph with our explicit calculation from the geometry side.

\subsection*{Definition of the box graph}

 We will focus on the $SU(4)$ model while it can also be applied to the other models studied in the present paper. Let $\varepsilon_i$ with $i=1,\cdots,4$ be the weights in the fundamental representation of $SU(4)$ and $C_i$ with $i=1,2,3$ be the simple roots of $SU(4)$. We have
\begin{align}
C_i = \varepsilon_i - \varepsilon_{i+1}.
\end{align}
The traceless condition of $SU(4)$ implies
\begin{align}
\sum_{i=1}^4 \varepsilon_i = 0.\label{trless}
\end{align}

Since we are interested in the general Tate form, both the fundamental $\bf 4$  and the antisymmetric representation $\bf 6$ matter fields are present. The (uncolored) $SU(4)$ box graph with $\bf 4$ and $\bf 6$ is shown in Figure \ref{ij}.
The box labeled by $(i,j)$ corresponds to the weight $\varepsilon_i + \varepsilon_j$.

\begin{figure}[h]
\begin{center}
\ytableausetup{nosmalltableaux, boxsize=2.7em}
\begin{ytableau}
 (1,1) &  (1,2) & (1,3)  &  (1,4) \\
 \none &  (2,2) & (2,3) &  (2,4) \\
 \none &    \none &  (3,3) & (3,4)\\
  \none &  \none &  \none & (4,4)
\end{ytableau}
\end{center}
\caption{The box graph for the $SU(4)$ model with both the fundamental and the antisymmetric representations. The box labeled by $(i,j)$ represents the weight $\varepsilon_i+\varepsilon_j$ where $\varepsilon_i$, $i=1,\cdots,4$, are the weights in the fundamental representation. The diagonal boxes $(i,i)$ stand for the weights in the fundamental representation $\bf4$ while the rest of the boxes are the weights in the antisymmetric representation $\bf6$.}\label{ij}
\end{figure}

\begin{figure}[h!]
\begin{center}
\includegraphics{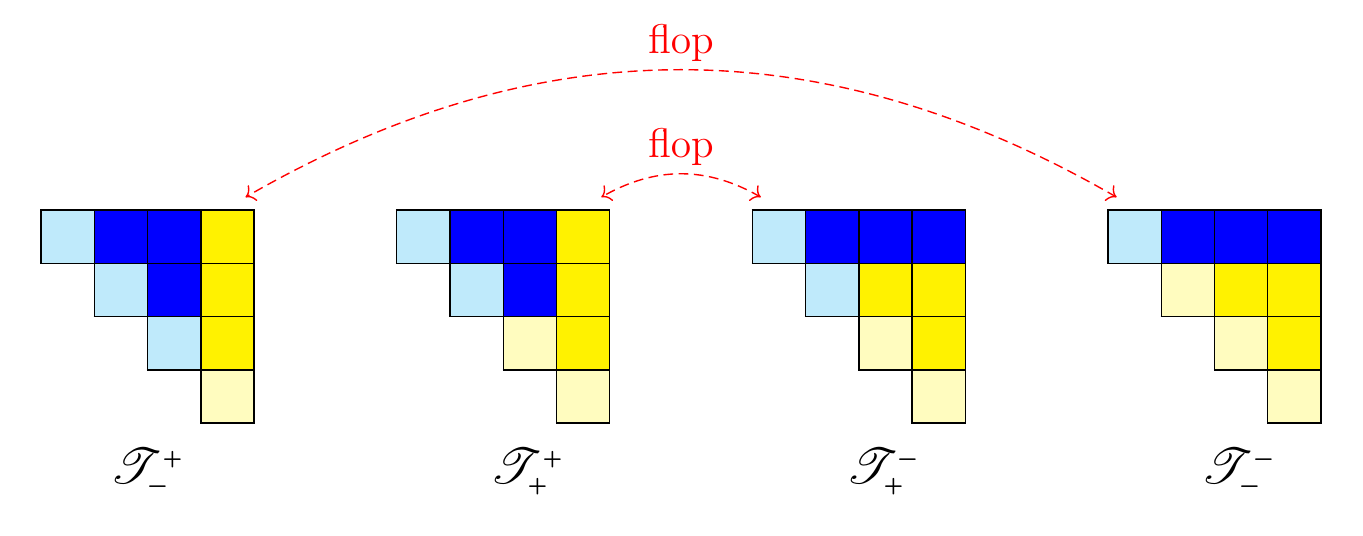}
\caption{The box graphs for the four resolutions $\mathscr{T}^\pm_\pm$ of the $SU(4)$ models. Each blue (yellow) box represents a weight $w$ with positive (negative) inner product with the real scalar $\phi$ in the vector multiplet. }\label{4T}
\end{center}
\end{figure}

Next, we will put color to each of the boxes in the box graph according to the sign of the inner product $\phi\cdot w$ between the corresponding weight $w$ with the real vector scalar $\phi$.  The blue (yellow) boxes stand for weights $w$ with positive (negative) inner products with $\phi$, which will be called the \textit{positive (negative) weights}. We use dark (light) color for the weights in the antisymmetric (fundamental) representation. A consistent assignment of signs to the boxes corresponds to a possible resolution, or equivalently, a subchamber on the Coulomb branch. The rules for the sign assignment was discussed in details in \cite{Hayashi:2014kca}. 
In the $SU(4)$ model, there are four consistent sign assignments for the box graphs shown in Figure \ref{4T}, corresponding to the four resolutions $\mathscr{T}^\pm_\pm$ in Figure \ref{SU4tree}.

\subsection*{Fibers from box graphs}

The fiber enhancement for each resolution can be reproduced from the box graph. Let us work out the case for $\mathscr{T}^+_+$. We start with the fiber enhancement over the codimension two locus associated with the fundamental representation. From the box graph, we see that $\phi \cdot \varepsilon_2 >0$ and $\phi \cdot \varepsilon_3<0$. Hence we can write the simple root $C_2$ as the sum of two positive weights,
\begin{align}
C_2 = \varepsilon_2 + ( - \varepsilon_3).\label{fiberenhance1}
\end{align}
Correspondingly on the geometry side, the node $C_2$, which we use the same notation as the associated simple root, splits into two nodes. This is indeed what we have seen in Table \ref{T++} where $C_2 \rightarrow C_4+C_5$ over the codimension two locus $w=P_4=0$.

Next moving on to the codimension two locus $w=a_1=0$ associated with the antisymmetric representation. From the box graph we see that the simple root $C_3$ can be written as the sum of three positive weights
\begin{align}
C_3  = (\varepsilon_2+\varepsilon_3) + ( -\varepsilon_1 - \varepsilon_3) + C_1.\label{fiberenhance2}
\end{align}
Indeed, from the direct blowup result shown in Table \ref{T++}, we see that the node $C_3$ splits into $C_3^{(1)}+C_3^{(2)}+C_1$.

The codimension three fiber enhancement can also be read off from the box graph. Over the codimension three locus $w=a_1=P_4=0$ where the $SU(5)$ and $SO(8)$ fibers collide, we have the fiber enhancement \eqref{fiberenhance1} and \eqref{fiberenhance2} at the same time. In fact, since $\varepsilon_2$ in $C_2$ can be written as the sum of two positive weights in this codimension three locus, 
\begin{align}
\varepsilon_2  = (  \varepsilon_2+\varepsilon_3)  + ( -\varepsilon_3),
\end{align}
it follows that $C_2$ splits into three nodes there. The fiber enhancements over the codimension three locus $w=a_1= P_4=0$ are then
\begin{align}
&C_2 = (\varepsilon_2+\varepsilon_3) + 2(- \varepsilon_3 ) ,\\
&C_3 =  (\varepsilon_2+\varepsilon_3) + ( -\varepsilon_1 - \varepsilon_3) + C_1.
\end{align}
Again, this matches with the fiber enhancement obtained directly from the blowup in Table \ref{T++}, $C_2\rightarrow C_2'+C_2^{(1)}+C_2^{(2)}$, $C_3\rightarrow C_2'+C_3^{(1)}+ C_1$.

\begin{figure}[h!]
\begin{center}
\includegraphics{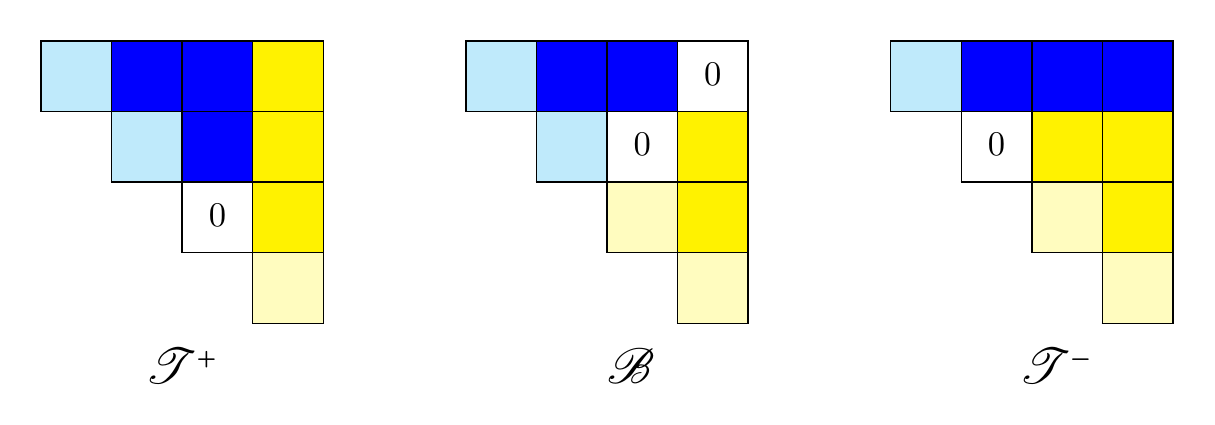}
\end{center}
\caption{The box graphs for the three partial resolutions $\mathscr{T}^\pm$ and $\mathscr{B}$ in the $SU(4)$ model. Each blue, yellow, or white box stands for a weight $w$ with positive, negative, or zero inner product with the real scalar $\phi$ in the vector multiplet, respectively. The two zeroes in $\mathscr{B}$ are correlated due to the traceless condition, $(\varepsilon_1+\varepsilon_4)+(\varepsilon_2+\varepsilon_3)=0$. On the corresponding codimension one loci on the Coulomb branch, the hypermultiplet scalars $Q_w$ and $\tilde Q_w$ corresponding to the weights $w$ labeled by 0 are massless, so one can active their vevs to go the Coulomb-Higgs branch. Hence these partial resolutions correspond to the Higgs branch roots where the Coulomb-Higgs branches intersect with the Coulomb branch.}\label{partial box}
\end{figure}

\subsection*{Box graphs for partial resolutions}

The partial resolutions can also be represented by the box graph by putting some of the weights to be zero (Figure \ref{partial box}). For example, the partial resolution $\mathscr{T}^+$ has $\varepsilon_3=0$ and $\mathscr{B}$ has $(\varepsilon_1+\varepsilon_4) = (\varepsilon_2+\varepsilon_3)=0$. It should be noted that the number of zeroes does not necessarily represent the codimension of the corresponding locus on the Coulomb branch because there are some relations between the weights in the box graph. For example for the partial resolution $\mathscr{B}$, setting $(\varepsilon_1+\varepsilon_4)=0$ implies $(\varepsilon_2+\varepsilon_3)=0$ due to the traceless condition \eqref{trless}.  Therefore, even though the box graph for $\mathscr{B}$ has two zeroes it still represents a codimension one wall $W^0$ on the Coulomb branch. 

The fiber enhancements for the partial resolutions can also be read off from the box graph in Figure \ref{partial box}. Let us start with the codimension two locus $w=P_4=0$ associated with the fundamental representation for the partial resolution $\mathscr{T}^+$. The weights in the fundamental representation $\bf4$ correspond to the light blue, light yellow, and white boxes on the diagonal line in Figure \ref{partial box}. In contrast to its final resolution $\mathscr{T}^+_+$, we can no longer write any simple root as the sum of two positive weights because the zero weight $\varepsilon_3$ is standing between the positive and negative weights. Indeed, as can be seen from Sec \ref{sec:T+}, the fiber for $\mathscr{T}^+$ does not enhance over this codimension two locus $w=P_4=0$. 

On the other hand, there are no zero weights standing in the way between positive and negative weights for the antisymmetric weights (dark blue and dark yellow boxes in Figure \ref{partial box}). Hence, the fiber enhancement over the codimension two locus $w=a_1=0$ associated with the antisymmetric representation should be the same as $\mathscr{T}^+_+$. This is indeed the case as one can check from Sec \ref{sec:T+}.

\begin{figure}[h!]
\begin{center}
\includegraphics[scale=1.1]{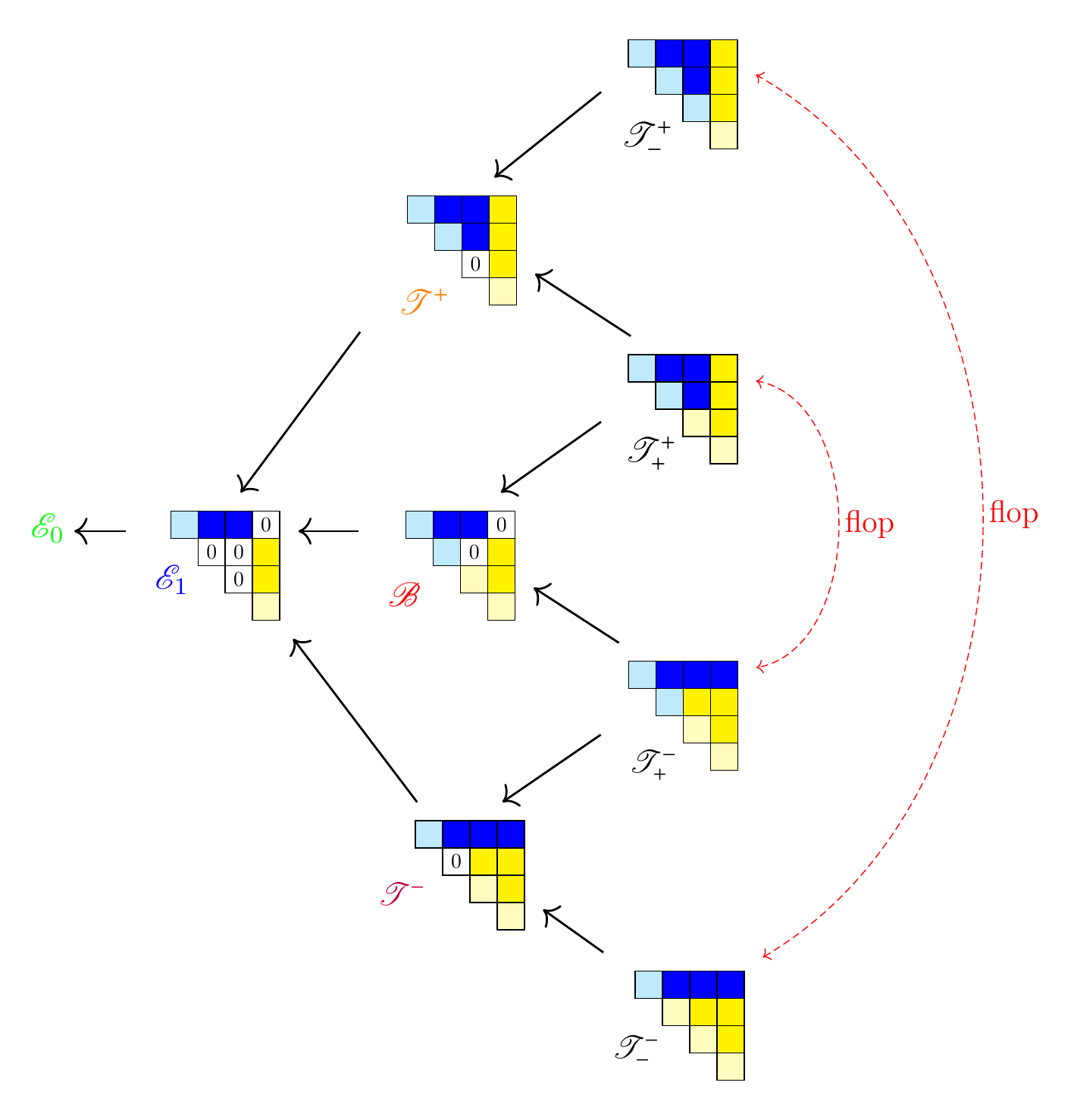}
\end{center}
\caption{The network of boxes for the $SU(4)$ model. Each box graph stands for a (partial) resolution of the $SU(4)$ model.  Each blue, yellow, or white box stands for a weight $w$ with positive, negative, or zero inner product with the real scalar $\phi$ in the vector multiplet, respectively.}\label{boxnetwork}
\end{figure}

Lastly, we can now relate each (partial) resolution to a box graph and draw the network of resolutions (see Figure \ref{SU4tree}) in terms of boxes in Figure \ref{boxnetwork}.

\newpage

\section{Discussion}

Let us summarize our results:
\begin{itemize}
\item We present a simple and systematic procedure to resolve   $SU(N)$  Weierstrass models by sequences of blow ups for $N=2,3,4$.
 The fiber enhancements in codimension two and three are analyzed for each case. We found the non-Kodaira type fiber I$_0^{*+}$ in codimension three in the $SU(4)$ model.  Such a fiber was observed  before in the study of elliptic threefolds with the assumption of normal crossing for the components of the discriminant of the fibration \cite{Miranda.Smooth}.  
It can also appear in codimension two or higher \cite{Szydlo}. See also \cite{Lawrie:2012gg} and \cite{Braun:2013cb}.
\item From the network of resolutions one can keep track of the way to blow down to various partial resolutions along the arrows. Furthermore, flops are manifest from the ramification of the branches in the network.  Since all the resolutions are obtained by sequences of blow ups, they are manifestly  projective varieties provided the base is projective too.

\item In connection with physics, the topology of the network of resolutions has an one-to-one correspondence with the  Coulomb branch of 5d $\mathcal{N}=1$ gauge theory. We explicitly match the subchambers, walls, and intersections of walls on the Coulomb branch with (partial) resolutions in the network for the Weierstrass model. In addition, flops are realized as reflections with respect to the walls. This provides a clean demonstration of phase transitions from a geometric point of view via M-theory compactification.
\item Since the singularity structure of the Weierstrass model does not depend on the choice of a specific fundamental line bundle,  the total space does not even have to be Calabi-Yau in particular. In that regard our correspondence goes beyond the context of string/M-theory compactification. It suggests a deep connection between small resolutions for singular Weierstrass models and representation theory.
\end{itemize}

It would be interesting to study explicitly the network of resolutions for the other Tate models.  For the $SU(5)$ model, a sub-network is already available from the six resolutions in \cite{Esole:2011sm} which are organized as an hexagon \cite{Esole:2011sm, Hayashi:2013lra}. 
 The full network of resolutions of the $SU(5)$ model should include all the partial resolutions as well as the known resolutions that are projective varieties. For example, in addition to the six resolutions of \cite{Esole:2011sm}, it would also include the ``toric resolutions" of \cite{Krause:2011xj,Grimm:2011fx,Hayashi:2013lra}, thus clarifying their definitions in terms of sequences of blow ups.  We would also like to extend this correspondence to the $D$- and $E$-series.

Throughout this paper we have only talked about the phase transitions within the Coulomb branch. It would also be interesting to understand the conifold transitions \cite{Strominger:1995cz,Greene:1995hu} from the Coulomb branch into the Coulomb-Higgs branch in this context to complete the picture. A similar story of deformation was recently discussed in \cite{GHS1,GHS2}.

\bigskip

\section*{Acknowledgments}

We are grateful to Paolo Aluffi,  Heng-Yu Chen, Clay C\'ordova, Patrick Jefferson, Hai Lin, David Morrison,  Sakura Sch\"afer-Nameki, Xi Yin, Jie Zhou for inspiring discussions. SHS would like to thank the Physics Department and Taida Institute for Mathematics Sciences of  the National Taiwan University for hospitality. SHS is supported by the Kao Fellowship and the An Wang Fellowship at Harvard University.

\appendix

\section{Second blow ups and flop for the $SU(3)$ model}\label{app1}

In this appendix we study the fiber enhancements for the resolved varieties $\mathscr{T}^\pm$ after the second blow up in the $SU(3)$ model. Recall that after the first blow up we arrive at the partial resolution $\mathscr{E}_1$
\begin{align}
\mathscr{E}_1:~ys = e_1 Q
\end{align}
where $s = y+ a_1 x+a_{3,1}e_0$ and $Q = x^3 +a_{2,1} e_0 x^2 +a_{4,2} e_0^2 x+a_{6,3}e_0^3$. To resolve the conifold singularity at $y=s=e_1=Q=0$ over $e_0e_1=P_3=0$ ($P_3$ is defined in (\ref{P3})) on the base $B$, we can either blow up along the ideal $(y,e_1)$ or the ideal $(s,e_1)$. These two  resolutions $\mathscr{T}^{\pm}$ are related by a flop.

\subsection{Resolution $\mathscr{T}^+:(y,e_1|e_2)$}

By blowing up along $(y,e_1)$ 
\begin{align}
(y,e_1)\rightarrow (e_2 y,e_2 e_1)
\end{align}
we obtain
\begin{align}
\mathscr{E}_0 \xleftarrow{(x,y,e_0|e_1)} \mathscr{E}_1 \xleftarrow{(y,e_1|e_2)} \mathscr{T}^+
\end{align}
\begin{align}
\mathscr{T}^+: y ( e_2  y + a_1 x +a_{3,1}e_0) =  e_1 (x^3 +a_{2,1}e_0x^2 + a_{4,2}e_0^2x+a_{6,3} e_0^3),\label{option1}
\end{align}
where we have written down the chain of blow ups for $\mathscr{T}^+$ to keep track which point in the {network of resolutions} we are at. The ambient space is parametrized by the following projective coordinates
\begin{align}
\,
[e_2e_1x:e_2^2e_1 y:z= 1][x:e_2 y:e_0][y:e_1].
\end{align}

The original divisor $e_0=0$ is now blown up to be $e_0e_1e_2=0$. The nodes in the fiber over the divisor ${e_2e_1e_0=0}$ are
\begin{align}
\begin{split}
& C_0:~ e_0=e_2 y^2 + a_1 x y-e_1 x^3=0,\\
&C_{1}:~e_1=e_2 y + a_1x +a_{3,1} e_0=0,\\
& C_{1}': ~ e_2= (a_1x  +a_{3,1} e_0)  y - e_1 (x^3 +a_{2,1}e_0x^2+a_{4,2}e_0^2x + a_{6,3} e_0^3)=0.
\end{split}
\end{align}
They intersect pairwise at three different points so the fiber is of type I$_3$. This corresponds to the affine Dynkin diagram for $SU(3)$.

Over the codimension two locus ${e_0e_1e_2=P_3=0}$ but $a_1,a_{3,1}$ nonzero, we have simultaneous solution to $s(y=0)=a_1x+a_{3,1}e_0=0$ and $Q(x,e_0)=0$, i.e. $Q(a_{3,1},-a_1)=0$. Above this locus, we can factor $Q$ as
\begin{align}
\begin{split}
Q(x,e_0)& = x^3+ a_{2,1}x^2e_0+a_{4,2}xe_0^2+a_{6,3}e_0^3\\
&=(a_1 x+a_{3,1} e_0) \left(  {1\over a_1}x^2 + {a_1a_{2,1}-a_{3,1}\over a_1^2}xe_0 + {a_{6,3}\over a_{3,1}}e_0^2\right)\\
&:= (a_1 x+a_{3,1} e_0) \tilde Q(x,e_0).\label{tildeq}
\end{split}
\end{align}
The defining equation for $C_{1}'$ thus becomes
\begin{align}
 C_{1}':~e_2= (a_1x+a_{3,1}e_0)\left[ y -e_1\tilde Q(x,e_0) \right]=0.
\end{align}
Hence $ C_{1}'$ splits into two nodes, which we will call $C_2,~ C_3$:
\begin{align}
\begin{split}
&C_{1}'\rightarrow C_2+C_3,\\
&~~~~ C_2: ~ [0:0:1][a_{3,1}:0:-a_1][ y:e_1],\\
& ~~~~C_3:~[0:0:1][x:0:e_0][\tilde Q(x,e_0):1],
\end{split}
\end{align}
From the intersections of $C_0, C_1,C_2,C_3$ we recognize the fiber to be of type I$_4$. This is the rank one enhancement from $SU(3)\rightarrow SU(4)$. Note that this codimension two locus $e_0e_1e_2=P_3=0$ is precisely the locus of the conifold singularity (\ref{conifold3}) of the partial resolution $\mathscr{E}_1$. After the second blow up, the singular point (\ref{conifold3}) is blown up to be a full $\mathbb{P}^1$ and gives rise to the rank one enhancement to $SU(4)$.

Over the codimension two locus ${e_0e_1e_2=a_1=0}$, the three nodes intersect at a single point, $ C_0 \cap  C_{1} \cap  C_{1}':~e_0=e_1= e_2=0$, so we have the IV fiber.

Over the codimension three locus ${e_0e_1e_2=a_1=a_{3,1}=0}$, we note that $C_{1}'$ splits into three components
\begin{align}
\begin{split}
&C_{1}'\rightarrow C_{1} +C_2^{(1)} +C_2^{(2)}+C_2^{(3)},\\
&~~~~C_{1}:~e_2=e_1=0,\\
&~~~~~~~[0:0:1][x:0:e_0][1:0],\\
&~~~~C_2^{(i)}:~e_2=Q(x,e_0) = 0,\\
&~~~~~~~[0:0:1][x^{(i)}:0:e_0^{(i)}][y:e_1],~~~i=1,2,3,
\end{split}
\end{align}
where $x^{(i)},e_0^{(i)}$ are the three roots to $Q(x,e_0)=0$. Also note that the multiplicity for $C_{1}$ is two now. From the intersections of $C_0,~ 2C_1,~ C_2^{(1)},~C_2^{(2)},~C_2^{(3)}$, we recognize the fiber to be of type I$_0^*$. 

The fiber enhancements for $\mathscr{T}^+$ are summarized in Table \ref{T+}.

\begin{table}[htb]
\begin{center}
\scalebox{.9}{\begin{tabular}{|c|c|c|c|c|}
\hline
$e_0e_1e_2=0$& $e_0e_1e_2=P_3=0$ &  $e_0e_1e_2=a_1=0$ & $e_0e_1e_2=a_1=a_{3,1}=0$ \\
\hline
& {\footnotesize $C_1'\to C_{2}+C_{3}$ }&
\scalebox{.8}{ \begin{tabular}{l}
\end{tabular} }
 & 
\scalebox{.8}{ \begin{tabular}{l}
$C_1'\to C_{1}+C_2^{(1)} + C_2^{(2)}+C_2^{(3)} $
\end{tabular} }

 \\
\hline
 I$_3$& I$_4$& IV & I$_0^*$\\
\hline
&&&\\
\scalebox{.95}{\begin{tikzpicture}[every node/.style={circle,draw, minimum size= 5 mm}]
\node (C3) at (120*3:1.4cm) {\tiny $C_{1}'$};
\node (C1) at (120*2:1.4cm) {\tiny $C_{1}$};
\node (C0) at (120*1:1.4cm) {\tiny $C_0$};
\draw (C1)--(C3)--(C0)--(C1);
\end{tikzpicture}}& 
\scalebox{.95}{\begin{tikzpicture}[every node/.style={circle,draw, minimum size= 5 mm}]
\node (C4) at (90*4:1.4cm) {\scalebox{.7}{ $C_{3}$}};
\node (C3) at (90*3:1.4cm) {\scalebox{.7}{$C_{2}$}};
\node (C1) at (90*2:1.4cm) {\tiny $C_{1}$};
\node (C0) at (90*1:1.4cm) {\tiny $C_0$};
\draw (C1)--(C3)--(C4)--(C0)--(C1);
\end{tikzpicture}}
& 
 \scalebox{1}{\begin{tikzpicture}[every node/.style={circle,draw, minimum size= 4 mm}, scale=.8]
\node (C2) at (0,0) {$C_{0}$};
\node (C5) [xshift=-1.8cm]  at (90:-1.8*0 cm) {$C_{1}$};
\draw (- 1.1cm,0)--+(0,-1.2cm);
\node  (C6) [xshift=-.9 cm] at (90:-1.8cm) { $C_{1}'$};
\draw (C2)--(C5);
\end{tikzpicture}}
& 
\scalebox{1}{\begin{tikzpicture}[every node/.style={circle,draw, minimum size= 5 mm}, scale=.8]
\node (C2) at (0,0) { \tiny$2C_1$};
\node (C0) at (90-35:1.8cm) {\tiny$C_2^{(1)}$} ;
\node (C1) at (90+35:1.8cm) {\small $C_0$};
\node  (C3) at  (-90-35:1.8cm){ \tiny$C_2^{(2)}$};
\node  (C4) at (-90+35:1.8cm) { \tiny$C_{2}^{(3)}$};
\draw (C2)--(C0);
\draw (C2)--(C1);
\draw (C2)--(C3);
\draw (C2)--(C4);
\end{tikzpicture}}
\\ 
\hline
\end{tabular}}
\end{center}
\caption{The fiber enhancements for $\mathscr{T}^+$ in the $SU(3)$ model. The trivalent point for IV means that the three nodes meet at the same point. Here $P_3 = a_{3,1}^3 - a_1 a_{2,1} a_{3,1}^2 + a_1^2a_{3,1} a_{4,2} - a_1^3 a_{6,3}=0$.}\label{T+}
\end{table}

\subsection{Resolution $\mathscr{T}^-:(s,e_1|e_2)$}

For $\mathscr{T}^-$ we choose to blow up along $(s,e_1)$ where $s=y+a_1x+a_{3,1} e_0$. By replacing 
\begin{align}
(s,e_1 ) \rightarrow (e_2 s, e_2 e_1)
\end{align}
and expressing $y= s-a_1 x-a_{3,1}e_0$, we arrive at $\mathscr{T}^-$,
\begin{align}
\mathscr{E}_0 \xleftarrow{(x,y,e_0|e_1)} \mathscr{E}_1 \xleftarrow{(s,e_1|e_2)} \mathscr{T}^-
\end{align}
\begin{align}
\mathscr{T}^- : ~ (e_2 s -a_1 x -a_{3,1} e_0) s =  e_1 (x^3 +a_{2,1}e_0x^2 + a_{4,2}e_0^2x+a_{6,3} e_0^3).
\end{align}
The ambient space is
\begin{align}
\, [e_2 e_1 x:e_2 e_1 (e_2s-a_1 x-a_{3,1} e_0):z=1][x:  (e_2s-a_1 x-a_{3,1} e_0): e_0 ] [s:e_1].
\end{align}

Over the divisor ${e_0e_1e_2=0}$, we have the following three nodes in the fiber,
\begin{align}
\begin{split}
C_0:~& e_0 = (e_2 s-a_1x)s -e_1 x^3= 0,\\
C_1 ':~&e_1=e_2 s-a_1 x-a_{3,1} e_0=0,\\
&[ 0:0:1] [ x: 0 :e_0 ] [1:0],\\
C_1:~& e_2 = (a_1 x+a_{3,1} e_0)s + e_1 (x^3 +a_{2,1}e_0x^2 + a_{4,2}e_0^2x+a_{6,3} e_0^3)
=0,\\
&[0:0:1][x:-a_1 x-a_{3,1} e_0:e_0][-(x^3 +a_{2,1}e_0x^2 + a_{4,2}e_0^2x+a_{6,3} e_0^3)
:a_1 x+a_{3,1} e_0].
\end{split}
\end{align}
Note that our labeling for the nodes is consistent with that for the $\mathscr{T}^+$ resolution, where $C_1$ comes from $e_1=s=0$ and $C_1'$ corresponds to $e_1=y=0$. Keeping track of the labeling will be important when we discuss the flop induced by the $\mathbb{Z}_2$ automorphism (\ref{MW}) from the Mordell-Weil group in the following.

From here we see that the analysis for $\mathscr{T}^-$ is identical to the analysis for $\mathscr{T}^+$ by exchanging $s$ with $-y$, the inverse action \eqref{MW}. One crucial point is that the role played by $C_1:~ e_1 =y=0$ and $C_1':~ e_1=s=0$ are switched when compared $\mathscr{T}^-$ with $\mathscr{T}^+$. For example, over $e_0e_1e_2=P_3 =0$, it is $C_1$ that splits into two, rather than $C_1'$ as would be the case of $\mathscr{T}^+$.

We here summarize the fiber enhancement for $\mathscr{T}^-$ in Table \ref{T-}. Note that it is obtained by exchanging $C_1$ with $C_1'$ from the fiber enhancement for $\mathscr{T}^+$ in Table \ref{T+}.

\begin{table}[htb]
\begin{center}
\scalebox{.9}{\begin{tabular}{|c|c|c|c|c|}
\hline
$e_0e_1e_2=0$& $e_0e_1e_2=P_3=0$ &  $e_0e_1e_2=a_1=0$ & $w=a_1=a_{3,1}=0$ \\
\hline
& {\footnotesize $C_1\to C_{2}+C_{3}$ }&
\scalebox{.8}{ \begin{tabular}{l}
\end{tabular} }
 & 
\scalebox{.8}{ \begin{tabular}{l}
$C_1\to C_{1}'+C_2^{(1)} + C_2^{(2)}+C_2^{(3)} $
\end{tabular} }

 \\
\hline
 I$_3$& I$_4$& IV & I$_0^*$\\
\hline
&&&\\
\scalebox{.95}{\begin{tikzpicture}[every node/.style={circle,draw, minimum size= 5 mm}]
\node (C3) at (120*3:1.4cm) {\tiny $C_{1}'$};
\node (C1) at (120*2:1.4cm) {\tiny $C_{1}$};
\node (C0) at (120*1:1.4cm) {\tiny $C_0$};
\draw (C1)--(C3)--(C0)--(C1);
\end{tikzpicture}}& 
\scalebox{.95}{\begin{tikzpicture}[every node/.style={circle,draw, minimum size= 5 mm}]
\node (C4) at (90*4:1.4cm) {\scalebox{.7}{ $C_{3}$}};
\node (C3) at (90*3:1.4cm) {\scalebox{.7}{$C_{2}$}};
\node (C1) at (90*2:1.4cm) {\tiny $C_{1}'$};
\node (C0) at (90*1:1.4cm) {\tiny $C_0$};
\draw (C1)--(C3)--(C4)--(C0)--(C1);
\end{tikzpicture}}
& 
 \scalebox{1}{\begin{tikzpicture}[every node/.style={circle,draw, minimum size= 4 mm}, scale=.8]
\node (C2) at (0,0) {$C_{0}$};
\node (C5) [xshift=-1.8cm]  at (90:-1.8*0 cm) {$C_{1}$};
\draw (- 1.1cm,0)--+(0,-1.2cm);
\node  (C6) [xshift=-.9 cm] at (90:-1.8cm) { $C_{1}'$};
\draw (C2)--(C5);
\end{tikzpicture}}
& 
\scalebox{1}{\begin{tikzpicture}[every node/.style={circle,draw, minimum size= 5 mm}, scale=.8]
\node (C2) at (0,0) { \tiny$2C_1'$};
\node (C0) at (90-35:1.8cm) {\tiny$C_2^{(1)}$} ;
\node (C1) at (90+35:1.8cm) {\small $C_0$};
\node  (C3) at  (-90-35:1.8cm){ \tiny$C_2^{(2)}$};
\node  (C4) at (-90+35:1.8cm) { \tiny$C_{2}^{(3)}$};
\draw (C2)--(C0);
\draw (C2)--(C1);
\draw (C2)--(C3);
\draw (C2)--(C4);
\end{tikzpicture}}
\\ 
\hline
\end{tabular}}
\end{center}
\caption{The fiber enhancements for $\mathscr{T}^-$ in the $SU(3)$ model. It can be obtained from the fiber enhancements for $\mathscr{T}^+$ (Table \ref{T+}) by switching $C_1\leftarrow C_1'$. The trivalent point for IV means that the three nodes meet at the same point. Here $P_3 = a_{3,1}^3 - a_1 a_{2,1} a_{3,1}^2 + a_1^2a_{3,1} a_{4,2} - a_1^3 a_{6,3}=0$.}\label{T-} 
\end{table}

\section{Small Resolutions  and flops for the $SU(4)$ model}\label{app2}

The $SU(4)$ model is defined by \cite{Bershadsky:1996nh,Katz:2011qp}
\begin{align}
y^2  + a_1 xy + a_{3,2}e_0^2 y = x^3 + a_{2,1} e_0x^2  + a_{4,2} e_0^2x + a_{6,4} e_0^4.\label{tate4}
\end{align}
 As before, one can check that the singularities of the total space  are supported on:
\begin{align}
x=y=e_0=0.
\end{align}

\subsection{Partial resolution $\mathscr{E}_1:(x,y,e_0|e_1)$}

We blow up along $(x,y,e_0)$ by replacing
\begin{align}
(x,y,e_0)\rightarrow (e_1 x,e_1y ,e_1e_0),
\end{align}
in $\mathscr{E}_0$ and factoring out the exceptional divisor $e_1$. The first partial resolution $\mathscr{E}_1$ is
\begin{equation}
\displaystyle{
\begin{tikzcd}[column sep=3cm]
\displaystyle{ \mathscr{E}_0}\arrow[leftarrow]{r}{\displaystyle (x,y,e_0|e_1)} &  \mathscr{E}_1,
\end{tikzcd}}
\end{equation}
\begin{align}
\mathscr{E}_1:~y^2 + a_1 x y + a_{3,2} e_1e_0^2y = e_1 x^3 + a_{2,1} e_1 e_0 x^2 + a_{4,2} e_1 e_0^2 x + a_{6,4} e_1^2 e_0^4.
\end{align}
The ambient space is parametrized by the following projective coordinates
\begin{align}
\, [e_1x:e_1y:z=1][x:y:e_0].
\end{align}

\paragraph{Description of the fiber}

We have the following three nodes $C_0,C_1,C_1'$ over the divisor ${e_1e_0=0}$:
\begin{align}
\begin{split}
&C_0 :~e_0 =y^2 +a_1xy -e_1x^3=0,\\
&C_{1}:~e_1=y+a_1x=0,\\
&C_{1}':~e_1= y=0.\label{SU4E1}
\end{split}
\end{align} 
From the intersections we see that it is a I$_3$ fiber. Recall that in the $SU(4)$ model there are supposed to be four nodes in the affine Dynkin diagram. 
In the partial resolution $\mathscr{E}_1$ above, we only have three nodes, which is one less than what we would have  in  the fiber of the fully resolved varieties.  
It follows that on the Coulomb branch, the line $L$ corresponding to $\mathscr{E}_1$ should be on the boundary of the fundamental chamber where part of the non-abelian symmetry is restored. This is indeed the case as can be  seen from the $SU(4)$ Coulomb branch in Figure \ref{SU4Coulomb}. 

After the second blow up, we will recover the vanishing node. On the gauge theory side, this corresponds to moving off the line $L$ to the bulk of the Coulomb branch.

\paragraph{Conifold singularity}

We can write $\mathscr{E}_1$ as
\begin{align}
&\mathscr{E}_1:~y s  = e_1 Q
\end{align}
where
\begin{align}
&s(x,y,e_0,e_1)=y+a_1x+a_{3,2}e_1e_0^2,\\
&Q(x,w,e_1)= x^3 + a_{2,1} e_0 x^2 + a_{4,2} e_0^2 x +a_{6,4} e_1 e_0^4.
\end{align}
There is a conifold singularity at
\begin{align}
y=e_1=a_1x=x^3 + a_{2,1} e_0x^2 + a_{4,2} e_0^2 x =0.
\end{align}
Over a general point on the divisor $e_0e_1=0$, the conifold singularity is at
\begin{align}
y=e_1=x=0.
\end{align}
Over the codimension two locus $e_0e_1= a_1 =0$, on the other hand, the conifold singularity is at
\begin{align}
y=e_1 = x(x^2+a_{2,1}e_0x+a_{4,2}e_0^2)=0.
\end{align}

To resolve the conifold singularity, we have the following three options for the second blow up: $\mathscr{T}^+:(y,e_1)$, $\mathscr{B}:(x,y,e_1)$, and $\mathscr{T}^-:(s,e_1)$. We will explore these options separately in the following sections.

\subsection{Partial resolution  $\mathscr{T}^+:(y,e_1|e_2)$}\label{sec:T+}

We start with the $\mathscr{T}^+$ resolution by replacing
\begin{align}
(y,e_1)\rightarrow (e_2 y,e_2 e_1).
\end{align}
The partially resolved variety $\mathscr{T}^+$ is then
\begin{align}
\mathscr{E}_0 \xleftarrow{(x,y,e_0|e_1)} \mathscr{E}_1 \xleftarrow{(y,e_1|e_2)} \mathscr{T}^+
\end{align}
\begin{align}
\mathscr{T}^+: y (e_2 y +a_1x + a_{3,2} e_0^2  e_1 e_2)
= e_1 (x^3 + a_{2,1} e_0 x^2 + a_{4,2} e_0^2 x +a_{6,4} e_1 e_2 e_0^4).
\end{align}
The ambient space is parametrized by the following projective coordinates
\begin{align}
\, [e_2e_1x:e_2^2e_1 y:z=1][x:e_2y:e_0][y:e_1].
\end{align}

\paragraph{Description of the fiber}

Over the divisor ${e_0e_1e_2=0}$, we have the following four nodes in the fiber,
\begin{align}
\begin{split}
&C_0: ~e_0= e_2  y^2 + a_1 x y -e_1 x^3=0,\\
&C_{1}:~e_1= e_2  y +a_1x =0,\\
&C_2:~e_2=x=0,\\
&C_3:~e_2 =   a_1 y-e_1(x^2 + a_{2,1} e_0 x+a_{4,2} e_0^2)=0.\label{affineSU4+}
\end{split}
\end{align}
From the intersections we recognize the fiber to be of type I$_4$, which is the affine Dynkin diagram for $SU(4)$. As advertised before, we recover all the affine Dynkin nodes in the second blow up. On the gauge theory side, the corresponding wall $W^+$ indeed lies in the bulk of the Coulomb branch (see Figure \ref{SU4Coulomb}) where all the Dynkin nodes are present.

\paragraph{Conifold singularity}

Let us rewrite the second blow up space $\mathscr{T}^+$ as
\begin{align}
\mathscr{T}^+:xr= e_2 t
\end{align}
where
\begin{align}
&r(x,y,e_0,e_1)=a_1  y -e_1 x^2-a_{2,1} e_1 e_0 x -a_{4,2} e_1 e_0^2,\\
&t(y,e_0,e_1)=- y^2-a_{3,2} e_0^2   e_1y+a_{6,4} e_1^2e_0^4.
\end{align}
There is a conifold singularity at
\begin{align}
x=e_2 = a_1  y -a_{4,2} e_1 e_0^2=- y^2-a_{3,2} e_0^2  y e_1+a_{6,4} e_1^2e_0^4=0.
\end{align}
It has solution only if
\begin{align}
P_4=
-a_{4,2}^2 -a_1 a_{3,2}a_{4,2}+a_1^2a_{6,4}=0.\label{P4}
\end{align}
(In deriving the above equation we assumed $a_1\neq0$. However if we assume $a_1=0$, this implies $a_{4,2}=0$ so also satisfies the above condition.) Recall that $P_4$ is the leading term in the second component of the discriminant for $\mathscr{E}_0$ (see \eqref{P2n}).

There are two options for the third blow up: $\mathscr{T}^+_+:(x,e_2)$ and $\mathscr{T}^-:(r,e_2)$. Naively, one might also want to blow up along the ideal $(x,r,e_2)$. However, this resolution is not small. In fact, one of the fiber component is a $\mathbb{P}^2$ rather than a $\mathbb{P}^1$ (node). We will therefore not consider this possibility.

\subsubsection{Resolution $\mathscr{T}^+_+(x,e_2|e_3)$}

To resolve the conifold singularity, we blow up along the ideal $(x,e_2)$
\begin{align}
(x,e_2 )\rightarrow (e_3 x,e_3 e_2),
\end{align}
arriving at the resolved variety $\mathscr{T}^+_+$,
\begin{align}
\mathscr{E}_0\xleftarrow{(x,y,e_0|e_1)} \mathscr{E}_1\xleftarrow{(y,e_1|e_2)} \mathscr{T}^+ \xleftarrow{(x,e_2|e_3)} \mathscr{T}^+_+
\end{align}
\begin{align}
\mathscr{T}^+_+:
x ( a_1  y -e_1 e_3^2 x^2 -a_{2,1} e_1e_0 e_3 x -a_{4,2} e_1e_0^2 )
= e_2( - y^2 - a_{3,2} e_0^2  e_1y +a_{6,4} e_1^2 e_0^4).\label{t++}
\end{align}
The ambient space is
\begin{align}
\,[e_3^2e_2e_1x:e_3^2e_2^2e_1y:z=1][e_3x:e_3e_2y:e_0][y:e_1][x:e_2].
\end{align}
One can check that $\mathscr{T}^+_+$ is a nonsingular variety for $\dim_{\mathbb{C}}B\le 3$ so we do not need to do any further blow up.

\paragraph{Fiber enhancements}

The divisor now is blown up to be ${e_0e_1e_2e_3=0}$, over which we have the following four nodes
\begin{align}
\begin{split}
&C_0:~e_0= a_1 x y - e_1e_3^2 x^3 + e_2  y^2 =0,\\
&C_{1}:~e_1= a_1 x+e_2  y=0,\\
&C_2:~e_3= (a_1   y -a_{4,2} e_1 e_0^2)x+ e_2 ( y^2 +a_{3,2} e_0^2 e_1y -a_{6,4} e_1^2e_0^4)=0,\\
&C_3:~e_2= a_1 y -e_1e_3^2 x^2 -a_{2,1} e_1 e_0e_3 x -a_{4,2} e_1 e_0^2=0.
\end{split}
\end{align}
 From the intersections we recognize the fiber to be of type I$_4$. This is the affine Dynkin diagram for $SU(4)$.

We label the nodes $C_i$ by their position in the affine $SU(4)$ Dynkin diagram rather than the order of blow ups. This is for later convenience when we compare the fibers between different resolutions. Note that $C_1:e_1=s=0$, $C_2: e_3=0$ (the exceptional divisor for $e_2=x=0$), $C_3:e_2=r=0$ (the exceptional divisor for $e_1=y=0$).

Over the codimension two locus ${e_0e_1e_2e_3=P_4=0}$ but $a_1,a_{4,2}\neq0$, we have the following factorization
\begin{align}
y^2 +a_{3,2} e_0^2  e_1y - a_{6,4}e_1^2e_0^4 
= {1\over a_1a_{4,2}} (a_1 y -a_{4,2} e_0^2e_1) (a_{4,2}  y +a_1a_{6,4} e_0^2e_1)
\end{align}
in $C_2$. Hence $C_2$ becomes
\begin{align}
C_2:~e_3= (a_1 y -a_{4,2} e_0^2e_1)\left[a_1a_{4,2} x +e_2(a_{4,2}  y+a_1a_{6,4}e_0^2e_1)\right]=0.
\end{align}
That is, $C_2$ splits into two components, $C_4,C_5$
\begin{align}
&C_2\rightarrow C_4+C_5,\\
&~~~~C_4:~e_3= a_1 y -a_{4,2} e_0^2e_1=0,\\
&~~~~C_5:~e_3=a_1a_{4,2} x +e_2(a_{4,2}  y+a_1a_{6,4}e_0^2e_1)=0.
\end{align}
Including other fibers, we have the following five nodes in the fiber over $e_0e_1e_2e_3=P_4=0$
\begin{align}
\begin{split}
&C_0:~e_0= a_1 x  y - e_3^2e_1 x^3 + e_2y^2 =0,\\
&C_{1}:~e_1= a_1 x+e_2  y=0,\\
&C_4:~e_3= a_1 y -a_{4,2} e_0^2e_1=0,\\
&C_5:~e_3=a_1a_{4,2} x +e_2(a_{4,2} y+a_1a_{6,4}e_0^2e_1)=0,\\
&C_3:~e_2= a_1 y -e_1e_3^2 x^2 -a_{2,1} e_1e_0e_3 x -a_{4,2} e_1 e_0^2=0,\label{T++I5}
\end{split}
\end{align}
From the intersections we recognize the fiber to be of type I$_5$. This corresponds to the rank one enhancement $SU(4)\rightarrow SU(5)$ in codimension two.

Over the codimension two locus ${e_0e_1e_2e_3=a_1=0}$, $C_3$ agains splits into three components, $C_1, C_3^{(i)}$, $i=1,2$. The fibers are
\begin{align}
\begin{split}
&C_0:~e_0= - e_3^2e_1 x^3 + e_2  y^2 =0,\\
&C_{1}:~e_1= e_2 =0~~\text{(with multiplicity 2)},\\
&C_2:~e_3= -a_{4,2} e_1 e_0^2 x+ e_2 ( y^2 +a_{3,2} e_0^2 y e_1 -a_{6,4} e_1^2e_0^4)=0,\\
&C_3^{(i)}:~e_2= e_3^2 x^2 +a_{2,1}e_0 e_3 x +a_{4,2} e_0^2=0,~i=1,2.
\end{split}
\end{align}
From the intersections we see that it is the I$_0^*$ fiber. This corresponds to the rank one enhancement $SU(4)\rightarrow SO(8)$ in codimension two.

Over ${ e_0e_1e_2e_3=a_1=a_{2,1}^2-4a_{4,2}=0}$, $e_3^2 x^2 +a_{2,1} e_0e_3 x +a_{4,2} e_0^2=0$ has a double root so the two $C_3^{(i)}$ of  coincide, and we end up with the non-Kodaira I$_0^{*+}$ fiber.

Over the codimension two locus ${e_0e_1e_2e_3=a_1=a_{4,2}=0}$, $C_2$ splits into three components, $C_2',C_2^{(1)},C_2^{(2)}$
\begin{align}
\begin{split}
&C_2\rightarrow C_2'+C_2^{(1)}+C_2^{(2)},\\
&~~~~C_2':~e_2 =e_3=0,\\
&~~~~C_2^{(i)}:~e_3= y^2 +a_{3,2} e_0^2y e_1-a_{6,4} e_1^2e_0^4=0,~i=1,2.
\end{split}
\end{align}
$C_3$ splits into $C_1,C_3^{(i)}$, where $C_3^{(i)}$ become
\begin{align}
C_3^{(1)}:~e_2 = e_3x +a_{2,1} e_0=0~~\text{and}~~C_3^{(2)}=C_2':~e_2=e_3=0.
\end{align}
In total, we have the following nodes in the fiber over $e_0e_1e_2e_3=a_1=a_{4,2}=0$,
\begin{align}
\begin{split}
&C_0:~e_0= - e_3^2e_1x^3 + e_2 y^2 =0,\\
&C_{1}:~e_1 =e_2 =0~~\text{(with multiplicity 2)},\\
&C_2':~e_2=e_3 =0~~\text{(with multiplicity 2)},\\
&C_2^{(i)}:~e_3= y^2 +a_{3,2} e_0^2e_1y -a_{6,4} e_1^2e_0^4=0,~i=1,2,\\
&C_3^{(1)}:~e_2= e_3x +a_{2,1} e_0=0.
\end{split}
\end{align}
From the intersections we recognize the fiber to be of type I$_1^*$.

We summarize the fiber enhancements for $\mathscr{T}^+_+$ in Table \ref{T++}.

\begin{table}[htb]
\begin{center}
\scalebox{.75}{\begin{tabular}{|c|c|c|c|c|}
\hline
$w=0$& $w=P_4=0$ &  $w=a_1=0$ & $w=a_1=a_{4,2}=0$ & $w=a_1=a_{2,1}^2-4a_{4,2}=0$ \\
\hline
& {\footnotesize $C_2\to C_{4}+C_{5}$ }&
\scalebox{.8}{ \begin{tabular}{l}
$C_3\rightarrow C_1 + C_3^{(1)}+C_3^{(2)}$
\end{tabular} }
 & 
\scalebox{.8}{ \begin{tabular}{l}
$C_2\to C_{2}'+C_{2}^{(1)}+C_2^{(2)}$\\
$C_3\to C_1 +C_{3}^{(1)}+C_2' $
\end{tabular} }
 & 
\scalebox{.8}{\begin{tabular}{l}
$C_3\rightarrow C_1 + 2C_3^{(1)}$

\end{tabular} }
 \\
\hline
 I$_4$& I$_5$& I$_0^*$ & I$_1^*$ & I$_0^{*+}$\\
\hline
&&&&\\
\scalebox{.95}{\begin{tikzpicture}[every node/.style={circle,draw, minimum size= 5 mm}]
\node (C2)  at (90*4:1.4cm)  {\tiny $C_{3}$};
\node (C3) at (90*3:1.4cm) {\tiny $C_{2}$};
\node (C1) at (90*2:1.4cm) {\tiny $C_{1}$};
\node (C0) at (90*1:1.4cm) {\tiny $C_0$};
\draw (C1)--(C3)--(C2)--(C0)--(C1);
\end{tikzpicture}}& 
\scalebox{.95}{\begin{tikzpicture}[every node/.style={circle,draw, minimum size= 5 mm}]
\node (C2)  at (18+72*5:1.4cm)  {\tiny $C_{3}$};
\node (C4) at (18+72*4:1.4cm) {\scalebox{.7}{ $C_{4}$}};
\node (C3) at (18+72*3:1.4cm) {\scalebox{.7}{$C_{5}$}};
\node (C1) at (18+72*2:1.4cm) {\tiny $C_{1}$};
\node (C0) at (18+72*1:1.4cm) {\tiny $C_0$};
\draw (C1)--(C3)--(C4)--(C2)--(C0)--(C1);
\end{tikzpicture}}& 
\scalebox{1}{\begin{tikzpicture}[every node/.style={circle,draw, minimum size= 5 mm}, scale=.8]
\node (C2) at (0,0) { \small$2C_{1}$};
\node (C0) at (90-35:1.8cm) {\small $C_2$} ;
\node (C1) at (90+35:1.8cm) {\small $C_0$};
\node  (C3) at  (-90-35:1.8cm){ \tiny$C_3^{(1)}$};
\node  (C4) at (-90+35:1.8cm) { \tiny$C_{3}^{(2)}$};
\draw (C2)--(C0);
\draw (C2)--(C1);
\draw (C2)--(C3);
\draw (C2)--(C4);
\end{tikzpicture}}
& 
\scalebox{1} {\begin{tikzpicture}[every node/.style={circle,draw, minimum size= 4 mm}, scale=.8]
\node (C2) at (0,0) {\tiny $2C_{1}$};
\node (C5) at (90:-1.8cm) { \scalebox{.7}{$2C_{2}'$}};
\node (C0) at (90-35:1.8cm) { \tiny$C_3^{(1)}$} ;
\node (C1) at (90+35:1.8cm) { $C_0$};
\node [yshift=-1.8cm] (C3) at  (-90-35:1.8cm){ \tiny $C_2^{(1)}$};
\node [yshift=-1.8cm] (C4) at (-90+35:1.8cm) {\tiny $C_{2}^{(2)}$};
\draw (C2)--(C0);
\draw (C2)--(C1);
\draw (C2)--(C5);
\draw (C5)--(C3);
\draw (C5)--(C4);
\end{tikzpicture}}
&
\scalebox{1}{\begin{tikzpicture}[every node/.style={circle,draw, minimum size= 5 mm}, scale=.8]
\node (C2) at (0,0) { \small$2C_{1}$};
\node (C0) at (90-35:1.8cm) {\small $C_2$} ;
\node (C1) at (90+35:1.8cm) {\small $C_0$};
\node  (C3) at  (-90:1.8cm){ \tiny$2C_3^{(1)}$};
\draw (C2)--(C0);
\draw (C2)--(C1);
\draw (C2)--(C3);
\end{tikzpicture}}

\\ 
\hline
\end{tabular}}
\end{center}
\caption{The fiber enhancements for $\mathscr{T}^+_+\cong\mathscr{B}^+$ in the $SU(4)$ model. The fiber enhancements for $\mathscr{T}^-_+\cong\mathscr{B}^-$ are obtained by exchanging $C_1$ with $C_3$ from $\mathscr{T}^+_+\cong \mathscr{B}^+$. Here $w=e_0e_1e_2e_3$ {and $P_4=
-a_{4,2}^2 -a_1 a_{3,2}a_{4,2}+a_1^2a_{6,4}=0$}. }\label{T++}
\end{table}

\subsubsection{Resolution $\mathscr{T}^+_-:(r,e_2|\alpha,\beta)$}

Recall that $\mathscr{T}^+$ takes the following form with manifest conifold singularity,
\begin{align}
\mathscr{T}^+:~xr= e_2 t,
\end{align}
where $r=a_1  y -e_1 x^2-a_{2,1} e_1 e_0 x -a_{4,2} e_1 e_0^2$ and $t=- y^2-a_{3,2} e_0^2   e_1y+a_{6,4} e_1^2e_0^4$. Let us now explore the other option for the third blow up, $\mathscr{T}^+_-:(r,e_2|\alpha,\beta)$.

Rather than introducing the parameter $e_3$ for the third exceptional divisor, we will explicitly use the homogeneous coordinates $[\alpha:\beta]$\footnote{$\alpha,\beta$ are the projective coordinates that we would have called $\bar r , \bar e_2$ according to our notations. However, to simplify the notation, we will use $\alpha, \beta$ instead.} for the extra $\mathbb{P}^1$ we introduce for the third blow up.
The blow up space $\mathscr{T}^+_-$ can then be described by
\begin{align}
\mathscr{E}_0 \xleftarrow{(x,y,e_0|e_1)} \mathscr{E}_1 \xleftarrow{(y,e_1|e_2)} \mathscr{T}^+\xleftarrow{(r,e_2|\alpha,\beta)}\mathscr{T}^+_-
\end{align}
\begin{align}
\begin{split}
\mathscr{T}^+_-:~
\begin{cases}
&\alpha e_2 - \beta (a_1  y -e_1 x^2-a_{2,1} e_1 e_0 x -a_{4,2} e_1 e_0^2)=0,\\
&\alpha x + \beta (y^2 + a_{3,2} e_0^2e_1 y-a_{6,4}e_1^2e_0^4)=0,
\end{cases}
\end{split}
\end{align}
where the second equation is the collinear condition between $\alpha,\beta$ and $r,e_2$. The ambient space is parametrized by the following projective coordinates
\begin{align}
\, [e_2e_1x:e_2^2e_1 y:z=1][x:e_2y:e_0][y:e_1][\alpha:\beta].
\end{align}
One can check that $\mathscr{T}^+_-$ is a nonsingular variety so we do not need to do any further blow up.

\paragraph{Fiber enhancements}

Over the codimension one divisor ${e_0e_1e_2=0}$, we have the following four fibers
\begin{align}
\begin{split}
&C_0:~e_0=\alpha e_2 - \beta a_1 y+\beta e_1 x^2=\alpha x +\beta y^2 =0,\\
&C_1:~e_1=\alpha e_2 - \beta a_1 y=\alpha x +\beta y^2=0,\\
&C_2:~e_2=\beta=x=0,\\
&C_3:~e_2= a_1y -e_1( x^2+a_{2,1} e_0 x+a_{4,2} e_0^2)=\alpha x+\beta (y^2+a_{3,2} e_0^2e_1y-a_{6,4}e_1^2e_0^4)=0.
\end{split}
\end{align}
From the intersections we see that it is the I$_4$ fiber. Note that we label the $C_i$ in the same way as $\mathscr{T}^+_+$.

Over ${e_0e_1e_2=P_4=0}$ but $a_1,a_{4,2}$ being nonzero, $C_3$ becomes
\begin{align}
C_3:~e_2=a_1 y-e_1(x^2+a_{2,1} e_0x+a_{4,2}e_0^2)= 
a_1a_{4,2}\alpha x + \beta(a_1y-a_{4,2}e_0^2e_1)(a_{4,2}y+a_1a_{6,4}e_0^2e_1)=0.
\end{align} 
We can rewrite it as
\begin{align}
C_3:~&e_2=a_1 y-e_1(x^2+a_{2,1} e_0x+a_{4,2}e_0^2)\\&= 
a_1^2a_{4,2}\alpha x + \beta e_1^2 x(x+a_{2,1} e_0) \left[a_{4,2}x^2 +a_{4,2}a_{2,1} e_0x+(a_{4,2}^2+a_1^2a_{6,4})e_0^2\right]=0.
\end{align}
Hence $C_3$ splits into three components
\begin{align}
\begin{split}
&C_3\rightarrow C_4+C_5,\\
&~~~~C_4:~e_2= a_1 y- a_{4,2} e_1e_0^2= x=0,\\
&~~~~C_5:~e_2= a_1 y-e_1(x^2+a_{2,1} e_0x+a_{4,2}e_0^2)\\&~~~~~~~= 
a_1^2a_{4,2}\alpha  + \beta e_1 ^2(x+a_{2,1} e_0) \left[a_{4,2}x^2  +a_{4,2}a_{2,1} e_0x+(a_{4,2}^2+a_1^2a_{6,4})e_0^2\right]=0.
\end{split}
\end{align}
Hence we have five nodes, $C_0, C_1, C_2, C_4,C_5$, in the fiber over the codimension two locus $e_0e_1e_2=P_4=0$. From the intersections we recognize the fiber to be of type I$_5$. This corresponds to the rank one enhancement $SU(4)\rightarrow SU(5)$.

Over ${e_0e_1e_2=a_1=0}$, $C_3:~e_2 =e_1(x^2 + a_{2,1} e_0x +a_{4,2} e_0^2)=\alpha x + \beta (y^2+a_{3,2} e_0^2 e_1 y-a_{6,4} e_1^2e_0^4)=0$ splits into three components
\begin{align}
\begin{split}
&C_3\rightarrow C_1+C_3^{(i)},\\
&~~~~C_1:~e_2 = e_1 = \alpha x +\beta y^2=0,\\
&~~~~C_3^{(i)}:~e_2= x ^2+a_{2,1}e_0x+a_{4,2}e_0^2= \alpha x+ \beta(y^2+ a_{3,2} e_0^2e_1y -a_{6,4} e_1^2e_0^4)=0,~i=1,2,
\end{split}
\end{align}
where $C_3^{(i)}$ corresponds to the two roots of $x^2 +a_{2,1} e_0x+a_{4,2}e_0^2=0$.
In total, we have the following nodes in the fiber over ${e_0e_1e_2=a_1=0}$
\begin{align}
\begin{split}
&C_0:~e_0=\alpha e_2 +\beta e_1 x^2=\alpha x +\beta y^2 =0,\\
&2C_1:~e_1= e_2 =\alpha x +\beta y^2=0~~,\\
&C_2:~e_2=\beta=x=0,\\
&C_3^{(i)}:~e_2= x ^2+a_{2,1}e_0x+a_{4,2}e_0^2= \alpha x+ \beta(y^2+ a_{3,2} e_0^2e_1y -a_{6,4} e_1^2e_0^4)=0,~i=1,2.
\end{split}
\end{align}
From the intersections we recognize the fiber to be of type I$_0^*$.

Over ${e_0e_1e_2=a_1=a_{2,1}^2-4a_{4,2}=0}$, $C_3^{(1)}=C_3^{(2)}$ and it becomes the non-Kodaira I$_0^{*+}$ fiber.

Over the codimension two locus ${e_0e_1e_2=a_1=a_{4,2}=0}$, $C_3:~e_2=e_1x(x+a_{2,1}e_0)= \alpha x+\beta (y^2+a_{3,2} e_0^2e_1y-a_{6,4} e_1^2e_0^4)=0$ becomes
\begin{align}
\begin{split}
&C_3 \rightarrow C_1+C_2 + C_3^{(1)'}+C_3^{(2)'}+C_3',\\
&~~~~C_1:~e_1=e_2 =\alpha x+\beta y^2=0,\\
&~~~~C_2:~e_2 =x=\beta=0 ,\\
&~~~~C_3^{(i)'}:~e_2=x=y^2+a_{3,2}e_0^2 e_1 y-a_{6,4}e_1^2e_0^4=0,~\\
&~~~~C_3':~e_2 =x+a_{2,1}e_0= \alpha x +\beta (y^2+a_{3,2} e_0^2e_1y-a_{6,4}e_1^2e_0^4)=0. 
\end{split}
\end{align}
In total, we have the following nodes in the fiber over $e_0e_1e_2=a_1=a_{4,2}=0$,
\begin{align}
\begin{split}
&C_0:~e_0 = \alpha e_2+\beta e_1x^2 = \alpha x+\beta y^2=0,\\
&C_1:~[0:0:1][x:0:e_0][1:0][1:-x]~~~~\text{with multiplicity 2},\\
&C_2:~ [0:0:1][0:0:1][y:e_1][1:0]~~~~\text{with multiplicity 2},\\
&C_3^{(i)'}:~ [0:0:1][0:0:1][y^{(i)}:e_1^{(i)}][\alpha:\beta],\\
&C_3':~[0:0:1][-a_{2,1}:0:1][y:e_1][y^2+a_{3,2}e_1y-a_{6,4}e_1^2:a_{2,1}].
\end{split}
\end{align}
From the intersections we recognize the fiber to be of the type I$_1^*$.

We summarize the fiber enhancements for $\mathscr{T}^+_-$ in Table \ref{T+-}.

\begin{table}[htb]
\begin{center}
\scalebox{.75}{\begin{tabular}{|c|c|c|c|c|}
\hline
${\displaystyle w=0}$& $\displaystyle w=P_4=0$ &  $\displaystyle w=a_1=0$ & $\displaystyle w=a_1=a_{4,2}=0$ & $\displaystyle w=a_1=a_{2,1}^2-4a_{4,2}=0$ \\
\hline
& { $C_3\to C_{4}+C_{5}$ }&
\scalebox{.8}{ \begin{tabular}{l}
$C_3\rightarrow C_1 + C_3^{(1)}+C_3^{(2)}$
\end{tabular} }
 & 
\scalebox{.9}{ \begin{tabular}{l}
{$C_3\to C_1+C_2+C_3^{(1)'}+C_3^{(2)'}+C_3'$}
\end{tabular} }
 & 
\scalebox{.8}{\begin{tabular}{l}
$C_3\rightarrow C_1 + 2C_3^{(1)}$

\end{tabular} }
 \\
\hline
 I$_4$& I$_5$& I$_0^*$ & I$_1^*$ &  I$_0^{*+}$\\
\hline
&&&&\\
\scalebox{.95}{\begin{tikzpicture}[every node/.style={circle,draw, minimum size= 5 mm}]
\node (C2)  at (90*4:1.4cm)  {\tiny $C_{3}$};
\node (C3) at (90*3:1.4cm) {\tiny $C_{2}$};
\node (C1) at (90*2:1.4cm) {\tiny $C_{1}$};
\node (C0) at (90*1:1.4cm) {\tiny $C_0$};
\draw (C1)--(C3)--(C2)--(C0)--(C1);
\end{tikzpicture}}& 
\scalebox{.95}{\begin{tikzpicture}[every node/.style={circle,draw, minimum size= 5 mm}]
\node (C2)  at (18+72*5:1.4cm)  {\tiny $C_{5}$};
\node (C4) at (18+72*4:1.4cm) {\scalebox{.7}{ $C_{4}$}};
\node (C3) at (18+72*3:1.4cm) {\scalebox{.7}{$C_{2}$}};
\node (C1) at (18+72*2:1.4cm) {\tiny $C_{1}$};
\node (C0) at (18+72*1:1.4cm) {\tiny $C_0$};
\draw (C1)--(C3)--(C4)--(C2)--(C0)--(C1);
\end{tikzpicture}}& 
\scalebox{1}{\begin{tikzpicture}[every node/.style={circle,draw, minimum size= 5 mm}, scale=.8]
\node (C2) at (0,0) { \small$2C_{1}$};
\node (C0) at (90-35:1.8cm) {\small $C_2$} ;
\node (C1) at (90+35:1.8cm) {\small $C_0$};
\node  (C3) at  (-90-35:1.8cm){ \tiny$C_3^{(1)}$};
\node  (C4) at (-90+35:1.8cm) { \tiny$C_{3}^{(2)}$};
\draw (C2)--(C0);
\draw (C2)--(C1);
\draw (C2)--(C3);
\draw (C2)--(C4);
\end{tikzpicture}}
& 
\scalebox{1} {\begin{tikzpicture}[every node/.style={circle,draw, minimum size= 4 mm}, scale=.8]
\node (C2) at (0,0) {\tiny $2C_{1}$};
\node (C5) at (90:-1.8cm) { \scalebox{.7}{$2C_{2}$}};
\node (C0) at (90-35:1.8cm) { $C_3'$} ;
\node (C1) at (90+35:1.8cm) { $C_0$};
\node [yshift=-1.8cm] (C3) at  (-90-35:1.8cm){ \tiny $C_3^{(1)'}$};
\node [yshift=-1.8cm] (C4) at (-90+35:1.8cm) {\tiny $C_3^{(2)'}$};
\draw (C2)--(C0);
\draw (C2)--(C1);
\draw (C2)--(C5);
\draw (C5)--(C3);
\draw (C5)--(C4);
\end{tikzpicture}}
&
\scalebox{1}{\begin{tikzpicture}[every node/.style={circle,draw, minimum size= 5 mm}, scale=.8]
\node (C2) at (0,0) { \small$2C_{1}$};
\node (C0) at (90-35:1.8cm) {\small $C_2$} ;
\node (C1) at (90+35:1.8cm) {\small $C_0$};
\node  (C3) at  (-90:1.8cm){ \tiny$2C_3^{(1)}$};
\draw (C2)--(C0);
\draw (C2)--(C1);
\draw (C2)--(C3);
\end{tikzpicture}}

\\ 
\hline
\end{tabular}}
\end{center}
\caption{The fiber enhancements for $\mathscr{T}^+_-$ in the $SU(4)$ model. The fiber enhancements for $\mathscr{T}^-_-$ are obtained by exchanging $C_1$ with $C_3$ from $\mathscr{T}^+_-$. Here $w=e_0e_1e_2$ because we did not introduce $e_3$ for $\mathscr{T}^+_-$  {and $P_4=
-a_{4,2}^2 -a_1 a_{3,2}a_{4,2}+a_1^2a_{6,4}=0$}.}\label{T+-}
\end{table}

This completes the analysis from the partial resolution $\mathscr{T}^+$. In the following we will return to another option for the second blow up, $\mathscr{T}^-:(s,e_1)$.

\subsection{Partial resolution $\mathscr{T}^-:(s,e_1|e_2)$}

The partial resolution $\mathscr{T}^-$ is related to $\mathscr{T}^+$ by the inverse action (\ref{MW}) induced by the $\mathbb{Z}_2$ automorphism in the Mordell-Weil group, hence the analysis will be identical to $\mathscr{T}^+$ by exchanging $y$ with $-s = -y - a_1 x - a_{3,2}e_1 e_0^2$. Geometrically, this corresponds to switching the nodes $C_1:~e_1=s=0$ with $C_3:~e_2 =r =0$ in $\mathscr{T}^+$. Note that $C_3$ in $\mathscr{T}^+$ comes from $e_1=y=0$ in $\mathscr{E}_1$, so the inverse action indeed exchanges $y$ with $-s$. We will not repeat the details of the analysis as it is similar to $\mathscr{T}^+$.

The partial resolution $\mathscr{T}^-$ is 
\begin{align}
\mathscr{E}_0 \xleftarrow{(x,y,e_0|e_1)} \mathscr{E}_1 \xleftarrow{(s,e_1|e_2)}\mathscr{T}^-
\end{align}
\begin{align}
\begin{split}
\mathscr{T}^-:~
\begin{cases}
& y s =e_1 ( x^3 +a_{2,1} e_0 x^2 +a_{4,2} e_0^2 x +a_{6,4} e_2e_1 e_0^4)\\
& y+ a_1 x +a_{3,2} e_2e_1 e_0^2 =e_2 s,
\end{cases}
\end{split}
\end{align}
with the ambient space parametrized by
\begin{align}
\, [ e_2 e_1 x : e_2 e_1 y:z=1] [x: y:e_0] [s:e_1].
\end{align}

\paragraph{Description of the fiber}
Over $e_0e_1e_2=0$, the nodes in the fiber are
\begin{align}
\begin{split}
C_0 :~&e_0= ys - e_1 x^3 = y+a_1 x -e_2 s= 0,\\
C_3:~& e_1 =y= a_1 x  -e_2 =0,\\
&~~[0:0:1] [x:0:e_0] [1:0],\\
C_2:~& e_2 =x =y=0,\\
&~~[0:0:1][0:0:1][s:e_1],\\
C_1:~& e_2 = a_1 s +e_1(x^2+a_{2,1} e_0x+a_{4,2} e_0^2)=y+a_1x=0,\\
&~[0:0:1][x: -a_1 x :e_0][x^2+a_{2,1}e_0x+a_{4,2} e_0^2:-a_1].\label{affineSU4-}
\end{split}
\end{align}
Note that our labeling is consistent with that for $\mathscr{T}^+$. From the intersections we see that it is the I$_4$ fiber. Note that we already have the affine $SU(4)$ Dynkin diagram at the second blow up. On the gauge theory side, this corresponds to the fact that the wall $W^-$ lies in the bulk of the Coulomb branch.

\paragraph{Conifold singularity}
If we rewrite $\mathscr{T}^-$ as
\begin{align}
\mathscr{T}^-:~  - x(a_1 s +e_1 x^2 +a_{2,1} e_1e_0 x +a_{4,2} e_1 e_0^2)=e_2 (-s^2 +a_{3,2} e_0^2e_1 s+a_{6,4}e_1^2e_0^4).
\end{align} 
The conifold singularity is clearly at
\begin{align}
x = e_2 = a_1 s +a_{4,2} e_1e_0^2 =-s^2 +a_{3,2} e_0^2e_1 s+a_{6,4}e_1^2e_0^4=0
\end{align}
which can be satisfied only over $P_4=0$ on the base $B$.

\subsubsection{Resolution $\mathscr{T}^-_+:(x,e_2|e_3)$}

 One option for the third blow up is obtained as below
\begin{align}
\mathscr{E}_0 \xleftarrow{(x,y,e_0|e_1)} \mathscr{E}_1 \xleftarrow{(s,e_1|e_2)} \mathscr{T}^-\xleftarrow{(x,e_2|e_3)} \mathscr{T}^-_+
\end{align}
\begin{align}
\begin{split}
\mathscr{T}^-_+:
\begin{cases}
 & - x (a_1 s +e_1 e_3^2x^2 + a_{2,1} e_3e_1e_0x+a_{4,2} e_1e_0^2)
=e_2 (-s^2 +a_{3,2} e_0^2e_1 s+a_{6,4}e_1^2 e_0^4),\\
&y  =  e_3 e_2 s-a_1 e_3x - a_{3,2} e_3 e_2 e_1 e_0^2,
\end{cases}
\end{split}
\end{align}
in the ambient space parametrized by
\begin{align}
\,[e_3^2 e_2 e_1 x:e_3 e_2 e_1 y:z=1][e_3 x: y:e_0] [s:e_1] [x:e_2].
\end{align}

Over the divisor $e_0e_1e_2e_3=0$, the nodes in the fiber are
\begin{align}
\begin{split}
C_0:~& e_0=-a_1 xs - e_1e_3^2 x^3 + e_2 s^2 = y - e_3e_2 s +a_1 e_3 x =0,\\
C_3:~& e_1 = a_1 x -e_2s =y-e_3 e_2s +a_1 e_3 x=0,\\
C_2:~& e_3 = y= - x(a_1 s +a_{4,2} e_1 e_0^2) + e_2 ( s^2 -a_{3,2} e_0^2 e_1 s -a_{6,4} e_1^2e_0^4)=  0,\\
C_1:~& e_2 =a_1 s +e_1 e_3^2x^2 + a_{2,1} e_3e_1e_0x+a_{4,2} e_1e_0^2=0.
\end{split}
\end{align}

The fiber enhancements are the same as $\mathscr{T}^+_+$ after exchanging $C_1$ with $C_3$.

\subsubsection{Resolution $\mathscr{T}^-_-:(r,e_2|\alpha,\beta)$}

The other option for the third blow up is
\begin{align}
\mathscr{E}_0 \xleftarrow{(x,y,e_0|e_1)} \mathscr{E}_1 \xleftarrow{(s,e_1|e_2)} \mathscr{T}^-\xleftarrow{(r,e_2|\alpha,\beta)} \mathscr{T}^-_-
\end{align}
\begin{align}
\begin{split}
\mathscr{T}^-_-:
\begin{cases}
& \alpha e_2 +\beta(a_1 s +e_1 x^2 +a_{2,1} e_1e_0 x +a_{4,2} e_1 e_0^2)
=0,\\
 &\alpha x + \beta (s^2 -a_{3,2} e_0^2e_1 s-a_{6,4}e_1^2e_0^4)=0,\\
 &y = e_2 s -a_1 x -a_{3,2} e_2 e_1 e_0^2,
 \end{cases}
 \end{split}
 \end{align}
 in the ambient space parametrized by
\begin{align}
\, [e_2 e_1 x:e_2e_1y:z=1][x:y:e_0][s:e_1][\alpha:\beta].
\end{align}

Over the divisor $e_0e_1e_2=0$, the nodes in the fiber are
\begin{align}
\begin{split}
C_0 : ~&e_0= \alpha e_2 + \beta (a_1 s +e_1 x^2)= \alpha x +\beta s^2= y-e_2s +a_1 x=0,\\
C_3:~& e_1 = \alpha e_2+\beta a_1 s=\alpha x + \beta s^2=0,\\
C_2:~&e_2 = \beta =x =0,\\
C_1:~& e_2 = a_1 s +e_1 x^2 +a_{2,1} e_1e_0 x +a_{4,2} e_1 e_0^2=\alpha x+\beta (y^2-a_{3,2} e_0^2e_1y-a_{6,4}e_1^2e_0^4)=0.
\end{split}
\end{align}
The fiber enhancements are the same as $\mathscr{T}^+_-$ after exchanging $C_1$ with $C_3$.

Next we will return to the last option for the second blow up, $\mathscr{B}:(x,y,e_1)$. As we will see shortly, the  resolutions we obtain from this branch will be identified with those in $\mathscr{T}^\pm$.

\subsection{Partial resolution $\mathscr{B}:(x,y,e_1|e_2)$}

 Recall that after the first blow up we end up with the following conifold singularity:
\begin{align}
\mathscr{E}_1:~y s  = e_1 Q
\end{align}
where $s=y+a_1x+a_{3,2}e_1e_0^2$ and $Q= x^3 + a_{2,1} e_0 x^2 + a_{4,2} e_0^2 x +a_{6,4} e_1 e_0^4$.

In the previous sections we blow up along $(y,e_1)$ obtaining $\mathscr{T}^+$. Now we blow up $(x,y,e_1)$ instead,
\begin{align}
(x,y,e_1)\rightarrow (e_2 x,e_2y, e_2 e_1),
\end{align}
obtaining the partially resolved variety $\mathscr{B}$,
\begin{align}
\mathscr{E}_0 \xleftarrow{(x,y,e_0|e_1)} \mathscr{E}_1 \xleftarrow{(x,y,e_1|e_2)}\mathscr{B}
\end{align}
\begin{align}
\mathscr{B}:y(y+a_1 x+a_{3,2} e_1e_0^2) = e_1 (e_2 ^2x^3 + a_{2,1} e_0e_2 x^2 +a_{4,2} e_0^2 x+a_{6,4} e_1e_0^4).
\end{align}
The ambient space is parametrized by the following projective coordinates
\begin{align}
\,
[e_2^2e_1 x:e_2^2e_1y:z=1][e_2 x:e_2y :e_0][x:y:e_1].
\end{align}

\paragraph{Description of the fiber}

The divisor is blown up to be ${e_0e_1e_2=0}$, over which we have four nodes in the fiber,
\begin{align}
\begin{split}
&C_0:~e_0 = y(y+a_1 x)- e_1e_2^2x^3=0,\\
&C_1:~e_1=y+a_1x=0,\\
&C_1':~e_1=y=0,\\
&C_2:~e_2=y(y+a_1 x+a_{3,2}e_1e_0^2) - e_1  (a_{4,2} e_0^2x+a_{6,4}e_1e_0^4)=0.\label{affineSU4B}
\end{split}
\end{align}
From the intersections we see that it is the I$_4$ fiber. Note that we already have the affine $SU(4)$ Dynkin diagram at the second blow up. On the gauge theory side, this corresponds to the fact that the wall $W^0$ lies in the bulk of the Coulomb branch.

\paragraph{Conifold singularity}

There is a conifold singularity in $\mathscr{B}$,
\begin{align}
\mathscr{B}:y(y+a_1 x+a_{3,2} e_1e_0^2) = e_1 (e_2 ^2x^3 + a_{2,1} e_0e_2 x^2 +a_{4,2} e_0^2 x+a_{6,4} e_1e_0^4),
\end{align}
 located at
\begin{align}
y=e_1 = e_2^2x^2 + a_{2,1}e_2 e_0 x + a_{4,2}e_0^2=0
\end{align}
over the codimension two locus $e_0e_1e_2=a_1=0$. 
In the following we will continue to the blow up this singularity. There are two options for the third blow up: $\mathscr{B}^+:(y,e_1)$ and $\mathscr{B}^-$. As we will see shortly, $\mathscr{B}^+$ and $\mathscr{B}^-$ are isomorphic to $\mathscr{T}^+_+$ and $\mathscr{T}^-_+$, respectively.

\subsubsection{Resolution $\mathscr{B}^+:(y,e_1|e_3) $}\label{isomorphism}

Let us blow up along $(y,e_1)$
\begin{align}
(y,e_1)\rightarrow (e_3 y,e_3e_1)
\end{align}
to obtain $\mathscr{B}^+$,
\begin{align}
\mathscr{E}_0\xleftarrow{(x,y,e_0|e_1)} \mathscr{E}_1\xleftarrow{(x,y,e_1|e_2)} \mathscr{B} \xleftarrow{(y,e_1|e_3)} \mathscr{B}^+
\end{align}
\begin{align}
\mathscr{B}^+:
y(e_3 y + a_1 x+ a_{3,2} e_3e_1 e_0^2) = e_1 (e_2^2x^3 + a_{2,1}e_0e_2 x^2 +a_{4,2} e_0^2 x+a_{6,4} e_3 e_1 e_0^4)\label{b+}.
\end{align}
The ambient space is parametrized by the following projective coordinates
\begin{align}
\, [e_3e_2^2e_1x: e_3^2e_2^2e_1y :z=1] [e_2x :e_2e_3y:e_0][x:e_3y:e_3e_1][y:e_1].
\end{align}

\paragraph{The isomorphism $\mathscr{B}^+\cong\mathscr{T}^+_+$}

By comparing $\mathscr{B}^+$ (\ref{b+}) with $\mathscr{T}^+_+$ (\ref{t++}), we see that the two defining equations are the same by exchanging $e_2$ with $e_3$. To claim that the two varieties are actually isomorphic to each other, one needs to further check the scalings of each of the variable and the restrictions on vanishing of the variables. From the chains of blow ups for $\mathscr{T}^+_+$ and $\mathscr{B}^+$,
\begin{align}
\mathscr{E}_0\xleftarrow{(x,y,e_0|e_1)} \mathscr{E}_1\xleftarrow{(y,e_1|e_2)} \mathscr{T}^+ \xleftarrow{(x,e_2|e_3)} \mathscr{T}^+_+,
\end{align}
\begin{align}
\mathscr{E}_0\xleftarrow{(x,y,e_0|e_1)} \mathscr{E}_1\xleftarrow{(x,y,e_1|e_2)} \mathscr{B} \xleftarrow{(y,e_1|e_3)} \mathscr{B}^+,\label{f1f2}
\end{align}
we can read off the scaling for each variable with respect to the ambient projective spaces:
\begin{align}
\mathscr{T}^+_+ : 
\left.\begin{tabular}{|c|c|c|c|c|c|c|}
\cline{2-7}
\multicolumn{1}{c|}{} & $x$ & $y$ & $e_0$ & $e_1$ & $e_2$ & $e_3$ \\   \hline $\mathbb{P}^2$  & ~$1$~ & ~1~ &~ 0~ &~ 0~ & 0 & 0 \\\hline \text{1st} & ~1~ & ~1~ & ~1 ~& -1 & ~0~ &~ 0~ \\\hline  \text{2nd} & 0 & 1 & 0 & 1 & -1 & 0 \\\hline  \text{3rd}& 1 & 0 & 0 & 0 & 1 & -1\\
 \hline
 \end{tabular}\right.
\end{align}
\begin{align}
\mathscr{B}^+:
\left.\begin{tabular}{|c|c|c|c|c|c|c|} 
\cline{2-7}
\multicolumn{1}{c|}{} & $x$ & $y$ & $e_0$ & $e_1$ & $e_2$ & $e_3$ \\\hline  $\mathbb{P}^2$ & ~1~ & ~1~ &~ 0~ &~ 0~ & 0 & 0 \\\hline \text{1st} & ~1~ & ~1~ & ~1 ~& -1 & ~0~ &~ 0~ \\\hline  \text{2nd} & 1 & 1 & 0 & 1 & -1 & 0 \\\hline  \text{3rd}& 0 & 1 & 0 & 1 & 0 & -1\\ \hline\end{tabular}\right.
\end{align}
The second rows in both tables labeled by $\mathbb{P}^2$ are the scalings from the original projective bundle $\mathbb{P}(\mathscr{O}_B\oplus \mathscr{L}^2\oplus \mathscr{L}^3)$. Note that $\mathscr{T}^+_+$ and $\mathscr{B}^+$ differ only in the last two blow ups.

Now we are going to show that the two sets of scalings above are actually the same. First we exchange the last two rows in $\mathscr{T}^+_+$ and then add the last row to the second last row. Lastly, we exchange the last two columns. In the end we found that this is the same scaling table as $\mathscr{B}^+$. Hence the scalings for the two sets of variables are the same.

Lastly, we need to check that the restrictions on the vanishing of variables are the same for the two varieties. The ambient spaces in the two cases are parametrized by
\begin{align}
&\mathscr{T}^+_+:~[e_3^2 e_2 e_1 x:e_3^2 e_2^2 e_1 y:z=1][e_3 x:e_3e_2 y:e_0][y:e_1][x:e_2],\\
&\mathscr{B}^+:~ [e_3e_2^2 e_1  x: e_3^2 e_2^2e_1y :z=1][e_2 x:e_2e_3y:e_0][x:e_3y:e_3e_1][y:e_1].
\end{align}
(Remember that $e_2$ has to be exchanged with $e_3$ to make the comparison.) It is easy to see that both varieties have the same restrictions on the vanishing of variables. For example, we cannot have $x=e_2=0$ in $\mathscr{T}^+_+$ due to the projective space $[x:e_2]$. On the other hand, $x=e_3$ (corresponding to $x=e_2=0$ in $\mathscr{T}^+_+$) is forbidden in $\mathscr{B}^+$ by the projective space $[x:e_3y:e_3e_1]$.

In summary, since $\mathscr{T}^+_+$ and $\mathscr{B}^+$ have the same defining equations, the same scalings for the variables, and also the same restrictions on the vanishing of the variables, they are indeed isomorphic to each other,
\begin{align}
\mathscr{B}^+\cong\mathscr{T}^+_+.
\end{align}

\paragraph{The identification of resolutions $\mathscr{B}^+$ and $\mathscr{T}^+_+$}
Above we have seen that the two resolutions $\mathscr{B}^+$ and $\mathscr{T}^+_+$ are isomorphic to each other. Now we are going to show that these two resolutions should be identified as one resolution, hence corresponding to a single subchamber on the Coulomb branch, i.e. $\mathcal{C}^+_+$. 

Let us begin with a general discussion. Given a variety $X$ and two of its resolutions
\begin{align}
&f_1: ~X_1 \rightarrow X,\\
&f_2:~X_2 \rightarrow X.
\end{align}
Suppose the two resolutions are isomorphic to each other, $X_1\cong X_2$, we arrive at the, not necessarily commutative, diagram shown in Figure \ref{commute}.
\begin{figure}[htb]
\begin{center}
\begin{tikzcd}[scale=1]
X_1\arrow[rightarrow]{dr}[swap]{f_1}\arrow[rightarrow]{rr}{\varphi}[swap]{\cong}&      &X_2\arrow[rightarrow]{dl}{f_2}\\
&X    &
\end{tikzcd} 
\end{center}
\caption{If two resolutions $f_1:~X_1\rightarrow X$ and $f_2:~X_2\rightarrow X$ are isomorphic to each other and the above diagram commutes, we identify them as a single resolution.}\label{commute}\end{figure}
In the case that the diagram is commutative, i.e.
\begin{align}
f_1= f_2\circ \varphi,
\end{align}
we identify the two resolutions $X_1$ with $X_2$ since the blow up maps are the same.

Let us consider an example where $X_1\cong X_2$ but we do \textit{not} identify the two resolutions. Consider the conifold in $\mathbb{C}^4$,
\begin{align}
X:~x_1x_2-x_3x_4=0~~~~\text{in}~~\mathbb{C}^4,
\end{align}
and let
\begin{align}
\begin{split}
&X_1:~\alpha x_2 - \beta x_4 = \alpha x_1 -\beta x_3=0,\\
&X_2:~\alpha x_1 - \beta x_4 = \alpha x_2 -\beta x_3=0,
\end{split}
\end{align}
in $\mathbb{C}^4\times \mathbb{P}^1$ be the two resolutions. Here $[\alpha:\beta]$ are the homogeneous coordinates for $\mathbb{P}^1$. In these coordinates, the blow up maps $f_1$ and $f_2$ are the same,
\begin{align}
\begin{split}
f_1: ~ &X_1\rightarrow X,\\
&(x_1,x_2,x_3,x_4)[\alpha:\beta]  \mapsto (x_1,x_2,x_3,x_4),\\
f_2: ~&X_2\rightarrow X,\\
& (x_1,x_2,x_3,x_4)[\alpha:\beta]  \mapsto (x_1,x_2,x_3,x_4).
\end{split}
\end{align}
The two resolutions $X_1$ and $X_2$ are isomorphic to each other by the isomorphism $\varphi$ exchanging $x_1$ with $x_2$,
\begin{align}
\begin{split}
\varphi:~&X_1\rightarrow X_2\\
&(x_1,x_2,x_3,x_4)[\alpha:\beta]\mapsto (x_2,x_1,x_3,x_4)[\alpha:\beta].
\end{split}
\end{align}
However, $f_1$ is not the same as $f_2\circ\varphi$,
\begin{align}
\begin{split}
f_1:~& X_1 \rightarrow X,\\
&(x_1,x_2,x_3,x_4)[\alpha:\beta]  \mapsto (x_1,x_2,x_3,x_4),\\
f_2 \circ\varphi:~&X_1\rightarrow X,\\
&(x_1,x_2,x_3,x_4)[\alpha:\beta]  \mapsto (x_2,x_1,x_3,x_4).
\end{split}
\end{align}
Therefore we do not identify the two resolutions. In fact, they are related by a flop.

Now back to the case for $\mathscr{B}^+$ and $\mathscr{T}^+_+$. The isomorphism $\varphi:~\mathscr{B}^+\rightarrow \mathscr{T}^+_+$ is given by
\begin{align}
\varphi: (e_2 ,e_3 ) \mapsto (e_3,e_2)
\end{align}
with other coordinates kept fixed. Consider the blow up maps $f_1$ and $f_2$ for $\mathscr{B}^+$ and $\mathscr{T}^+_+$, respectively,
\begin{align}
\begin{split}
&f_1:~\mathscr{B}^+\rightarrow \mathscr{E}_0,\\
&f_2:~\mathscr{T}^+_+\rightarrow \mathscr{E}_0,
\end{split}
\end{align}
where $f_1$ and $f_2$ are both sequences of three blow ups shown in \eqref{f1f2}. 
However, since $e_2$ and $e_3$ are not variables in $\mathscr{E}_0$, they are projected out by $f_1$ and $f_2$. It follows that $f_1 = f_2\circ \varphi$, i.e. the following diagram is commutative
\begin{center}
\begin{tikzcd}[scale=2]
\mathscr{B}^+\arrow[rightarrow]{dr}[swap]{f_1=f_2\circ \varphi}\arrow[rightarrow]{rr}{\varphi}[swap]{\cong}&      &\mathscr{T}^+_+\arrow[rightarrow]{dl}{f_2}\\
&\mathscr{E}_0   &
\end{tikzcd} \end{center}
 and we identify the two resolutions (indicated by the blue line as in Figure \ref{tree4}),
\begin{center}
\begin{tikzcd}[column sep=huge]
\mathscr{B}^+ \arrow[leftrightarrow,blue, thick]{r} &\mathscr{T}^+_+.
\end{tikzcd}
\end{center}
The two resolutions therefore correspond to a single subchamber $\mathcal{C}^+_+$ on the Coulomb branch (see Figure \ref{SU4Coulomb}).

\subsubsection{Resolution $\mathscr{B}^-:(s,e_1|e_3)$}

The other option for the third blow up from the partial resolution $\mathscr{B}$ is $\mathscr{B}^-:(s,e_1)$,
\begin{align}
\mathscr{E}_0\xleftarrow{(x,y,e_0|e_1)} \mathscr{E}_1\xleftarrow{(x,y,e_1|e_2)} \mathscr{B} \xleftarrow{(s,e_1 | e_3)} \mathscr{B}^-
\end{align}
\begin{align}
\begin{split}
\mathscr{B}^-:
\begin{cases}
&( e_3 s -a_1 x-a_{3,2} e_3e_1 e_0^2 ) s = e_1 (e_2^2 x^3 +a_{2,1} e_0 e_2 x^2 +a_{4,2} e_0^2 x+a_{6,4}e_3 e_1 e_0^4),\\
& y = e_3 s -a_1 x-a_{3,2} e_3e_1 e_0^2,
\end{cases}
\end{split}
\end{align}
with the ambient space parametrized by
\begin{align}
\, [ e_3 e_2^2e_1 x :e_3 e_2^2e_1 y:z=1] [e_2 x : e_2 y :e_0 ] [x:y:e_3e_1] [s:e_1].
\end{align}

Since $\mathscr{B}^-$ and $\mathscr{T}^-_+$ are related to $\mathscr{B}^+$ and $\mathscr{T}^+_+$ by the inverse action (\ref{MW}) respectively and $\mathscr{B}^+\cong \mathscr{T}^+_+$, we immediately conclude that they should be identified as one single resolution,
\begin{center}
\begin{tikzcd}[column sep=huge]
\mathscr{B}^- \arrow[leftrightarrow,blue, thick]{r} &\mathscr{T}^-_+.
\end{tikzcd}
\end{center}

\end{document}